\documentclass{aa}  

\usepackage{graphicx}
\usepackage{txfonts}
\usepackage{multirow}
\usepackage{pythonhighlight}
\usepackage{longtable}
\usepackage{lscape}
\usepackage{hyperref}
\begin{document}

   \title{The GALAH survey: Improving chemical abundances using star clusters}

   \subtitle{}

   \author{Janez Kos\inst{1}
          \and
          Sven Buder\inst{2,3,4}
          \and
          Kevin L. Beeson\inst{1}
          \and
          Joss Bland-Hawthorn\inst{5,3}
          \and
          Gayandhi~M.~De~Silva\inst{6,13}
          \and
          Valentina D'Orazi\inst{7}
          \and
          Ken Freeman\inst{2,3}
          \and
          Michael Hayden\inst{8,9,5,3}
          \and
          Geraint F. Lewis\inst{5}
          \and
          Karin Lind\inst{10}
          \and
          Sarah L. Martell\inst{9,3}
          \and
          Sanjib Sharma\inst{11}
          \and
          Daniel B. Zucker\inst{12,13,3}
          \and
          Toma\v{z} Zwitter\inst{1}
          \and
          Gary S. Da Costa\inst{2,3}
          \and
          Richard de Grijs\inst{12,13,14}
          \and
        Madeline Howell\inst{15,3}
        \and
          Madeleine McKenzie\inst{2,3}
          \and
          Thomas Nordlander\inst{2, 3, 16}
          \and
          Dennis Stello\inst{9,5,3,17}
          \and
          Gregor Traven\inst{1}
          }

   \institute{Faculty of mathematics and physics, University of Ljubljana, Jadranska 19, 1000 Ljubljana, Slovenia\\
              \email{janez.kos@fmf.uni-lj.si}
         \and
    Research School of Astronomy and Astrophysics, Australian National University, Canberra, ACT 2611, Australia
    \and
    ARC Centre of Excellence for All Sky Astrophysics in 3 Dimensions (ASTRO 3D), Australia
    \and
    ACCESS-NRI, Australian National University, Canberra, ACT2601, Australia
    \and
    Sydney Institute for Astronomy, School of Physics, A28, The University of Sydney, NSW 2006, Australia
    \and
    Australian Astronomical Optics, Faculty of Science and Engineering, Macquarie University, Macquarie Park, NSW 2113, Australia
    \and
    Department of Physics, University of Rome Tor Vergata, via della ricerca scientifica 1, 00133, Rome, Italy
    \and
    Homer L. Dodge Department of Physics \& Astronomy, University of Oklahoma, 440 W. Brooks St., Norman, OK 73019, USA
    \and
    School of Physics, University of New South Wales, Sydney, NSW 2052, Australia
    \and
    Department of Astronomy, Stockholm University, AlbaNova University Centre, SE-106 91 Stockholm, Sweden
    \and
    Space Telescope Science Institute, 3700 San Martin Drive, Baltimore, MD, 21218, USA
    \and
    School of Mathematical and Physical Sciences, Macquarie University, Balaclava Road, Sydney, NSW 2109, Australia
    \and
    Astrophysics and Space Technologies Research Centre, Macquarie University, Balaclava Road, Sydney, NSW 2109, Australia
    \and
    International Space Science Institute Beijing, 1 Nanertiao, Zhongguancun, Beijing 100190, China
    \and
    School of Physics and Astronomy, Monash University, Clayton, VIC 3800, Australia
    \and
    Theoretical Astrophysics, Department of Physics and Astronomy, Uppsala University, Box 516, SE-751 20 Uppsala, Sweden
    \and
     Stellar Astrophysics Centre, Aarhus University, Ny Munkegade 120, DK-8000 Aarhus C, Denmark
}

   \date{Received September 15, 1996; accepted March 16, 1997}

 
  \abstract
  {Large spectroscopic surveys aim to consistently compute stellar parameters of very diverse stars while minimizing systematic errors. We explore the use of stellar clusters as benchmarks to verify the precision of spectroscopic parameters in the fourth data release (DR4) of the GALAH survey. We examine 58 open and globular clusters and associations to validate measurements of temperature, gravity, chemical abundances, and stellar ages. We focus on identifying systematic errors and understanding
trends between stellar parameters, particularly temperature and chemical abundances. We identify trends by stacking measurements of chemical abundances against effective temperature and modelling them with splines. We also refit spectra in three clusters with the Spectroscopy Made Easy and Korg packages to reproduce the trends in DR4 and to search for their origin by varying temperature and gravity priors, linelists, and spectral continuum. Trends are consistent between clusters of different ages and metallicities, can reach amplitudes of ~0.5 dex and differ for dwarfs and giants. We use the derived trends to correct the DR4 abundances of 24 and 31 chemical elements for dwarfs and giants, respectively, and publish a detrended catalogue. While the origin of the trends could not be pinpointed, we found that: i) photometric priors affect derived abundances, ii) temperature, metallicity, and continuum levels are degenerate in spectral fitting, and it is hard to break the degeneracy even by using independent measurements, iii) the completeness of the linelist used in spectral synthesis is essential for cool stars, and iv) different spectral fitting codes produce significantly different iron abundances for stars of all temperatures. We conclude that clusters can be used to characterise the systematic errors of parameters produced in large surveys, but further research is needed to explain the origin of the trends.}

   \keywords{methods: data analysis --
                techniques: spectroscopic --
                surveys --
                stars: abundances --
                open clusters and associations: general --
                globular clusters: general
               }

   \maketitle
%
\section{Introduction}

Large spectroscopic surveys of stars observe hundreds of thousands or even millions of stars, and these numbers are expected to increase significantly in the coming years with the commissioning of new, highly multiplexed instruments like WEAVE \citep{jin24} and 4MOST \citep{jong19}. The scientific goals of these surveys encompass a wide range of topics in stellar physics and Galactic science, necessitating observations of a diverse population of stars found throughout the Galaxy and beyond. 

It is the job of the analysis pipelines to derive stellar parameters from diverse spectra consistently and homogeneously. Parameters for different stellar types and spectral quality must be comparable with minimal systematic uncertainties. The performance of these pipelines is typically monitored using a set of benchmark stars —- stars that have high-quality, well-studied, spectra and whose parameters are measured independently through various methods \citep[e.g.][]{heiter15, jofre18, adibekyan20}. Typically, a relatively small number of benchmark stars are observed with each instrument to assess the accuracy of the instruments and the analysis process, as well as to determine the zero points for certain parameters and to homogenize them across different surveys or instruments when needed. However, benchmark stars can rarely measure the precision of derived parameters, which would require a statistically large sample of benchmark stars that are representative of the whole survey. Additionally, benchmark stars are not suitable for probing the influence of atmospheric models, linelists, and algorithms in spectral fitting codes if both benchmark stars and science targets are analyzed using the same analysis approach. While spectral fitting codes (such as SME \citep{valenti96, piskunov17} and Turbospectrum \citep{plez12, gerber23}), atmospheric models \citep[e.g.][]{gustafsson08, magic13} and linelists \citep{smith21, heiter21} have been well tested, any issues affecting both benchmark stars and science stars alike might go undetected, or worse, misinterpreted as genuine physical phenomena.

In this work, we explore the use of open clusters, young associations, and some globular clusters as benchmarks to derive more precise stellar parameters, specifically focusing on the abundances of chemical elements. Star clusters as benchmarking objects are complementary to the benchmark stars. Stars in clusters, particularly in open clusters, are resolved conatal, coeval populations, for which we can assume high level of chemical homogeneity \citep{krumholz19, kos21}. We can then test whether the spectral fitting codes indeed produce consistent values of abundances of chemical elements across a large range of temperatures and stellar types. This approach is particularly beneficial, as we can often observe more stars in a single cluster than the total number of benchmark stars available in the entire survey. We can independently test the consistency of measured abundances across various stellar types and temperature ranges, which is not feasible for benchmark stars. Because open clusters are chemically homogeneous \citep{bovy16, poovelil20, sinha24} they are best suited for our analysis. Many globular clusters, however, contain multiple stellar populations \citep{milone22, gratton19} and thus we cannot assume they are chemically homogeneous. They can still be used to search for trends in the analysis, if those trends are consistent across multiple populations or if they are larger than the chemical inhomogeneities. Stars in very young clusters or associations can also show chemical patterns that resemble systematic trends, for example, due to fast rotation, strong magnetic fields or pre-main-sequence evolution \citep{spina20}. Even for open clusters, we cannot assume that stars of all types will show chemical homogeneity for all elements due to atomic diffusion \citep{dotter17, moedas22}, convective motion \citep{grassitelli15} or dredge-up effects \citep{smith90, loaiza23}. Hence, we only analyse systematic trends that are consistent in clusters in a large age and metallicity range and we take caution for elements like Li, C, N, and O.

Using the observations of clusters, we evaluate the precision of the recent fourth data release (DR4) of the GALAH survey \citep{buder24}. It provides measurements of abundances of up to 32 elements from spectra of $917\,588$ stars. The main goal of the survey is to measure the abundances of chemical elements in a wide range of Galactic stars to support studies in the fields of stellar physics and galactic archaeology \citep{desilva15, martell17}. In DR4, we observe trends -- systematic patterns -- when plotting the abundances of chemical elements against effective temperature. Similar trends were observed recently in other surveys using visible and near infrared spectra, low and high resolution \citep{casali20, fu22, magrini23, grilo24, reyes24}. Non-physical discrepancies between the abundances measured in giants and dwarfs were observed as well \citep{dutra16, delponte24}. This study aims to characterize these trends, assess their consistency, and explore potential empirical origins. We also publish a detrended catalogue of abundances of chemical elements.

In addition to examining the precision and systematic errors in the measurements of chemical element abundances, we investigate the discrepancies in effective temperatures, surface gravities, and ages of stars as reported in DR4. In clusters, these can be measured very precisely, due to high-quality photometric and astrometric observations of almost all GALAH stars, and because coeval cluster stars must lie on the same isochrone, which can be accurately determined from a large number of stars in a cluster \citep[see previous work based on GALAH DR4 in][]{beeson24}. The systematic discrepancies between temperatures, gravities, and ages determined from isochrones compared to those from spectra may originate from similar sources as the trends observed in the measurements of chemical abundances -- which is the hypothesis explored in this work. While we present several issues regarding the parameters in GALAH DR4, it is important to note that significant improvements have been made since the third data release in terms of data reduction and analysis. This has resulted in improved precision, accuracy and new parameters reported in DR4.

This paper is structured as follows. Section \ref{sec:data} presents the spectroscopic parameters from GALAH DR4 and clusters that are used in this work. In Section \ref{sec:deriving_trends}, we derive the trends in the measurements of the abundances of chemical elements and explore their potential origins in Section \ref{sec:origins}. Then, we dedicate two sections to the verification of stellar ages reported in GALAH DR4 (Section \ref{sec:age_verification}) and temperature and surface gravity (Section \ref{sec:teff_verification}). The paper concludes with some discussions and implications of this study in Section \ref{sec:discussion}. Functional forms of the derived trends and instructions on how to use them are given in the appendix.

\section{Data}
\label{sec:data}

In this work we combine the spectroscopic observations and parameters obtained in the GALAH survey with parameters of open clusters, young associations, and globular clusters sourced from the literature. 

\subsection{Spectroscopic data and GALAH DR4}

The GALAH fourth data release is the latest by the GALAH collaboration. Stars are observed with the 3.9 m AAT telescope at the Siding Spring Observatory in Australia with the 2dF instrument \citep{lewis02}. Around 350 targets we observe simultaneously, with spectra obtained with the HERMES spectrograph \citep{sheinis15}. The spectrograph consists of four arms, producing spectra in roughly $250\,\text{\AA}$ wide bands in blue, green, red, and infra-red channels. The wavelength ranges of the channels were selected such that they cover prominent absorption lines of as many chemical elements as possible. Hydrogen's $H\alpha$ and $H\beta$ Balmer lines are also included, as well as several interstellar lines (KI doublet and some strong diffuse interstellar bands). The same setup is used by several observing programs, the largest being the main GALAH survey, which we complement with dedicated observing programs of clusters, young associations, and K2 targets, turn-off stars, bright fields observed during twilight, and some public data produced by third party observers throughout the last ten years of operations. Public data were only used if the observers did not deviate from the observing procedures required by the GALAH survey (choice of calibration fields, targeting constraints, magnitude limits etc., see \citep{buder24}). All these observations are combined in GALAH DR4. They are reduced with a common pipeline dedicated to the requirements of the GALAH survey \citep{kos17, buder21, buder24}. The pipeline produces reduced and normalised spectra, and calculates the first estimates for stellar radial velocities and basic spectroscopic parameters (temperature, gravity, metallicity, rotational broadening, and microturbulence velocity). The pipeline also produces a resolution profile for each observed spectrum. 

The reduced data are processed by an analysis pipeline. The main component of the pipeline is the fitting of observed spectra with synthetic spectra. The difference is minimised between an observed and synthesised spectrum with some labels, which represent the parameters we are trying to fit. The labels (including abundances of 31 chemical elements) are fitted across the whole wavelength range simultaneously. To make this computationally feasible, synthetic spectra are produced only on a grid of discrete $T_\mathrm{eff}$, $\log g$, and $[\mathrm{Fe/H}]$ using \textsc{Spectroscopy Made Easy (SME)}\citep{valenti96, piskunov17}, \textsc{MARCS} atmosphere models \citep{gustafsson08}, a linelist adapted from \citet{heiter21}, \citet{gravesse07} Solar abundances, and non-LTE departure coefficients for elements H, Li, C, N, O, Na, Mg, Al, Si, K, Ca, Mn, and Ba \citep{amarsi20}. A neural network similar to \textsc{The Payne}\citep{ting19} is then used to interpolate spectra from a grid to any (physically sensible) set of labels. With this approach, a synthetic spectrum can be produced in a fraction of a second, which enables a quick global fitting process. On the fly continuum normalisation and resolution matching is done while fitting.

First, the individual observations are analysed, which produces an initial set of labels (dataset with \texttt{allspec} in the name of GALAH DR4 products). Radial velocities are inspected, and observations of different epochs are then combined into a single spectrum. SB2 binaries are flagged, and SB1 binaries have spectra shifted based on their measured radial velocities. Combined spectra are re-analysed, this time also by including photometric information to constrain $\log g$ and help resolve degeneracies when fitting many labels at once. Labels derived for combined spectra (dataset with \texttt{allstar} in the name of GALAH DR4 products) are the ones we use in this work. This means that we use only one set of labels for each star, regardless of the number of repeated observations. Any subsequent quality cuts used in this work are described further in the text. 

\subsection{Open cluster data}

We use \citet{cantat20} as a primary source of stars in open clusters, which we append with \citet{hunt24}. We find both catalogues to be very reliable and sufficiently extensive, however, for some clusters one catalogue includes members that the other does not. We made a cross match between these catalogues and GALAH DR4 stars to obtain a list of 3353 stars in 93 open clusters. Most clusters contain only a small number of stars observed in the GALAH survey. These were observed serendipitously throughout the survey. A small fraction of the clusters were observed intentionally, and they contain a larger number of observations.  Here we only consider clusters with five or more observed stars. A list of open clusters used in this work is given in Table \ref{tab:open_clusters}. 

\begin{table}
 \caption{Open clusters used in this work. $N_\mathrm{lit}$ is the number of members in the literature (\citet{cantat20} or \citet{hunt24}), $N_\mathrm{GALAH}$ is the number of members observed in the GALAH survey. We give the number of all stars, and a number of stars with unflagged spectra in the GALAH survey. Age and extinction ($A_\mathrm{V}$) are obtained from \citet{cantat20} or \citet{hunt24} if cluster is not in the former catalogue.}
    \setlength{\tabcolsep}{4pt}
    \renewcommand{\arraystretch}{0.95}
    \begin{tabular}{lcccccc}
    \hline
        Cluster & $N_\mathrm{lit}$ & \multicolumn{2}{c}{$N_\mathrm{GALAH}$} & age & $A_\mathrm{V}$ & notes \\\hline
        & & all & no flag & [Myr] & [mag] & \\\hline
ASCC 16 &175& 69 & 20 &13&0.20& $^\ast$\\
ASCC 19 &149& 40 & 9 &10&0.13& $^\ast$\\
ASCC 20 &194& 45 & 11 &12&0.11& $^\ast$\\
ASCC 21 &90& 42 & 8 &8&0.10& $^\ast$\\
ASCC 58 &133& 5 & 3 &52&0.25& \\
ASCC 99 &63& 7 & 5 &354&0.47& \\
Alessi 24 &156& 20 & 5 &72&0.34& \\
Alessi 44 &175& 11 & 1 &223&0.71& \\
Alessi 9 &183& 5 & 4 &281&0.00& \\
Alessi-Teutsch 12 &104& 5 & 1 &104&0.20& \\
BH 99 &371& 5 & 2 &95&0.34& \\
Berkeley 32 &361& 23 & 14 &4897&0.34& 2\\
Berkeley 33 &126& 10 & 1 &234&1.75& \\
Berkeley 73 &45& 9 & 2 &1412&0.69& \\
Blanco 1 &380& 70 & 47 &104&0.01& 1\\
Briceno 1 &171& 9 & 4 &8&0.18& $^\ast$\\
Collinder 110 &881& 11 & 6 &1819&1.14& \\
Collinder 135 &324& 44 & 26 &26&0.01& $^\dagger$, 1\\
Collinder 140 &124& 30 & 16 &26&0.00& $^\dagger$, 1\\
Collinder 261 &1799& 245 & 111 &6309&0.81& 2\\
Collinder 277 &195& 10 & 3 &977&0.72& \\
Collinder 338 &63& 6 & 2 &144&0.45& \\
Collinder 359 &269& 11 & 1 &37&0.51& 1\\
Collinder 69 &620& 131 & 38 &12&0.25& $^\ast$, 1\\
Collinder 74 &107& 6 & 3 &1905&0.85& \\
CWNU 1044 &377& 6 & 3 &34&0.15& $^\dagger$\\
ESO 521-38 &84& 8 & 1 &208&0.83& \\
FSR 1197 &140& 5 & 1 &145&0.14& \\
FSR 1361 &199& 6 & 1 &288&1.73& \\
Gulliver 6 &318& 65 & 16 &16&0.25& $^\ast$, 1\\
HSC 1865 &246& 5 & 1 &7&0.16& \\
HSC 2505 &119& 27 & 1 &6&0.21& $^\ddagger$\\
HSC 2636 &284& 33 & 2 &9&0.30& $^\ddagger$\\
HSC 2733 &165& 22 & 1 &9&0.14& $^\ddagger$\\
HSC 2907 &349& 134 & 19 &9&0.55& $^\ddagger$\\
HSC 2919 &175& 95 & 7 &27&3.22& $^\ddagger$\\
IC 2391 &222& 31 & 11 &28&0.04& 1\\
IC 2395 &291& 102 & 1 &20&0.51& \\
IC 2602 &311& 13 & 6 &36&0.03& $^\ddagger$, 1\\
IC 4651 &806& 22 & 16 &1659&0.42& \\
IC 4665 &167& 30 & 15 &33&0.45& 1\\
IC 4756 &466& 8 & 8 &1288&0.29& \\
LP 2117 &526& 23 & 4 &630&2.08& \\
LP 5 &342& 8 & 5 &2691&0.27& \\
Mamajek 4 &180& 15 & 12 &371&0.31& 1\\
Melotte 101 &472& 10 & 2 &165&0.91& \\
Melotte 22 &952& 113 & 59 &77&0.18& 1\\
Melotte 25 &515& 94 & 37 &794&0.00& 1\\
NGC 1647 &604& 9 & 1 &363&0.64& \\
NGC 1750 &417& 11 & 1 &257&0.79& \\
NGC 1817 &413& 34 & 30 &1122&0.59& 2\\
NGC 1901 &69& 23 & 9 &891&0.21& \\
NGC 1977 &115& 16 & 3 &97&1.64& $^\ast$\\
NGC 1980 &121& 37 & 18 &13&0.70& $^\ast$\\
NGC 2112 &687& 42 & 31 &2089&1.71& 2\\
NGC 2184 &93& 8 & 2 &645&0.48& \\
\hline
\end{tabular}
\end{table}

\begin{table}
\addtocounter{table}{-1}
 \caption{Contd.}
    \setlength{\tabcolsep}{4pt}
    \renewcommand{\arraystretch}{0.95}
    \begin{tabular}{lcccccc}
    \hline
        Cluster & $N_\mathrm{lit}$ & \multicolumn{2}{c}{$N_\mathrm{GALAH}$} & age & $A_\mathrm{V}$ & notes \\\hline
        & & all & no flag & [Myr] & [mag] & \\\hline
NGC 2204 &531& 98 & 25 &2089&0.01& 1, 2\\
NGC 2232 &188& 22 & 4 &17&0.01& \\
NGC 2243 &506& 8 & 6 &4365&0.02& 2\\
NGC 2354 &265& 13 & 2 &1412&0.35& \\
NGC 2360 &680& 15 & 4 &1023&0.39& \\
NGC 2451A &331& 26 & 13 &35&0.00& 1\\
NGC 2451B &283& 61 & 31 &40&0.18& $^\dagger$, 1\\
NGC 2506 &1468& 5 & 5 &1659&0.09& \\
NGC 2516 &652& 125 & 28 &239&0.11& 1\\
NGC 2539 &485& 22 & 10 &691&0.11& 2\\
NGC 2548 &454& 60 & 8 &389&0.15& 1\\
NGC 2632 &685& 136 & 91 &676&0.00& 1\\
NGC 2670 &191& 7 & 1 &102&1.08& \\
NGC 2671 &123& 40 & 1 &331&2.31& \\
NGC 2682 &598& 243 & 186 &4265&0.07& 1, 2\\
NGC 3114 &1222& 20 & 3 &144&0.27& \\
NGC 3324 &41& 6 & 1 &10&1.19& \\
NGC 5460 &148& 25 & 1 &158&0.30& \\
NGC 5617 &565& 17 & 1 &104&1.55& \\
NGC 5662 &238& 6 & 1 &199&0.60& \\
NGC 5822 &634& 23 & 2 &912&0.39& \\
NGC 6124 &1273& 45 & 5 &190&1.93& 1\\
NGC 6208 &314& 10 & 3 &1412&0.66& \\
NGC 6253 &324& 39 & 22 &3235&0.78& 2\\
NGC 6281 &497& 9 & 5 &512&0.30& \\
NGC 6530 &680& 6 & 1 &3&1.48& \\
NGC 6583 &170& 18 & 7 &1202&1.52& 2\\
NGC 6705 &1183& 10 & 1 &309&1.20& \\
NGC 6755 &217& 6 & 1 &144&2.34& \\
OCSN 100 &87& 59 & 2 &3&0.41& $^\ddagger$\\
OCSN 61 &147& 16 & 8 &11&0.19& $^\ast$\\
OCSN 65 &70& 6 & 1 &13&0.36& $^\ast$\\
OCSN 96 &223& 134 & 17 &4&0.40& $^\ddagger$\\
OC 0354 &83& 5 & 1 &238&1.60& \\
OC 0470 &415& 25 & 10 &7&0.26& $^\dagger$\\
Pozzo 1 &341& 14 & 8 &9&0.07& $^\dagger$\\
Ruprecht 1 &115& 7 & 1 &288&0.61& \\
Ruprecht 121 &330& 7 & 2 &177&2.54& \\
Ruprecht 145 &219& 17 & 2 &812&0.55& \\
Ruprecht 147 &169& 87 & 70 &3019&0.06& 1, 2\\
Ruprecht 171 &719& 5 & 3 &2754&0.68& \\
Sigma Ori &114& 40 & 7 & &nan& $^\ast$\\
Trumpler 10 &446& 101 & 60 &32&0.00& 1\\
Trumpler 20 &738& 29 & 10 &1862&0.88& 2\\
UBC 17a &180& 66 & 20 &18&0.80& $^\ast$, 1\\
UBC 17b &103& 31 & 7 &11&0.05& $^\ast$\\
UBC 199 &55& 14 & 4 &1148&0.93& \\
UPK 12 &47& 9 & 1 &199&0.22& \\
UPK 436 &98& 7 & 1 &25&0.23& \\
UPK 45 &50& 5 & 1 &104&0.83& \\
UPK 50 &97& 11 & 1 &218&1.07& \\
UPK 579 &110& 11 & 2 &120&0.11& \\
UPK 606 &63& 29 & 1 & &nan& $^\ddagger$, 1\\
UPK 612 &235& 36 & 18 &100&0.05& 1\\
UPK 640 &616& 177 & 17 &25&0.40& $^\ddagger$, 1\\
\hline
    \end{tabular}
    \\$^\ast$ Cluster is also included in Ori OB1\\
    $^\dagger$ Cluster is also included in Vela OB2\\
    $^\ddagger$ Cluster is also included in Sco-Cen-Cru association\\
1: Cluster was used for calculating trends (dwarfs)\\
2: Cluster was used for calculating trends (giants)
    \label{tab:open_clusters}
\end{table}

\subsection{Globular cluster data}

Here we use the same selection of globular cluster stars as in the main DR4 paper \citep{buder24}. Stars were searched for in a six-dimensional parameter space (sky positions, distances, and proper motions obtained from \textit{Gaia} DR3, and radial velocity from GALAH DR4) around the positions of known globular clusters sourced from \citet{vasiliev21}. Stars in 28 globular clusters were identified. The summary of globular clusters with more than 10 observed stars, used in this work, is listed in Table \ref{tab:globular_clusters}. For the purpose of this work, we use globular cluster ages from \citet{franch09} (Table 4, column $\mathrm{G00_{CG}}$) and from \citet{cabrera22}. Typical uncertainties are 0.6 Gyr. We also use metallicities from \citet{schiavon24, massari17, omalley18, monaco18, munoz21}.

\begin{table}
 \caption{Globular clusters used in this work. $N_\mathrm{GALAH}$ is the number of members observed in the GALAH survey. Age and metallicity are literature values.}
 \renewcommand{\arraystretch}{0.95}
    \begin{tabular}{lccccc}
    \hline
        Cluster & \multicolumn{2}{c}{$N_\mathrm{GALAH}$} & age$^\mathsection$ & metallicity & notes \\\hline
        & all & no flag & [Gyr] & $[Fe/H]$ & \\\hline
        47 Tuc & 568 & 417 & 13.4 & -0.74$^\ast$ & 1\\
        E 3 & 16 & 5 & 12.8 & -0.89$^\parallel$ & \\
        NGC 288 & 107 & 58 & 10.6 & -1.27$^\ast$ & 1\\
        NGC 362 & 52 & 47 & 10.4 & -1.11$^\ast$ & 1\\
        NGC 1261 & 32 & 10 & 10.2 & -1.37$^{\ast\ast}$ & \\
        NGC 1851 & 20 & 13 & 9.6 & -1.13$^\ast$ & 1\\
        NGC 2808 & 53 & 42 & 10.9 & -1.07$^\ast$ & \\
        NGC 3201 & 62 & 1 & 10.2 & -1.39$^\ast$ & \\
        NGC 5904 & 86 & 54 & 10.6 & -1.21$^\ast$ & \\
        NGC 6121 & 295 & 226 & 12.5 & -1.07$^\ast$ & 1\\
        NGC 6254 & 14 & 11 & 11.4 & -1.51$^\ast$ & \\
        NGC 6362 & 12 & 5 & 13.7 & -1.07$^\dagger$ &\\
        NGC 6397 & 195 & 91 & 12.7 & -2.02$^\ast$ & 1\\
        NGC 6544 & 78 & 51 & 11.8$^\P$ & -1.52$^\ast$ & 1\\
        NGC 6656 & 238 & 186 & 12.7 & -1.70$^\ast$ & 1\\
        NGC 6752 & 123 & 99 & 11.8 & -1.47$^\ast$ & \\
        NGC 6809 & 63 & 62 & 12.3 & -1.76$^\ast$ & \\
        NGC 7089 & 12 & 0 & 11.8 & -1.47$^\ast$ & 1\\
        NGC 7099 & 24 & 23 & 12.9 & -2.29$^\ddagger$ & 1\\
        $\omega$ Cen & 264 & 247 & 11.4 & -1.60$^\ast$ &\\\hline
    \end{tabular}
    \\1: Cluster was used for calculating trends (giants)\\
    $^\ast$ \citet{schiavon24}\\
    $^\dagger$\citet{massari17}\\
    $^\ddagger$\citet{omalley18}\\
    $^\mathsection$\citet{franch09}\\
    $^\P$\cite{cabrera22}\\
    $^\parallel$\citet{monaco18}\\
    $^{\ast\ast}$\citet{munoz21}
    \label{tab:globular_clusters}
\end{table}

\subsection{Young association data}

We treat stars in young associations separately from open clusters, although some stars in young associations are assembled into clusters included as part of open clusters in Table \ref{tab:open_clusters}. Many stars in young associations are not bound into dense clusters, as they have dispersed significantly since their birth. They can still be identified as members of the association, which can also be partitioned into smaller building blocks \citep[e.g.][]{zari18, chen20, ratzenbock23}. These, however, do not necessarily satisfy the conditions for being open clusters, and are thus not properly surveyed in open cluster catalogues. We make no distinction between open clusters and associations, apart from using additional catalogues to cross-match stars in young associations. Our main focus is on the elemental abundances and ages of the stars, and if we can confirm the common origin of stars, it is irrelevant whether they are part of a gravitationally bound cluster or a dispersing association. 

GALAH observed numerous stars in three nearby young associations (Orion OB1, Vela OB2, and the Scorpius-Centaurus-Crux). We use third party catalogues, which we cross-match with GALAH observations to identify stars in the associations. The selected catalogues were also required to provide subdivisions of the association stars into smaller groups. These have to form meaningful groups based on their kinematics and ages, and at the same time be large enough that the GALAH survey was able to observe a statistically relevant number of stars in a group (we aimed for groups with at least 20 stars observed in GALAH). If a hierarchical structure of an association is provided in the literature \citep{larson81, vazques17, sun18, ratzenbock23}, we did not necessary partition an association into the smallest possible groups to avoid the undersampling described above. The summary of the associations and their subdivisions used in this work is given in Table \ref{tab:asc}. For the names of subdivisions, we use the same names as the original papers. 

\subsubsection{Vela}

We used \citet{cantat-gaudin19} to select stars in the Vela OB2 association and its sub-groups. All regions in Vela were observed serendipitously, except for population 7 (including cluster $\gamma^2$ Vel) and population 4 (including Collinder 135 and 140, and NGC 2451B). 

\subsubsection{Orion}

For the Orion OB1 association and the $\lambda$ Orionis group, we use regions defined in \citet{kos19, kos21}. For the purpose of this work, we use subdivisions known as the $\lambda$ Orionis group, Orion OB1a, OB1b, a single region including Orion OB1c and OB1d, and the region of the NGC 1788 cluster. All stars in Orion were targeted intentionally. 

\subsubsection{Scorpius-Centaurus-Crux}

Clustering for the Scorpius-Centaurus-Crux region was taken from \citet{zerjal23}. Some regions in Sco-Cen-Cru were observed intentionally within different observing programs. Some regions were observed serendipitously, hence a large difference between the number of stars observed in each region. 

\begin{table}
 \caption{Young associations used in this work. Data for Vela OB2 association is taken from \citet{cantat-gaudin19}, data for Orion from \citet{kos19} and \citet{kos21} (with the exception for the age range of Ori OB1cd), and data for Sco-Cen-Crux is taken from \citet{zerjal23}. If a group or population is made of subgroups with different ages, an age range is given.}
 \renewcommand{\arraystretch}{0.95}
    \begin{tabular}{lcccc}
    \hline
        Cluster & \multicolumn{2}{c}{$N_\mathrm{GALAH}$} & age & notes \\\hline
        & all & no flag & [Myr] &\\\hline
        \multicolumn{5}{c}{Vela OB2}\\
        Vela pop 1 & 6 & 5 & $46.8\pm3.5$\\
        Vela pop 2 & 24 & 9 & 44.1 -- 46.7\\
        Vela pop 3 & 19 & 9 & 34.0 -- 43.9\\
        Vela pop 4 & 183 & 92 & 35.0 -- 40.3\\
        Vela pop 5 & 16 & 11 & 23.3 -- 33.0\\
        Vela pop 6 & 12 & 5 & 20.1 -- 21.0\\
        Vela pop 7 & 101 & 40 & 8.7 -- 13.3\\
        All & 355 & 167 & 8.7 -- 46.8 & 1\\[0.2cm]
        \multicolumn{5}{c}{Orion OB1 + $\lambda$ Ori}\\
        lambda Ori & 206 & 52 & 6.5 -- 9.2\\
        Ori OB1a & 496 & 125 & 11.0 -- 21.2\\
        Ori OB1b & 473 & 122 & 9.0 -- 17.0\\
        Ori OB1cd & 314 & 71 & 4 -- 11$^\ast$\\
        NGC 1788 & 35 & 8 & $8.5\pm2.1$\\
        All & 1537 & 382 & 4 -- 21.2 & 1\\[0.2cm]
        \multicolumn{5}{c}{Sco-Cen-Cru}\\
        Sco Cen a & 134 & 13 & 7\\
        Sco Cen c & 805 & 73 & 4\\
        Sco Cen d & 164 & 17 & 2\\
        Sco Cen e & 205 & 19 & 11\\
        Sco Cen f & 3 & 0 & 15\\
        Sco Cen g & 219 & 24 & 13\\
        Sco Cen h & 12 & 5 & 15\\
        Sco Cen i & 4 & 2 & 2\\
        Sco Cen t & 112 & 12 & 15\\
        Sco Cen u & 63 & 3 & 9\\
        All & 1721 & 168 & 2 -- 15 & 1\\\hline
    \end{tabular}
    \\$^\ast$ \citet{zari19}\\
    1: Used for calculating trends (dwarfs)
    \label{tab:asc}
\end{table}

\section{Deriving trends in the abundances of chemical elements}
\label{sec:deriving_trends}

Any deviation from chemical homogeneity within a cluster should be of a physical nature, meaning that the measured abundances of chemical elements should be identical for all stars within a cluster, regardless of factors such as the quality of the spectra, observing time, resolution, or the observed wavelength bands. Any deviations from this expectation should be of a physical nature. However, a quick examination of the abundance measurements across clusters reveals that this is not the case in practice; various factors are likely causing systematic deviations or trends. In this section, we demonstrate that many of these trends are consistent across different clusters and provide a way to parametrize them.

When initially exploring the trends, we considered  the trends of abundances as a function of several stellar and spectroscopic parameters (temperature, gravity, projected rotational velocity, microturbulence velocity and the signal-to-noise ratio of the spectra). We realised that the trends were most coherent when displayed as a function of temperature. In addition, when deriving the trends with temperature and de-trending the elemental abundances, we found no significant residual trends with other parameters. We conclude that any trends initially observed with other parameters must have existed due to the correlations of those parameters and temperature. Hence, in the rest of this work we only present trends as a function of stellar effective temperature. 

The aim is to calculate trends for as many elements as possible and for the largest possible range of temperatures. We separate the trends based on the spectral type of the stars (giants and dwarfs, where the distinction between them is described in the next section) and further dissect them for the trends valid for unflagged stars only and all stars (both flagged and unflagged stars). GALAH DR4 provides two main sets of flags that indicate issues with the given data products. A \texttt{flag\_sp} value indicates issues with a star or its spectra. In this work we consider all stars with $\texttt{flag\_sp}<8$ as unflagged stars. This means we ignore flags indicating a detected emission in $H\alpha$ or $H\beta$, flags indicating spectra where one of the four channels is missing (the IR channel is missing for a small number of observations), and flags marking the SB1 binaries. All other issues indicated by \texttt{flag\_sp} are considered as valid flags in this work. Additionally, a \texttt{flag\_x\_fe} (\texttt{x} being an element) indicates issues with the measurements of the abundance of one particular element. If this flag is raised, we consider the measurements for that element as flagged. 

\subsection{Giants and dwarfs distinction}
\label{sec:dwarfs_vs_giants}

Our initial inspection of the trends showed that they can be different for giants and dwarfs. This can be expected for different reasons, including differences in model atmospheres used in spectral synthesis, different linelists, sensitivity to the isochrone fitting and photometric priors (see \citealt{beeson24}), or the correlation between the temperature and signal-to-noise ratio per pixel ($S/N$, in general correlated for dwarfs and anti-correlated for giants). Hence, we split the stars into giants, dwarfs, and a small region around the turn-off (subgiants), which we call the overlap region. The overlap is necessary to verify the trends for giants and dwarfs in the region where one trend should smoothly transition into the other. It is possible that giants and dwarfs experience completely different trends with no smooth transition. However, for the purpose of detrending the data, a smooth transition is preferred.

\begin{figure}
    \centering
    \includegraphics[width=\columnwidth]{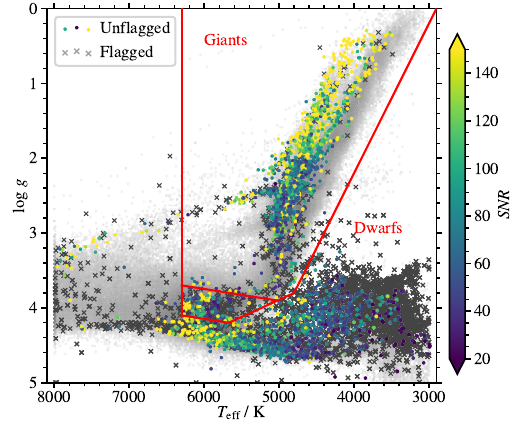}
    \caption{Kiel diagram of the stars used to compute the trends. Unflagged stars are plotted with colours and flagged stars with grey crosses. Gray background distribution are all stars in GALAH DR4. Regions delineating dwarfs, giants, and the overlap region are marked in red. The equations of all red lines are specified in Equation \ref{eq:lines}.}
    \label{fig:kiel}
\end{figure}

The distinction between dwarfs and giants was done in the Kiel diagram. We chose two regions: one defining the giants and one defining the overlap. Stars outside these two regions are considered to be dwarfs. The regions are defined (arbitrary) as
\begin{equation}
\label{eq:lines}
    \begin{cases}
    \multirow{ 4}{*}{$\mathrm{giants},$} & \mathrm{if}\ T_\mathrm{eff}<6300\ \mathrm{K}\ \mathrm{and}\\
                    & \log g<(-0.000158(\, T_\mathrm{eff}/\mathrm{K}-5670))+3.8\ \mathrm{and}\\
                    &  \log g<(0.000459(\, T_\mathrm{eff}/\mathrm{K}-5670))+4.2\ \mathrm{and}\\
                    &  \log g<(0.002(\, T_\mathrm{eff}/\mathrm{K}-4800))+3.8.\\[0.18cm]
    \multirow{ 4}{*}{$\mathrm{overlap},$} & \mathrm{if}\ T_\mathrm{eff}<6300\ \mathrm{K}\ \mathrm{and}\\
                      & \log g<(-0.000158(\, T_\mathrm{eff}/\mathrm{K}-5670))+4.2\ \mathrm{and}\\
                      & \log g<(0.000459(\, T_\mathrm{eff}/\mathrm{K}-5670))+4.2\ \mathrm{and}\\
                      & \log g\geq(-0.000158(\, T_\mathrm{eff}/\mathrm{K}-5670))+3.8.\\[0.18cm]
    \mathrm{dwarfs},              & \text{otherwise.}
    
\end{cases}
\end{equation}
Figure \ref{fig:kiel} shows the Kiel diagram with all stars used to determine the trends in this work. Regions defining dwarfs, giants. and the overlap are marked as well.

The only exception to this classification are globular clusters where we observed a small number of stars on the horizontal branch. The distribution of the horizontal branch stars stretches into the dwarfs region, as defined above. Therefore, we designated all stars in globular clusters as giants. There are no dwarfs observed in any globular cluster in the GALAH survey. 

\subsection{Parametrising the trends}

After the initial inspection of the trends, we had decided to parameterise the trends with cubic splines. The goal is to describe smooth trends at higher temperatures and a sharp transition at lower temperatures with a single function. After initial experimentation, any polynomials were ruled out, because very large orders would be needed to describe the aforementioned features. For some elements and for some clusters we do not have enough data-points to reliably fit the trends with high order polynomials. We use cubic splines instead. 

A univariate cubic spline is a piecewise function of the form:
\begin{equation}
    f_i(x)=A\,x^3+B\,x^2+C\,x+D,
\end{equation}
where $i$ are sections of the function between consecutive knots. In the knots, the following conditions of continuity up to the second derivative must be satisfied:
\begin{align}
    f_i(x_i)&=f_{i+1}(x_i),\nonumber\\
    f_i'(x_i)&=f_{i+1}'(x_i),\ \mathrm{and}\\
    f_i''(x_i)&=f_{i+1}''(x_i),\nonumber
\end{align}
where $x_i$ is the position of a knot. We selected the positions of the knots, so they separate different regimes of the trends. We fixed the knots at $3800$, $4200$, $4750$, and $5700\ \mathrm{K}$. A cubic spline was fitted to the trends by a chi-square minimisation method. In practice, both the definition of the spline function and the fitting procedure were handled by the \texttt{scipy}'s \texttt{LSQUnivariateSpline} function \citep{virtanen20}. Additionally, we skipped a knot $x_i$, if the function $f_{i+1}$ was to be constrained by just 5 or fewer data-points.

Instead of coefficients $A$ to $D$ for each segment, \texttt{scipy} produces coefficients of B-spline base functions. How the fitted function can be generated from coefficients, positions of knots, and the order of the spline is explained in Appendix \ref{sec:splines}. 

\subsection{Stacking trends of individual clusters}

We wish to define trends that are uniformly valid over as large parts of the temperature range as possible. This is hardly feasible by using observations in a single cluster, as only stars in a narrow range of temperatures are usually observed. Hence, we combine observations of multiple clusters, so we can fit a single trend representative of all clusters. Because stars in clusters have different compositions, one should not fit trends onto the joined measurements of abundance vs. effective temperature. Instead, we first fit a trend to the measurements of each individual cluster, and then join measurements from individual clusters by shifting them accordingly. This way we can later fit the trends onto joint data-sets with minimal scatter. We determined the required shift independently for each cluster and for each chemical element. We shifted the data by the value of the trend at $T_\mathrm{eff}=5400\ \mathrm{K}$ for the dwarfs and at $T_\mathrm{eff}=4800\ \mathrm{K}$ for the giants. We wanted to select the value for the dwarfs close to the temperature of the Sun, where we believe the abundances are the most accurate. However, a lot of clusters do not have any stars observed at the temperature of the Sun. Instead we chose a temperature of $T_\mathrm{eff}=5400\ \mathrm{K}$ where we have observations for most clusters, therefore we do not have to extrapolate the trend just to determine the required shift. For giants we selected the temperature close to the red clump, where GALAH abundances should be most accurate. If the extrapolation of the trend beyond the extent of the data is still needed, we assume a constant value at the edge of the data range. 

\begin{figure*}
    \centering
    Dwarfs\\[0.0cm]
    \includegraphics[width=0.98\textwidth]{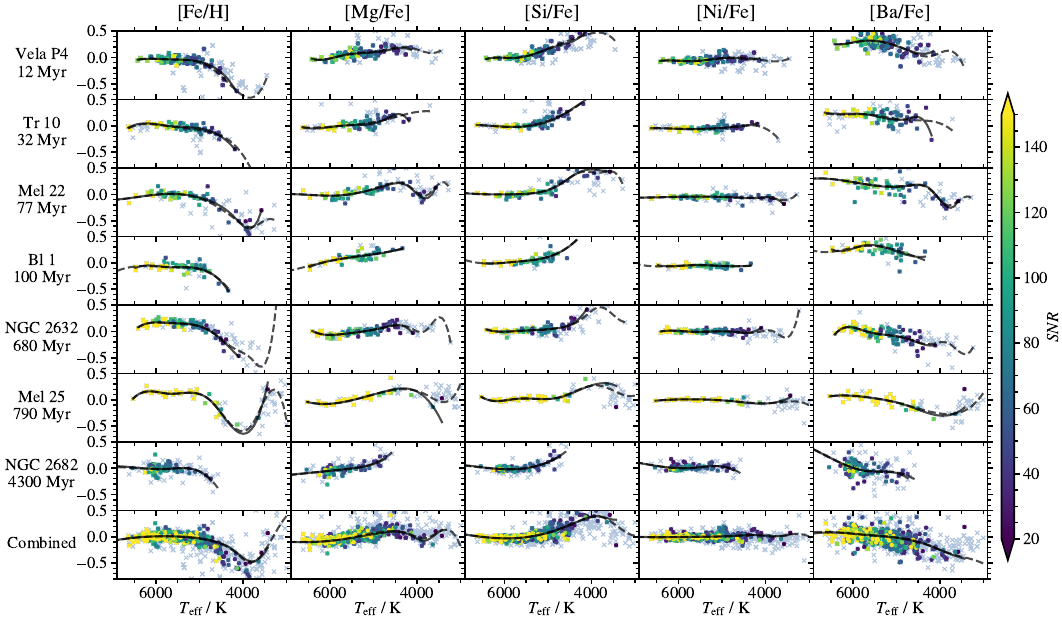}\\
    Giants\\[0.0cm]
    \includegraphics[width=0.98\textwidth]{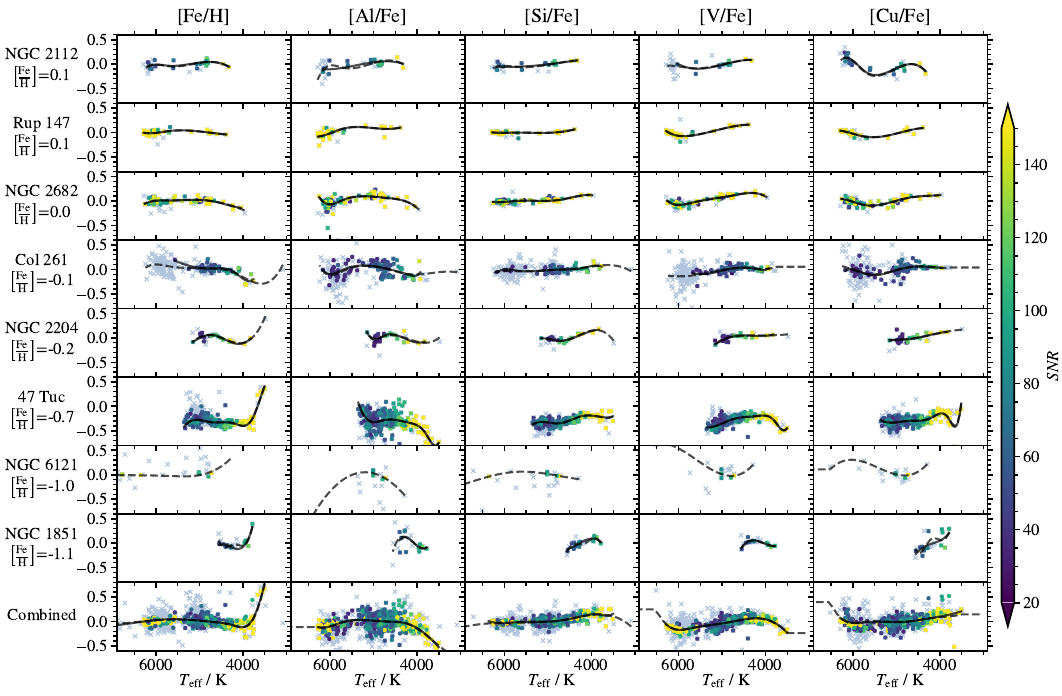}
    \caption{Trends for several clusters illustrated for a small selection of elements. Top: trends for seven open clusters and associations. Last row shows the combined measurements. Bottom: trends for eight open and globular clusters. Quantities in the first column are shifted by the value of $[Fe/H]$ given under the cluster's name, so they fit into the same plotting range. Colored points are unflagged stars, grey crosses are flagged stars. Solid line is a fit for unflagged stars, dashed line is a fit for all stars.}
    \label{fig:trends_individual}
\end{figure*}

Figure \ref{fig:trends_individual} illustrates the trends for a small (morphologically representative) number of elements observed in seven open clusters of different ages (top panel, showing the trends for dwarfs) and seven clusters with different metallicities (bottom panel, showing the trends for giants). Note that the trends observed in unrelated clusters display a similar shape. This led us to believe that the trends are not peculiar but rather general, i. e. they are not caused by the properties of clusters (like an abundance of active stars in young clusters, fast rotators etc.) or data reduction (like correlation between the $S/N$ and temperature on the main sequence, continuum fitting, etc.). Some features, like the steep trend in iron abundance of the coldest stars, is common to both dwarfs and giants. 

\subsection{Fitted trends}

\label{sec:fitted_trends}

We fitted the trends separately for dwarfs and giants, with the stars from the overlap region being included in both domains. For some elements the uncertainties in their abundance measurement were too large to fit meaningful trends. 

For all elements, for dwarfs and giants, the trends were fitted with cubic splines, with the same algorithm as for individual clusters in the previous two sections. $S/N$ of each star measured in the red channel was used as a weight in the fitting algorithm. Additionally, we find the fit to perform best with an arbitrary increase of the weights of unflagged stars by a factor of 3. The first reason for giving unflagged stars a higher weight is that we trust unflagged measurements more. The second reason is that we want to use a single function for detrending. Using a different function for flagged and unflagged stars would produce discontinuities in the final table of abundances. We can avoid this by making the function for flagged stars follow the function for unflagged very tightly in the overlapping temperature range. Figures \ref{fig:trends_all_dwarfs} and \ref{fig:trends_all_giants} in Appendix \ref{sec:all_trends} show the fitted trends for all elements. 

\subsection{Detrending}

A product of this work is a table of abundances of chemical elements as measured in GALAH DR4, but detrended with the relations calculated here. The other parameters given in Tables \ref{tab:coef_dwarfs} and \ref{tab:coef_giants} were not modified. 

For detrending we used the trends derived from the combined fit of flagged and unflagged stars. Consequently, the trends are constrained only by observations of flagged stars in some temperature regions. This is not ideal, but allows us to detrend measurements for valuable stars that might have temperatures outside the temperature ranges of our unflagged cluster stars. One can also notice from figures \ref{fig:trends_all_dwarfs} and \ref{fig:trends_all_giants} that such a generalisation is justified, as the trends derived with the inclusion of flagged stars are usually a good extrapolation of the trends derived from unflagged stars only. 

GALAH stars (regardless of their $S/N$ or flags values) were first split into dwarfs, giants, and stars from the overlap, as defined in Section \ref{sec:dwarfs_vs_giants}. Elements that need to be detrended were selected. For dwarfs we only detrended the measurements of the abundances of [Fe/H] and [X/Fe] for elements X (O, Na, Mg, Al, Si, K, Ca, Sc, Ti, V, Cr, Mn, Co, Ni, Cu, Zn, Rb, Sr, Y, Zr, Ba, Nd, and Sm). For giants we also included elements C, N, Mo, Tu, La, Ce, and Eu. Other elements show trends that are much smaller than the scatter of the abundance measurements in our cluster stars, or only have abundances measured in such small numbers of stars that the trends cannot be reliably calculated. Next, we calculated the correction (value of the derived trend at some temperature) for each measurement, if the temperature of the star lies within the boundaries where the trend is defined. The value of the correction was subtracted from the original measurement to produce a detrended measurement. For the stars in the overlap region between dwarfs and giants, we subtracted the mean of both trends. Whether the stars lie in the dwarfs or giants region was determined based solely on the DR4 $T_\mathrm{eff}$ and $\log g$.

The schema for a table of detrended abundances is the same as the one for the \texttt{allstars} table in GALAH DR4 \citep[see schema in Table~7 of ][]{buder24} with the following modifications. Every measurement that was not detrended (either for a particular star or the element as a whole) has an additional flag in the \texttt{flag\_X\_Fe} column of the stellar parameters table. The flag was increased by a value of 128 (8th bit was raised in the flag value). An additional column (\texttt{detrend\_model}) was added. The value in this column is set to 0, if trends for dwarfs were used to detrend the measurements for a star, 1, if trends for giants were used, and 2, if a star belongs to the overlap region, so a combination of trends for giants and dwarfs was used. When using the detrended table, one can treat the flags similarly as in the original table; small flag values mean reliable data, while large flag values indicate caution or poor data. 

\section{Origin of the trends}
\label{sec:origins}

In this section we discuss some possible causes for the trends observed in the relation between abundances of chemical elements and stellar effective temperature. We try to explain two general features observed for many elements. First, is the gentle trend above $T_\mathrm{eff}>4800\ \mathrm{K}$ (see Figure \ref{fig:trends_individual}), which becomes steeper closer to $T_\mathrm{eff}=4800\ \mathrm{K}$ for elements like [Fe/H]. The behaviour for other elements can be different, but the trend in this temperature range is almost always monotonic below $T_\mathrm{eff}<6800\ \mathrm{K}$. Second, is a sharp upturn observed for [Fe/H] at around $T_\mathrm{eff}=3800\ \mathrm{K}$ and a steep trend observed for some elements at the lowest temperatures. Another obvious observation is that the trends are weaker in giants than dwarfs. 

The fact that the trends are different in giants and dwarfs led us to some immediate conclusions. Firstly, we can rule out a number of stellar astrophysical processes as the cause of these apparent chemical inhomogeneities. Dredge-up would affect just a subset of elements and cause inhomogeneities only for giants (which show milder trends than dwarfs). Atomic diffusion would affect stars close to the main-sequence turn-off. This could potentially cause a trend in the temperature range $5500\ \mathrm{K}<T_\mathrm{eff}<6500\ \mathrm{K}$ where we indeed observe trends. However, the trends are not as expected from the effect of atomic diffusion \citep{dotter17}. Secondly, we conclude that although the trends could be caused by the atmosphere interpolation and spectrum fitting procedures -- for example where the stellar atmospheres transition from convective to radiative -- we cannot imagine a particular trend that this could cause. However, the spectral fitting codes might be responsible for the trends, which we explore below. We also explore the effect of different linelists and photometric priors used in spectral fitting.

It is also obvious that the trends are not induced by the $S/N$ variations. Although $S/N$ is correlated with $T_\mathrm{eff}$ on the main sequence, the trends as a function of $S/N$ are less pronounced than with $T_\mathrm{eff}$ and can be explained solely as a consequence of the correlation. After detrending the dataset with respect to $T_\mathrm{eff}$, no trends remain as a function of $S/N$. On the giant branch, the $S/N$ is anticorrelated with $T_\mathrm{eff}$. However, the strongest trends at $T_\mathrm{eff}<4500\ \mathrm{K}$ are similar for giants and dwarfs. This further confirms that the abundances cannot be highly influenced by the $S/N$ of our spectra. Below we also explore the influence of continuum fitting.

The above arguments rule out some possible origins of the trends, but give no answer about the possible cause. In this work we focus on the empirical causes, and leave physical causes for a future paper. 

\subsection{Refitting the spectra}
\label{sec:refitting}

We refitted the red-channel (CCD3) DR4 spectra of stars in three open clusters with stars observed in a wide $T_\mathrm{eff}$ range (Melotte 22, Melotte 25, and NGC 2632), first to verify that the trends can be reproduced by direct fitting, and then to test the effect of various input parameters on the obtained iron abundance. Among our tests presented below, $T_\mathrm{eff}$ and $\log g$ were only treated as free parameters for some tests. Overall metallicity, iron abundance, projected rotational velocity ($v \sin i$), and microturbulence ($v_\mathrm{mic}$) were free parameters for all tests. The resolving power of all spectra was degraded to $R=22\,000$, which is the lowest common $R$. Therefore we were able to keep the resolving power fixed during fitting. Continuum was either adapted during fitting or fixed, depending on the test. We initially fitted abundances of other elements than iron, but because it turned out that explaining the trend of iron abundance is difficult already, we did not expand the analysis for other elements. The tests presented below were therefore done with the abundance pattern fixed to the solar value. We did use a mask that removed the $H\alpha$ line ($6552.8\ \text{\AA}<\lambda<6572.8\,\text{\AA}$) and the lithium line ($6707.5\ \text{\AA}<\lambda<6709.5\,\text{\AA}$) from the fit. We used two different spectral fitting codes, SME and Korg.

To avoid fitting and possibly misinterpreting spectra of peculiar stars, we took additional care in selecting which cluster members to fit and use in the exploration of the trends. Careful manual selection of cluster stars was not done for the trends derived from all clusters in the previous section, but is manageable for just three clusters which spectra are reanalysed hereafter. We rejected stars based on a few qualitative criteria and on the flags produced in the DR4 analysis. Stars were rejected if:
\begin{itemize}
    \item $S/N<10$ ($S/N$ is given per pixel, averaged over the red arm),
    \item $v \sin i>30\ \mathrm{km\,s^{-1}}$,
    \item error of radial velocity $\sigma(v_\mathrm{r})>5\ \mathrm{km\, s^{-1}}$,
    \item a spectroscopic binary in DR4 and the error of the secondary star radial velocity is $<5\times$ the error of the primary star radial velocity,
    \item reduced $\chi^2$ of the DR4 fitting process is $\chi^2>3.0$
    \item star lies closer to the binary sequence than the main sequence in the HRD,
    \item $RUWE>1.4$, or
    \item visual inspection of the spectrum shows peculiarities.
\end{itemize}
Some diagnostics plots illustrating the rejection criteria and a list of stars is given in Appendix \ref{sec:rejection}. More than half of the stars with $T_\mathrm{eff}<3500\ \mathrm{K}$ were rejected, but only a small fraction of hotter stars. Large number of rejected cool stars is due to the observing strategy. After allocating the fibres to the targeted stars, we assigned the remaining fibres to the cluster stars that might be too dark for the complete analysis. Hence many of the coolest stars have $S/N<10$. Rejected stars were not used in the subsequent analysis and are not shown in the following figures in this section. 

\subsubsection{SME spectral fitting}

Spectroscopy Made Easy \citep[SME,][]{valenti96, piskunov17} is used in the GALAH DR4 analysis to produce synthetic spectra, which are used as a training set for a machine-learning synthetic spectrum interpolator. We used the same framework, specifically the
PySME package \citep{wehrhahn23}, to refit the spectra directly with SME. To synthesise spectra in SME, we used the VALD3 linelists \citep{piskunov95, ryabchikova15}, MARCS models of stellar atmospheres \citep{gustafsson08}, values for solar abundances of elements by \citet{asplund09}, and NLTE corrections for the following elements: Iron \citep{asplund21}, Lithium \citep{lind09}, Magnesium \citep{osorio16}, Sodium \citep{lind11}, Silicon \citep{amarsi17}, Barium \citep{mashonkina99}, Calcium \citep{mashonkina07}, Titanium \citep{sitnova16}, and Oxygen \citep{amarsi16}. Lines of molecules that include C, N, O, or Ti in their structure were computed in 1D LTE approximation. The macroturbulence velocity was fixed at $v_\mathrm{mac}=3\ \mathrm{km\,s^{-1}}$, which is a typical value for these stars in DR4 \citep{buder24}.

Our iron abundance trend is flatter than the one in the DR4 data at $T_\mathrm{eff}>6500\ \mathrm{K}$, and we are also not able to clearly reproduce the upturn at $T_\mathrm{eff}=3800\ \mathrm{K}$. In the subsequent analysis we focused our effort into the region in-between these temperatures, where our trends appear similar to those in DR4.

\subsubsection{Korg fitting}

Korg \citep{wheeler23, wheeler24} is a modern 1D LTE spectral synthesis and spectral fitting code. We fitted GALAH DR4 spectra using an expanded MARCS grid of stellar atmospheres \citep{meszaros12}, which include changing alpha and Carbon abundances. We used the same linelists as for the SME fitting, and solar abundances from \citet{asplund21}. Korg has no support for NLTE corrections, so we did not use any. It also does not fit the macroturbulence velocity, which means this broadening velocity is included in the fitting of the rotational broadening $v \sin i$ in Korg.

The iron abundance trend produced by Korg's values is much flatter than any trend we get from SME's values or in GALAH DR4. The difference between photometric and spectroscopic $T_\mathrm{eff}$ and $\log g$ is also much smaller for Korg values as opposed to DR4 or our SME values. See Section \ref{sec:models} for a further analysis.

\subsection{Tests}

\subsubsection{Photometric and spectroscopic $T_\mathrm{eff}$ and $\log g$}

In GALAH DR4, photometric and distance information from \textit{Gaia}'s 3rd data release \citep{lindegren21, bailer-jones21} and 2MASS \citep{skrutskie06} are used to estimate $T_\mathrm{eff}$ and $\log g$, which are then used to better constrain the spectral fit \citep{buder24}. For clusters, we are able to get very precise values for $T_\mathrm{eff}$ and $\log g$ \citep{beeson24}, which we call photometric temperature and gravity. These are different in DR4 and in this work, because in DR4 all stars in a cluster are treated independently, and here we assume that they must lie on the same isochrone. We test whether the use of precise photometric $T_\mathrm{eff}$ and $\log g$ has any impact on our observed trends. We fitted the spectra with all four combinations, where $T_\mathrm{eff}$ and $\log g$ are either fixed at photometric values or added as completely free parameters. In the latter case, the initial values for the fitting procedure were the photometric values. 

\begin{figure}[!h]
    \centering
    \includegraphics[width=\columnwidth]{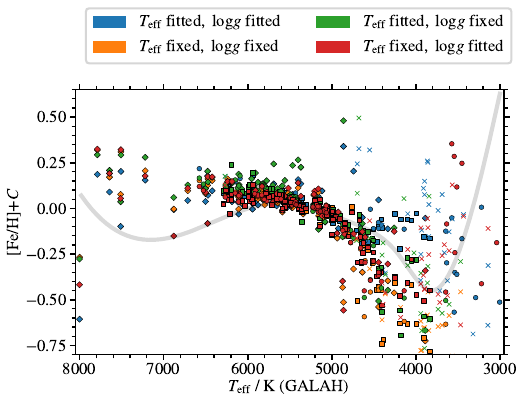}
    \caption{Effects of keeping $T_\mathrm{eff}$ and $\log g$ free parameters or keeping them fixed at the photometric value. Symbols mark the stars belonging to clusters Melotte 25 (circles), Melotte 22 (diamonds), and NGC 2632 (squares). Crosses are measurements for the stars without a trustworthy fit. The solid line is the trend derived in Section \ref{sec:fitted_trends} for the GALAH DR4 iron abundance in dwarfs.}
    \label{fig:phot_vs_fit}
\end{figure}

In Figure \ref{fig:phot_vs_fit} we see that the trends for cool stars follow the shape in DR4, except in the case where both $T_\mathrm{eff}$ and $\log g$ are free parameters (blue points). The discrepancy between the four fits sharply diverges at $T_\mathrm{eff}=4500\ \mathrm{K}$. There is also a small difference at around $T_\mathrm{eff}=6000\ \mathrm{K}$ between the fit where only $\log g$ is fixed and the other three cases. For stars with $T_\mathrm{eff}>6000\ \mathrm{K}$, our results are consistent, but do not follow the trend in DR4. In DR4, the trend in this region is constrained mostly by flagged stars, and we only have a small number of observations, so we can not make any meaningful conclusions.

\subsubsection{Continuum fitting}

The GALAH reduction pipeline provides a normalised spectrum, which has been normalised without any prior knowledge of stellar parameters and without the use of a synthetic spectrum. A low order polynomial is fitted to the observed spectrum with several iterations of asymmetric sigma clipping, so the fit represents the upper envelope of the spectrum. Such continuum normalisation is inadequate for abundance analysis, but useful for radial velocity analysis and is used to derive initial stellar parameters with a neural network. With better known stellar parameters and radial velocity, the continuum can be refitted later with no loss of information. In the DR4 analysis, the spectra are renormalised at each iteration of the fitting process. This works very well for most stars. However, for cool stars a problem can arise, where there might be no regions in the spectrum where the actual continuum can be sampled (spectra only consist of blended lines). If the continuum levels, temperature, and metallicity are correlated, a wrong combination of these three parameters can produce the best fit and consequently wrong spectroscopic stellar parameters. If the photometric values for $T_\mathrm{eff}$ can be trusted, and using the literature values for the metallicity of the three well studied clusters, the degeneracy can be broken. 

\begin{figure}[!h]
    \centering
    \includegraphics[width=\columnwidth]{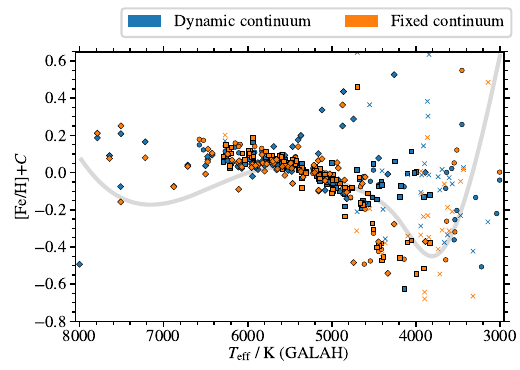}
    \caption{Effect of two different continuum fits on the measured iron abundances. Symbols mark the stars belonging to clusters Melotte 25 (circles), Melotte 22 (diamonds), and NGC 2632 (squares). Crosses are measurements for the stars without a trustworthy fit. The solid line is the trend derived in Section \ref{sec:fitted_trends} for the GALAH DR4 iron abundance in dwarfs.}
    \label{fig:continuum}
\end{figure}

We test whether the iron abundances are different if we dynamically modify the continuum levels (in this case with a polynomial of the 12th order), as in DR4, or if we fit the continuum using photometric $T_\mathrm{eff}$, $\log g$, literature metallicity, and DR4 $v \sin i$ and $v_\mathrm{mic}$. The continuum in the latter case is then kept fixed in the spectral fitting process. In both cases we kept $T_\mathrm{eff}$ and $\log g$ as free parameters. In Figure \ref{fig:continuum}, we see significant differences below $T_\mathrm{eff}<4500\ \mathrm{K}$. Fixing the continuum produces stronger trends. A dynamic continuum appears to  produce trends with a much smaller amplitude than in DR4. This might be a solution to the dip in the trends at around $T_\mathrm{eff}=3800\ \mathrm{K}$, however a dynamic continuum also results in the fitted $T_\mathrm{eff}$ and $\log g$ to deviate significantly from the expected values (see analysis in Section \ref{sec:teff_verification} and Figure \ref{fig:phot_spec}). The degeneracy between $T_\mathrm{eff}$, $\log g$ and [Fe/H] manifests itself in apparently better [Fe/H] measurements when using a dynamic continuum, but causes issues elsewhere. Hence a dynamic continuum is not a universal solution to any of the problems explored here.

\subsubsection{Linelists}

The linelist used in GALAH DR4 was adapted from \citet{heiter21}, which was ultimately produced by VALD \citep{ryabchikova15}. In this work we produced our own VALD linelists to have more freedom when experimenting with line depth thresholds as opposed to the curated DR4 linelist with 81768 entries between $6450$ and $6750\,\text{\AA}$, which are decreased based on a depth threshold (above 0.001) calculation for each region of the HRD. We produced the linelists by selecting lines based on their depths in spectra of stars with temperatures $3500$, $4500$, $5500$, and $6500\ \mathrm{K}$, $v_\mathrm{mic}=2\ \mathrm{km\,s^{-1}}$, and $\log g=4.5$ and $0.5$ (at $T_\mathrm{eff}=3500\ \mathrm{K}$), $\log g=4.5$ and $2.5$ (at $T_\mathrm{eff}=4500\ \mathrm{K}$), $\log g=4.5$ and $4.0$ (at $T_\mathrm{eff}=5500\ \mathrm{K}$), and $\log g=4.0$ (at $T_\mathrm{eff}=6500\ \mathrm{K}$). We varied the depth thresholds as given in Table \ref{tab:thresholds} to produce linelists of different sizes.

\begin{table}
\caption{Thresholds for the creation of linelists. Thresholds show the minimum depth of the line selected from the VALD linelist of a specific temperature. Last column gives the number of lines in the linelist created for the red channel ($6450$ to $6750\,\text{\AA}$).}
\begin{tabular}{l  cccc  c}
\hline
Linelist & \multicolumn{4}{c}{Thresholds at $T_\mathrm{eff}\ /\ \mathrm{K}$} & Nr. of lines\\ 
 & 3500 & 4500 & 5500 & 6500 & \\\hline
     XS & 0.4 & 0.2 & 0.1 & 0.1 & 1093\\
     S & 0.15 & 0.075 & 0.03 & 0.03 & 6629\\
     M & 0.1 & 0.05 & 0.03 & 0.03 & 10060\\
     L & 0.05 & 0.02 & 0.01 & 0.01 & 17015\\
     XL & 0.02 & 0.01 & 0.005 & 0.005 & 35498\\
     \multicolumn{5}{c}{No limitations} & 81768\\\hline
\end{tabular}
\label{tab:thresholds}
\end{table}

\begin{figure}[!h]
    \centering
    \includegraphics[width=\columnwidth]{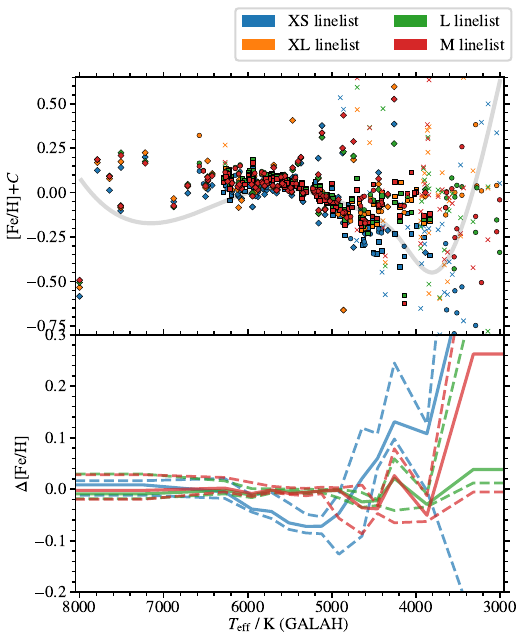}
    \caption{Top: Iron abundances computed with linelists of different fidelities. Symbols mark the stars belonging to clusters Melotte 25 (circles), Melotte 22 (diamonds), and NGC 2632 (squares). Crosses are measurements for the stars without a trustworthy fit. The solid line is the trend derived in Section \ref{sec:fitted_trends} for the GALAH DR4 iron abundance in dwarfs. Bottom: Difference of iron abundances between the computations using the XL linelist and other three linelists. Solid line shows the median difference in each temperature bin, and the dashed lines show the 16th and 84th percentiles of the differences in each bin.}
    \label{fig:lineslists_sizes}
\end{figure}

Many weak lines, individually too weak to pass the threshold for a small linelist, can blend together to produce a significant continuum against which the stronger lines are formed. If this continuum of weak lines is not accounted for in the spectral synthesis, the fit to the observed spectra will be compensated elsewhere, mainly by increasing the metallicity. We expect that by using linelists with insufficient numbers of weak lines, the metallicity and iron abundance will be increased for cool stars. We tested this effect by fitting the spectra using four linelists of different sizes. The results are presented in Figure \ref{fig:lineslists_sizes}. Limitations of linelists may explain the trends below $T_\mathrm{eff}<3800\ \mathrm{K}$ but not for hotter stars. Above $T_\mathrm{eff}>5000\ \mathrm{K}$ we only see significant deviations when using the smallest (XS) linelist.

\begin{figure}[!h]
    \centering
    \includegraphics[width=\columnwidth]{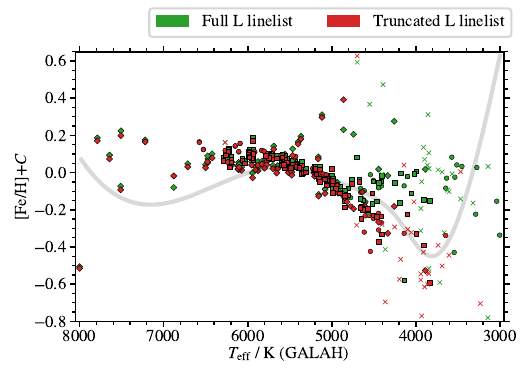}
    \caption{Effects of a truncated linelist on the measured iron abundance. Symbols mark the stars belonging to clusters Melotte 25 (circles), Melotte 22 (diamonds), and NGC 2632 (squares). Crosses are measurements for the stars without a trustworthy fit. The solid line is the trend derived in Section \ref{sec:fitted_trends} for the GALAH DR4 iron abundance in dwarfs.}
    \label{fig:truncated}
\end{figure}

A similar issue can arise if linelists are created from lines that only appear in hotter stars. We fitted the spectra using a truncated large linelist (L), produced by omitting lines that would otherwise pass the threshold at $3500\ \mathrm{K}$. We compare iron abundances with the ones obtained with a fit using the full L linelist. Figure \ref{fig:truncated} shows that the differences appear for stars with $T_\mathrm{eff}<4500\ \mathrm{K}$. The effect of a truncated linelist is similar to the effect of a small linelist. 

\subsubsection{Models and spectral fitting codes}
\label{sec:models}

\begin{figure}[!h]
    \centering
    \includegraphics[width=\columnwidth]{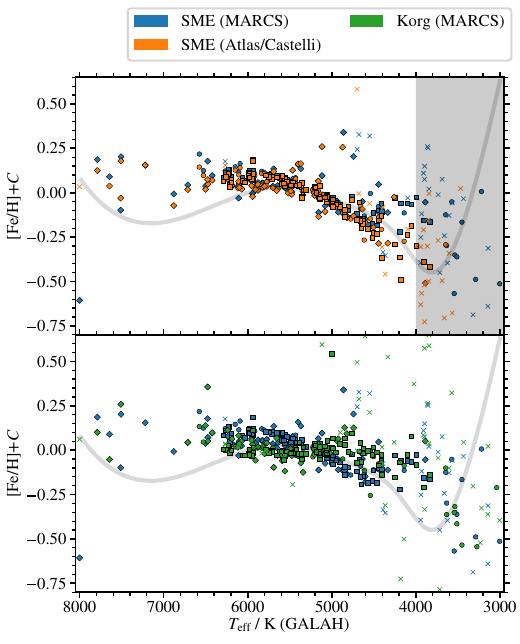}
    \caption{Top: Iron abundance obtained with fitting using SME with either MARCS atmospheres or Atlas/Castelli atmospheres. Atlas/Castelli atmospheres are extrapolated in the gray region. Bottom: Comparison of the iron abundance obtained by fitting with SME and with korg. Both use MARCS atmospheres. Symbols mark the stars belonging to clusters Melotte 25 (circles), Melotte 22 (diamonds), and NGC 2632 (squares). Crosses are measurements for the stars without a trustworthy fit. The solid line is the trend derived in Section \ref{sec:fitted_trends} for the GALAH DR4 iron abundance in dwarfs.}
    \label{fig:sme_vs_korg}
\end{figure}

SME comes ready with two different stellar atmosphere model grids. MARCS \citep{gustafsson08} is valid for stars with $T_\mathrm{eff}$ between 3000 and 8000 K, and Atlas/Castelli \citep{castelli03} for $T_\mathrm{eff}$ between 4000 and $10\,000$ K. We compared the abundances computed with both grids and found small differences. Iron abundances are lower between $4000<T_\mathrm{eff}<4500\ \mathrm{K}$ when calculated using the Atlas/Castelli grid (see Figure \ref{fig:sme_vs_korg}, top panel). 

\begin{figure}[!h]
    \centering
    \includegraphics[width=\columnwidth]{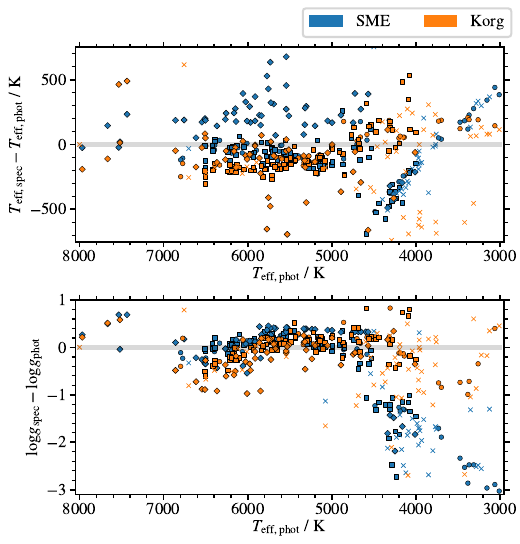}
    \caption{Difference between the spectroscopic and photometric $T_\mathrm{eff}$ (top) and $\log g$ (bottom) for the spectroscopic parameters fitted by SME or Korg. Symbols mark the stars belonging to clusters Melotte 25 (circles), Melotte 22 (diamonds), and NGC 2632 (squares). Crosses are measurements for the stars without a trustworthy fit.}
    \label{fig:teff_logg_korg}
\end{figure}

The trend for iron abundace derived by Korg is remarkably flat. Only below $T_\mathrm{eff}<3700\ \mathrm{K}$ does the iron abundance measured by Korg decrease significantly. The most telling indicator of the difference between the fits produced by SME and Korg are plots of photometric $T_\mathrm{eff}$ and $\log g$ compared to the fitted spectroscopic values as illustrated in Figure \ref{fig:teff_logg_korg}. Korg's parameters follow the photometric values much closer and also show no ``phase shift'' at around $T_\mathrm{eff}=4500\ \mathrm{K}$. For some stars the spectral fitting did not converge, regardless of the initial conditions. This happens more often with Korg, as one cannot control the boundaries of the fitted parameters in its current implementation. Usually the computed $v \sin i$ is much too high. Such stars are marked with crosses in figures discussed in this section, and produce the obvious outliers among the cool stars (e.g. in Figure \ref{fig:teff_logg_korg}, top panel). Note that we are comparing NLTE abundances derived by SME with LTE abundances derived by Korg. For dwarfs with approximately solar metallicity, the NLTE correction is small, much smaller than the amplitude of the trends that we observe in SME results. 

\section{Age verification}
\label{sec:age_verification}

\subsection{Age determination in the GALAH survey}

The analysis of stellar parameters and abundances in GALAH DR4 relies heavily on the photometric information from the \textit{Gaia} mission. In particular, the $\log g$ is estimated from the basic principles, i.e. from the measurements of stellar luminosities ($L$) (combining \textit{Gaia}'s photometry and parallaxes), fitted spectroscopic temperatures ($T_\mathrm{eff}$), stellar masses ($M$) and corresponding Solar values:
\begin{equation}
    \log g=\log g_\odot + \log \frac{M}{M_\odot}+4 \log \frac{T_\mathrm{eff}}{T_\mathrm{eff, \odot}}-\log \frac{L}{L_\odot}.
\end{equation}
Stellar masses are found from the isochrones that best match the observed stellar properties. In DR4 they are interpolated over isochrones with ages between $8<\log(\tau/\mathrm{Gyr})<10.8$ in steps of $\log(\Delta_\tau/\mathrm{Gyr})=0.1$, metallicities between $-2.75<\mathrm{[M/H]}<0.75$ in steps of $0.25\ \mathrm{dex}$, and then in steps of $0.1\ \mathrm{dex}$ to $\mathrm{[M/H]}=0.7$. Hot stars, evolved white dwarfs and extremely luminous giants are excluded, as they are outside the range of atmospheric models used in spectral fitting. The procedure is performed by the \texttt{ELLI} code \citep{lin18}, which also provides stellar ages as a by-product.

For stars in globular clusters, the age has a lower limit of $\tau>4.5\ \mathrm{Gyr}$, and for stars in young open clusters, we extend the age range of possible isochrones to $\log(\tau/\mathrm{Gyr})>6.19$; see \citet{buder24} for more information.

\subsection{Comparison with isochronal cluster ages}

Ages of clusters can be calculated by isochrone fitting, where we assume that all stars within a cluster have the same age. So called isochronal ages are much more precise than the age estimates for individual stars as calculated in GALAH DR4. In this work we use isochronal ages of open clusters from \citet{cantat20}, and \citet{hunt24} for clusters not in the former catalogue. Ages of globular clusters were taken from \citet{franch09}. For young associations and globular clusters we use ages from various literature sources given in Tables \ref{tab:asc} and \ref{tab:globular_clusters}. 

To verify stellar ages, we used all clusters with at least 8 stars with a GALAH age. We also exclude $\omega$ Cen due to the large age spread of its stellar populations.  

\begin{figure}
    \centering
    \includegraphics[width=\columnwidth]{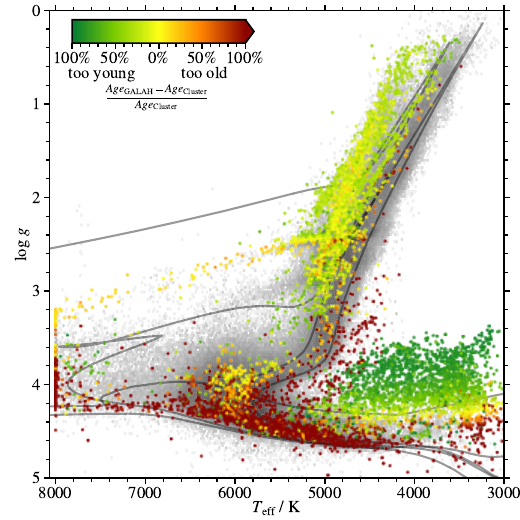}
    \caption{GALAH's ages of cluster stars compared to isochronal ages. `Too young' means that the GALAH ages are smaller than the isochronal ages, analogous for `too old'. Background distribution shows all GALAH DR4 stars. Isochrones for ages $10\ \mathrm{Myr}$, $100\ \mathrm{Myr}$, $1\ \mathrm{Gyr}$, and $10\ \mathrm{Gyr}$ and Solar metallicity are plotted with black lines. An isochrone for the age of $10\ \mathrm{Gyr}$ and metallicity of $[M/H]=-0.5$ is also plotted. }
    \label{fig:ages_kiel}
\end{figure}

We find the GALAH ages are well determined for stars at the evolutionary stages at and beyond the main sequence turn-off. GALAH and literature ages agree best at the turn-off, and in the red clump (Figure \ref{fig:ages_kiel}). On the red giant branch and the asymptotic giant branch, the GALAH ages are typically slightly underestimated (and slightly overestimated on the horizontal branch). On the pre-main sequence there are stars with well estimated ages. However, most stars on the pre-main sequence have ages under- or over-estimated, depending on the position of the pre-main sequence. Stars with correctly estimated ages do not belong to the same clusters, but are rather serendipitous stars that happen to have well matching photometric and spectroscopic parameters, so they get assigned the correct age. 

\begin{figure}[!ht]
    \centering
    \includegraphics[width=\columnwidth]{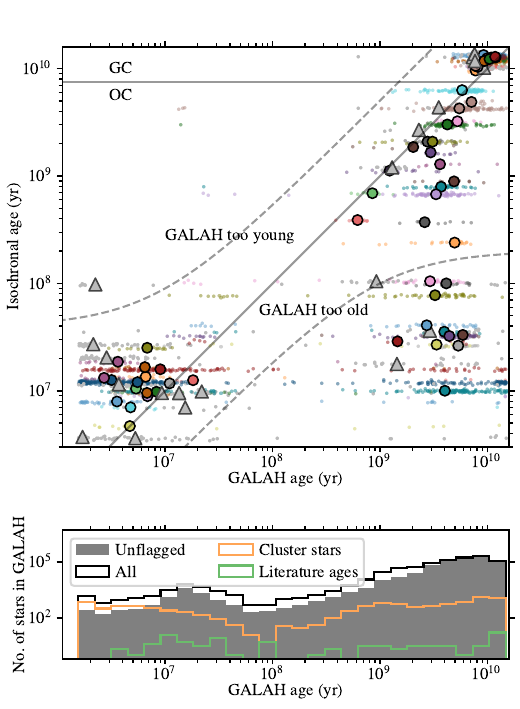}
    \caption{Comparison between GALAH's ages and isochronal ages. Top: Ages of individual stars in clusters are plotted as small dots (artificially slightly scattered vertically for better perception). The median for each cluster is plotted with a circle of the same colour. Medians marked with a triangle are for clusters with only flagged stars. A one-to-one relation is indicated. A horizontal line separates open clusters from globular clusters. Note a bimodal distribution of GALAH ages for some clusters; one peak is at around $10^7\ \mathrm{yr}$ and another at around $5.0\times10^9\ \mathrm{yr}$. Dashed lines mark the region used in the calculation of statistical uncertainties of stellar ages. Bottom: Distribution of stars in log-spaced age bins. Black line shows all stars and the gray area shows unflagged stars. Both distributions are for the whole GALAH DR4 sample. Orange histogram shows the distrubution of GALAH ages for cluster members, and the green histogram for individual clusters' literature ages. A diagram with added cluster names is given in Appendix \ref{sec:age_ap}.}
    \label{fig:ages_comp}
\end{figure}

\begin{figure}[!ht]
    \centering
    \includegraphics[width=\columnwidth]{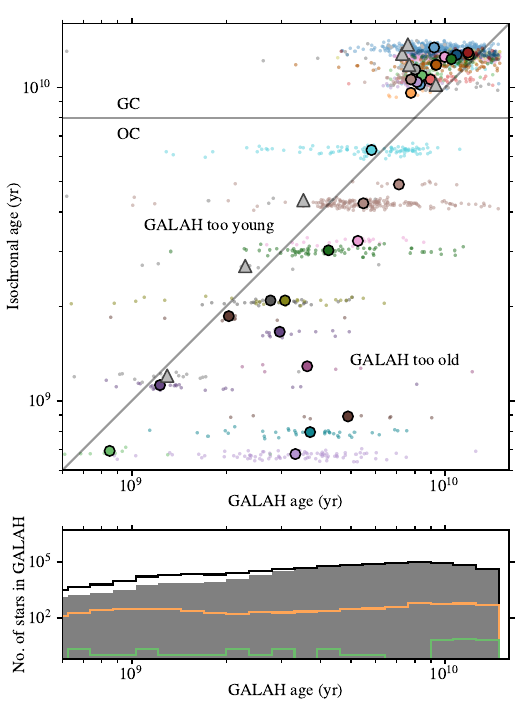}
    \caption{Same as Figure \ref{fig:ages_comp}, but zoomed into the oldest part of the diagram. The same diagram with added cluster names is given in Appendix \ref{sec:age_ap}.}
    \label{fig:ages_old_comp}
\end{figure}

The main lesson learned from the diagrams in Figures \ref{fig:ages_comp} and \ref{fig:ages_old_comp} is that seemingly old stars in the GALAH survey are not necessary old. A selection based solely on the age will be polluted by young stars. This might be negligible in practice because there are fewer young than old stars in GALAH DR4, especially among the stars evolved past the main sequence. Although some care is needed when selecting main-sequence stars.

Similarly, by selecting young stars based on the GALAH ages alone, we can miss a large fraction (around a quarter) of the young population. The pollution of such a selection by older stars is mostly negligible in this case. 

\begin{figure}[!ht]
    \centering
    \includegraphics[width=\columnwidth]{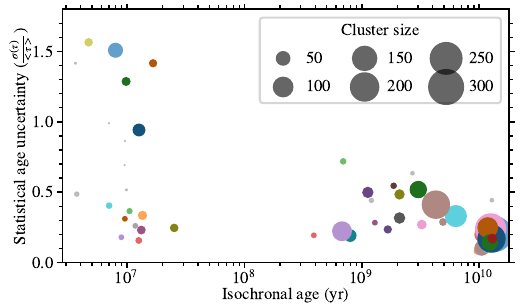}
    \caption{Relative standard deviations of stars in clusters. Colors match the clusters in Figures \ref{fig:ages_comp} and \ref{fig:ages_old_comp}. Symbol size is proportional to the number of GALAH stars in a cluster. Clusters with standard deviation above 1.0 are Gulliver 6, HSC 2907, OCSN 100, OCSN 96, and $\lambda$ Ori. Large cluster just below 1.0 is Collinder 69.}
    \label{fig:ages_std}
\end{figure}

In addition to systematic errors of ages, we also estimated the statistical uncertainty. To be able to measure the statistical uncertainty without being affected by large systematics, we only used stars within the region marked by the dashed lines in Figure \ref{fig:ages_comp}. Additionally, we only used clusters that have their mean GALAH age within that region as well. We kept cluster only if at least eight stars remained after the above cuts. We define the one sigma statistical uncertainty as the standard deviation of ages of stars in a cluster divided by the mean GALAH age. We illustrate the results in Figure \ref{fig:ages_std}. Globular clusters have typical statistical uncertainties between 10 and 30\%. The uncertainty for open clusters varies between 20 and 60\% with some outliers among the young clusters. Ignoring the outliers, ages of stars in young clusters have slightly lower uncertainty than in the older clusters. An upper and a lower limit of stellar ages can artificially reduce the measured standard deviations, so the uncertainties in young clusters and in globular clusters might be biased. Unlike the precision, the statistical uncertainty is lower by just around 5\% for stars at the turn-off. 

\section{Verification of $T_\mathrm{eff}$ and $\log g$}
\label{sec:teff_verification}

Temperature and surface gravity in GALAH DR4 are computed by spectral fitting using photometric priors. An isochrone that best describes the position of a star on an HR diagram is used to produce a prior for $T_\mathrm{eff}$ and $\log\, g$. Because stars are treated independently from each other, a different isochrone might be used to get priors for stars in the same cluster. This could be easily avoided for stars known to be part of a cluster, but most stars in the GALAH survey are not in clusters, so a consistent approach is used where even cluster stars are treated independently. 

\subsection{Photometric parameters}

Here we check how the temperature and surface gravity calculated by isochrone fitting compare to spectroscopic (GALAH DR4) values. To obtain the photometric parameters of stars in clusters, we used literature ages, metallicities, and extinctions to produce\footnote{\url{http://stev.oapd.inaf.it/cmd}} a Padova isochrone \citep{bressan12, chen14, chen15, tang14, marigo17, pastorelli19} for each cluster. A distance from the isochrone was calculated for each star in a cluster, using \textit{Gaia} magnitudes. The distance is defined as
\begin{equation}
    d_\mathrm{CMD}=\left[ (M_{G, \mathrm{s}}-M_{G, \mathrm{i}})^2 + \left(3 \left( (BP-RP)_\mathrm{s}-(BP-RP)_\mathrm{i}\right)\right)^2 \right]^{1/2},
\end{equation}
where $s$ indicates the values for a star, and $i$ a value on the isochrone. Absolute magnitudes $M_G$ were calculated from \textit{Gaia} parallaxes. Extinctions of individual stars were not taken into account when calculating distances. Extinction was used to produce the correct isochrone, though. Stars closer to the binary sequence than to the main sequence or pre-main sequence were rejected for having poorly determined photometric parameters. Stars further than 0.5 magnitudes from any part of the isochrone were rejected as outliers. We used the nearest point on the isochrone to determine the photometric parameters for each star.  

\subsection{Comparison of spectroscopic and photometric parameters}

\begin{figure*}[!ht]
    \centering
    \includegraphics[width=\textwidth]{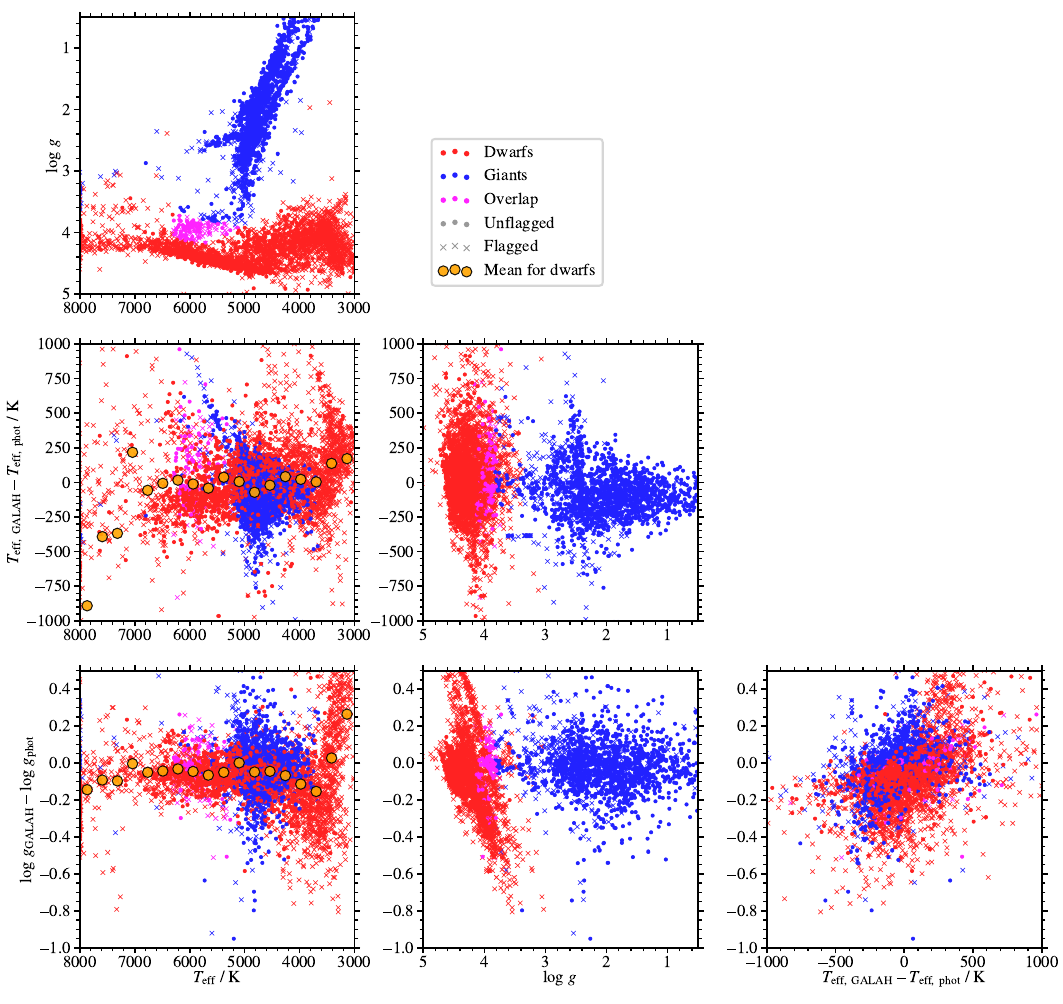}
    \caption{Relationship between photometric $T_\mathrm{eff}$ and $\log g$, spectroscopic $T_\mathrm{eff}$ and $\log g$ and their respective differences. Red symbols show dwarfs, blue symbols giants, and purple symbols the overlap region. Orange symbols show the averaged values for dwarfs. Only stars with good photometric data are shown.}
    \label{fig:phot_spec}
\end{figure*}

In the comparison described below, we included all clusters that display a well fitted isochrone. The isochrone is defined by the literature values for age, extinction, and metallicity. We further analysed only stars that lie reasonably close to the isochrone (this excludes outliers, blue stragglers, binaries, and stars with peculiar extinction). The results of the comparison are summarised in Figure \ref{fig:phot_spec}.

\begin{figure}[!ht]
    \centering
    \includegraphics[width=\columnwidth]{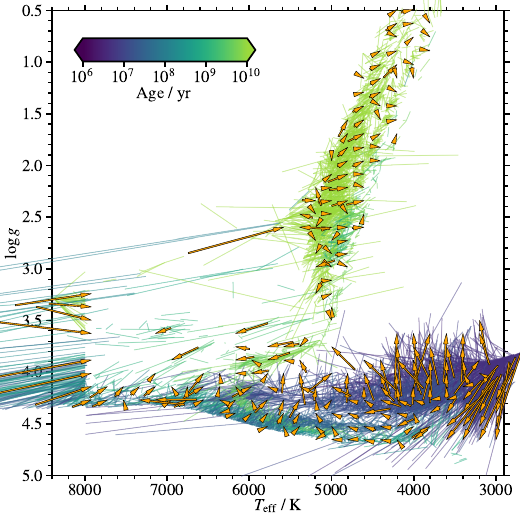}
    \caption{Illustration of the photometric and spectroscopic $T_\mathrm{eff}$ and $\log g$. Arrows point from the photometric values to the spectroscopic values. Thin lines show the value pairs for each star, and orange arrows for the average flow in that region. Photometric temperatures are sometimes far outside the plotted range, hence the lines coming from outside of the panel.}
    \label{fig:tension}
\end{figure}

A similar pattern is observed for the difference between photometric and spectroscopic parameters vs. temperature as in the trends of abundances of chemical elements. Stars that have photometric temperatures outside the $8000\ \mathrm{K}>T_\mathrm{eff}>3000\ \mathrm{K}$ range get squished within these boundaries when a spectroscopic temperature is calculated. This is expected, because, if a star has a temperature in GALAH DR4, it must be within the boundaries of the models used for spectral synthesis. Most of these stars are correctly flagged in DR4. This tension between the photometric and spectroscopic values is illustrated in Figure \ref{fig:tension}. Vectors point from photometric $(T_\mathrm{eff}, \log g)$ to spectroscopic $(T_\mathrm{eff}, \log g)$, which displays a ''flow'' in the parameter space. Some coherency in the flow is observed on the RGB and AGB, however the flow is chaotic on the HB. Correlations are obvious for the dwarfs, as is also evident from Figure \ref{fig:phot_spec}.

\section{Discussion and Conclusions}
\label{sec:discussion}

The fourth data release by the GALAH survey marks more than 10 years of operations. DR4 served $1\,085\,520$ spectra of $917\,588$ stars with parameters that include abundances of up to 32 chemical elements. Although the precision and accuracy of these parameters have improved compared to previous data releases, we observed systematic deviations from expected values. This issue is particularly evident in clusters, which are expected to contain coeval, conatal, and chemically homogeneous populations. $T_\mathrm{eff}$ and $\log g$ are parameters that can be accurately derived from photometric observations of cluster stars with known age,  allowing for well-constrained isochrones that the cluster stars should closely follow. Deriving $T_\mathrm{eff}$ and $\log g$ is a complex process that involves finding the isochrone that best describes each star, which defines a photometric prior for $T_\mathrm{eff}$ and $\log g$. Our initial assumption was that fixing the isochrone for the cluster stars would lead to more consistent measurements of their chemical element abundances. However, this assumption proved to be incorrect because we saw clear systematic abundance trends, which were unrelated to stellar astrophysical processes.

We were unable to identify a single reason for these systematic trends. Our primary focus was on potential issues arising from data processing rather than on possible problems with the understanding and modelling of stellar physics, which will be explored in a future paper. However, we did identify several factors contributing to the systematic errors. The degeneracy between the continuum levels and stellar parameters could be addressed more thoroughly in future analyses, but it remains unclear if this can be resolved at the level of the analysis of the whole survey. More concerning are different results produced by the two spectral fitting codes used in this study, which indicate a significant disagreement between them, even when using the same input parameters.

Consistent measurements of abundances of chemical elements across a wide temperature range are essential for studies where parameters of stars of different temperatures must be compared. Examples are observations of clusters to trace chemical gradients in the Galaxy \citep{spina21} or chemical cartography of the Galaxy \citep{recioblanco23}. In such studies, where stars in a large range of distances must be observed, it is inevitable to combine measurements of stars with vastly different temperatures. We showed that abundances, particularly of cool stars, can have systematic errors in the order of $0.5$ dex, which is equivalent to the radial metallicity gradient in the Galactic disk over $5\ \mathrm{kpc}$ \citep{spina21}. It is also essential to be aware of any systematic trends if only a small number of stars are observed in each cluster, like in \citet{zhang22}, because the systematic trends might not be as easy to derive from sparse observations.

Machine learning techniques used to derive stellar parameters from spectra \citep[e.g.][]{ness15, fabbro18, leung19,buder18} are very sensitive to any systematic errors in the training set. Fixing any non-physical trends, as we show is possible with clusters, could improve the range and reliability of machine learning techniques for parameter derivation. In applications where labels from one dataset are passed and applied to another, like training a large low-resolution survey on parameters derived in a small overlap with a high-resolution survey \citep{das24, ambrosch23, nepal23}, some care for the accuracy and zero-points should be made. Consistency of abundance measurements is also essential for any non-supervised machine learning methods, like clustering or chemical tagging studies. Chemical tagging in the Galactic disk is challenging, if even possible, with the precision of abundances of chemical elements at the level of chemical homogeneity of open clusters \citep[$\sim0.03\ \mathrm{dex,}$][]{casamiquela21, spina22}. Hence, consistent measurements with systematic errors corrected at the same level should ideally be used to chemically cluster or tag stars across a large temperature range \citep{poovelil20}.

The application of chemical clocks is another example where it is of utmost importance that abundances of chemical elements are measured consistently among stars at different evolutionary stages \citep{storm23, hayden22}. Connecting the chemical signatures observed in clusters with the theory of star formation similarly requires observations of a range of stars across the HRD \citep{kos21, griggio22}. 

In this study, we adopted an empirical approach to address the challenge of achieving consistent measurements of spectroscopic parameters. An open problem remains: how to achieve consistency and precision without empirical ad hoc corrections, but with models and stellar physics \citep[e.g.][]{spina20} used in spectral synthesis codes \citep{wheeler23}. This is undoubtedly a complex and extensive problem that will require collaboration and a diverse range of expertise. What is also needed are methods for verifying the codes and models in real data, taken with a variety of instruments in an increasingly larger region of the universe, in the Galaxy and beyond. We demonstrate that clusters serve as an effective testing ground; in fact, they may be even more powerful than what we have shown here, as the observations of clusters in the GALAH survey were not originally designed for benchmarking purposes. 

\subsection*{Data availability}

In this work, we produced fitted functions that represent the observed trends in the measurements of abundances of chemical elements in the fourth data release of the GALAH survey. The parameters of the functions are given in Appendix \ref{sec:aptrends} and are valid for GALAH DR4 only. We also produced a detrended DR4 catalogue of stellar parameters. It is available at \url{https://www.galah-survey.org/dr4/overview/} or via private communication.

\begin{acknowledgements}
We acknowledge the traditional owners of the land on which the AAT and ANU stand, the Gamilaraay, the Ngunnawal and the Ngambri peoples. We pay our respects to Elders, past and present, and are proud to continue their tradition of surveying the night sky in the Southern hemisphere. JK is supported by the Slovenian Research Agency ARIS grants P1-018. This work has made use of the VALD database, operated at Uppsala University, the Institute of Astronomy RAS in Moscow, and the University of Vienna.
\end{acknowledgements}

\bibliographystyle{aa}
\bibliography{bib}

\begin{thebibliography}{107}
\expandafter\ifx\csname natexlab\endcsname\relax\def\natexlab#1{#1}\fi

\bibitem[{{Adibekyan} {et~al.}(2020){Adibekyan}, {Sousa}, {Santos}, {Figueira},
  {Allende Prieto}, {Delgado Mena}, {Gonz{\'a}lez Hern{\'a}ndez}, {de Laverny},
  {Recio-Blanco}, {Campante}, {Tsantaki}, {Hakobyan}, {Oshagh}, {Faria},
  {Bergemann}, {Israelian}, \& {Boulet}}]{adibekyan20}
{Adibekyan}, V., {Sousa}, S.~G., {Santos}, N.~C., {et~al.} 2020, \aap, 642,
  A182

\bibitem[{{Amarsi} \& {Asplund}(2017)}]{amarsi17}
{Amarsi}, A.~M. \& {Asplund}, M. 2017, \mnras, 464, 264

\bibitem[{{Amarsi} {et~al.}(2016){Amarsi}, {Asplund}, {Collet}, \&
  {Leenaarts}}]{amarsi16}
{Amarsi}, A.~M., {Asplund}, M., {Collet}, R., \& {Leenaarts}, J. 2016, \mnras,
  455, 3735

\bibitem[{{Amarsi} {et~al.}(2020){Amarsi}, {Lind}, {Osorio}, {Nordlander},
  {Bergemann}, {Reggiani}, {Wang}, {Buder}, {Asplund}, {Barklem}, {Wehrhahn},
  {Sk{\'u}lad{\'o}ttir}, {Kobayashi}, {Karakas}, {Gao}, {Bland-Hawthorn}, {de
  Silva}, {Kos}, {Lewis}, {Martell}, {Sharma}, {Simpson}, {Zucker},
  {{\v{C}}otar}, {Horner}, \& {GALAH Collaboration}}]{amarsi20}
{Amarsi}, A.~M., {Lind}, K., {Osorio}, Y., {et~al.} 2020, \aap, 642, A62

\bibitem[{{Ambrosch} {et~al.}(2023){Ambrosch}, {Guiglion}, {Mikolaitis},
  {Chiappini}, {Tautvai{\v{s}}ien{\.{e}}}, {Nepal}, {Gilmore}, {Randich},
  {Bensby}, {Bayo}, {Bergemann}, {Morbidelli}, {Pancino}, {Sacco}, {Smiljanic},
  {Zaggia}, {Jofr{\'e}}, \& {Jim{\'e}nez-Esteban}}]{ambrosch23}
{Ambrosch}, M., {Guiglion}, G., {Mikolaitis}, {\v{S}}., {et~al.} 2023, \aap,
  672, A46

\bibitem[{{Asplund} {et~al.}(2021){Asplund}, {Amarsi}, \&
  {Grevesse}}]{asplund21}
{Asplund}, M., {Amarsi}, A.~M., \& {Grevesse}, N. 2021, \aap, 653, A141

\bibitem[{{Asplund} {et~al.}(2009){Asplund}, {Grevesse}, {Sauval}, \&
  {Scott}}]{asplund09}
{Asplund}, M., {Grevesse}, N., {Sauval}, A.~J., \& {Scott}, P. 2009, \araa, 47,
  481

\bibitem[{{Bailer-Jones} {et~al.}(2021){Bailer-Jones}, {Rybizki}, {Fouesneau},
  {Demleitner}, \& {Andrae}}]{bailer-jones21}
{Bailer-Jones}, C.~A.~L., {Rybizki}, J., {Fouesneau}, M., {Demleitner}, M., \&
  {Andrae}, R. 2021, \aj, 161, 147

\bibitem[{{Beeson} {et~al.}(2024){Beeson}, {Kos}, {de Grijs}, {Martell},
  {Buder}, {Traven}, {Lewis}, {Zafar}, {Bland-Hawthorn}, {Freeman}, {Hayden},
  {Sharma}, \& {De Silva}}]{beeson24}
{Beeson}, K.~L., {Kos}, J., {de Grijs}, R., {et~al.} 2024, \mnras, 529, 2483

\bibitem[{{Bovy}(2016)}]{bovy16}
{Bovy}, J. 2016, \apj, 817, 49

\bibitem[{{Bressan} {et~al.}(2012){Bressan}, {Marigo}, {Girardi}, {Salasnich},
  {Dal Cero}, {Rubele}, \& {Nanni}}]{bressan12}
{Bressan}, A., {Marigo}, P., {Girardi}, L., {et~al.} 2012, \mnras, 427, 127

\bibitem[{{Buder} {et~al.}(2018){Buder}, {Asplund}, {Duong}, {Kos}, {Lind},
  {Ness}, {Sharma}, {Bland-Hawthorn}, {Casey}, {de Silva}, {D'Orazi},
  {Freeman}, {Lewis}, {Lin}, {Martell}, {Schlesinger}, {Simpson}, {Zucker},
  {Zwitter}, {Amarsi}, {Anguiano}, {Carollo}, {Casagrande}, {{\v{C}}otar},
  {Cottrell}, {da Costa}, {Gao}, {Hayden}, {Horner}, {Ireland}, {Kafle},
  {Munari}, {Nataf}, {Nordlander}, {Stello}, {Ting}, {Traven}, {Watson},
  {Wittenmyer}, {Wyse}, {Yong}, {Zinn}, {{\v{Z}}erjal}, \& {Galah
  Collaboration}}]{buder18}
{Buder}, S., {Asplund}, M., {Duong}, L., {et~al.} 2018, \mnras, 478, 4513

\bibitem[{{Buder} {et~al.}(2024){Buder}, {Kos}, {Wang}, {McKenzie}, {Howell},
  {Martell}, {Hayden}, {Zucker}, {Nordlander}, {Montet}, {Traven},
  {Bland-Hawthorn}, {De Silva}, {Freeman}, {Lewis}, {Lind}, {Sharma},
  {Simpson}, {Stello}, {Zwitter}, {Amarsi}, {Armstrong}, {Banks}, {Beavis},
  {Beeson}, {Chen}, {Ciuc{\u{a}}}, {Da Costa}, {de Grijs}, {Martin}, {Nataf},
  {Ness}, {Rains}, {Scarr}, {Vogrin{\v{c}}i{\v{c}}}, {Wang}, {Wittenmyer},
  {Xie}, \& {The GALAH Collaboration}}]{buder24}
{Buder}, S., {Kos}, J., {Wang}, E.~X., {et~al.} 2024, arXiv e-prints,
  arXiv:2409.19858

\bibitem[{{Buder} {et~al.}(2021){Buder}, {Sharma}, {Kos}, {Amarsi},
  {Nordlander}, {Lind}, {Martell}, {Asplund}, {Bland-Hawthorn}, {Casey}, {de
  Silva}, {D'Orazi}, {Freeman}, {Hayden}, {Lewis}, {Lin}, {Schlesinger},
  {Simpson}, {Stello}, {Zucker}, {Zwitter}, {Beeson}, {Buck}, {Casagrande},
  {Clark}, {{\v{C}}otar}, {da Costa}, {de Grijs}, {Feuillet}, {Horner},
  {Kafle}, {Khanna}, {Kobayashi}, {Liu}, {Montet}, {Nandakumar}, {Nataf},
  {Ness}, {Spina}, {Tepper-Garc{\'\i}a}, {Ting}, {Traven},
  {Vogrin{\v{c}}i{\v{c}}}, {Wittenmyer}, {Wyse}, {{\v{Z}}erjal}, \& {Galah
  Collaboration}}]{buder21}
{Buder}, S., {Sharma}, S., {Kos}, J., {et~al.} 2021, \mnras, 506, 150

\bibitem[{{Cabrera-Ziri} \& {Conroy}(2022)}]{cabrera22}
{Cabrera-Ziri}, I. \& {Conroy}, C. 2022, \mnras, 511, 341

\bibitem[{{Cantat-Gaudin} {et~al.}(2020){Cantat-Gaudin}, {Anders},
  {Castro-Ginard}, {Jordi}, {Romero-G{\'o}mez}, {Soubiran}, {Casamiquela},
  {Tarricq}, {Moitinho}, {Vallenari}, {Bragaglia}, {Krone-Martins}, \&
  {Kounkel}}]{cantat20}
{Cantat-Gaudin}, T., {Anders}, F., {Castro-Ginard}, A., {et~al.} 2020, \aap,
  640, A1

\bibitem[{{Cantat-Gaudin} {et~al.}(2019){Cantat-Gaudin}, {Mapelli},
  {Balaguer-N{\'u}{\~n}ez}, {Jordi}, {Sacco}, \& {Vallenari}}]{cantat-gaudin19}
{Cantat-Gaudin}, T., {Mapelli}, M., {Balaguer-N{\'u}{\~n}ez}, L., {et~al.}
  2019, \aap, 621, A115

\bibitem[{{Casali} {et~al.}(2020){Casali}, {Magrini}, {Frasca}, {Bragaglia},
  {Catanzaro}, {D'Orazi}, {Sordo}, {Carretta}, {Origlia}, {Andreuzzi}, {Fu}, \&
  {Vallenari}}]{casali20}
{Casali}, G., {Magrini}, L., {Frasca}, A., {et~al.} 2020, \aap, 643, A12

\bibitem[{{Casamiquela} {et~al.}(2021){Casamiquela}, {Castro-Ginard}, {Anders},
  \& {Soubiran}}]{casamiquela21}
{Casamiquela}, L., {Castro-Ginard}, A., {Anders}, F., \& {Soubiran}, C. 2021,
  \aap, 654, A151

\bibitem[{{Castelli} \& {Kurucz}(2003)}]{castelli03}
{Castelli}, F. \& {Kurucz}, R.~L. 2003, in IAU Symposium, Vol. 210, Modelling
  of Stellar Atmospheres, ed. N.~{Piskunov}, W.~W. {Weiss}, \& D.~F. {Gray},
  A20

\bibitem[{{Chen} {et~al.}(2020){Chen}, {D'Onghia}, {Alves}, \&
  {Adamo}}]{chen20}
{Chen}, B., {D'Onghia}, E., {Alves}, J., \& {Adamo}, A. 2020, \aap, 643, A114

\bibitem[{{Chen} {et~al.}(2015){Chen}, {Bressan}, {Girardi}, {Marigo}, {Kong},
  \& {Lanza}}]{chen15}
{Chen}, Y., {Bressan}, A., {Girardi}, L., {et~al.} 2015, \mnras, 452, 1068

\bibitem[{{Chen} {et~al.}(2014){Chen}, {Girardi}, {Bressan}, {Marigo},
  {Barbieri}, \& {Kong}}]{chen14}
{Chen}, Y., {Girardi}, L., {Bressan}, A., {et~al.} 2014, \mnras, 444, 2525

\bibitem[{{Dal Ponte} {et~al.}(2024){Dal Ponte}, {D'Orazi}, \&
  {Bragaglia}}]{delponte24}
{Dal Ponte}, M., {D'Orazi}, V., \& {Bragaglia}, A. 2024, in EAS2024, European
  Astronomical Society Annual Meeting, 1978

\bibitem[{{Das} {et~al.}(2024){Das}, {Zucker}, {De Silva}, {Borsato},
  {Mura-Guzm{\'a}n}, {Buder}, {Ness}, {Nordlander}, {Casey}, {Martell},
  {Bland-Hawthorn}, {de Grijs}, {Freeman}, {Kos}, {Stello}, {Lewis}, {Hayden},
  \& {Sharma}}]{das24}
{Das}, P.~B., {Zucker}, D.~B., {De Silva}, G.~M., {et~al.} 2024, arXiv
  e-prints, arXiv:2410.12272

\bibitem[{{de Jong} {et~al.}(2019){de Jong}, {Agertz}, {Berbel}, {Aird},
  {Alexander}, {Amarsi}, {Anders}, {Andrae}, {Ansarinejad}, {Ansorge},
  {Antilogus}, {Anwand-Heerwart}, {Arentsen}, {Arnadottir}, {Asplund}, {Auger},
  {Azais}, {Baade}, {Baker}, {Baker}, {Balbinot}, {Baldry}, {Banerji},
  {Barden}, {Barklem}, {Barth{\'e}l{\'e}my-Mazot}, {Battistini}, {Bauer},
  {Bell}, {Bellido-Tirado}, {Bellstedt}, {Belokurov}, {Bensby}, {Bergemann},
  {Bestenlehner}, {Bielby}, {Bilicki}, {Blake}, {Bland-Hawthorn}, {Boeche},
  {Boland}, {Boller}, {Bongard}, {Bongiorno}, {Bonifacio}, {Boudon}, {Brooks},
  {Brown}, {Brown}, {Br{\"u}ggen}, {Brynnel}, {Brzeski}, {Buchert},
  {Buschkamp}, {Caffau}, {Caillier}, {Carrick}, {Casagrande}, {Case}, {Casey},
  {Cesarini}, {Cescutti}, {Chapuis}, {Chiappini}, {Childress}, {Christlieb},
  {Church}, {Cioni}, {Cluver}, {Colless}, {Collett}, {Comparat}, {Cooper},
  {Couch}, {Courbin}, {Croom}, {Croton}, {Daguis{\'e}}, {Dalton}, {Davies},
  {Davis}, {de Laverny}, {Deason}, {Dionies}, {Disseau}, {Doel}, {D{\"o}scher},
  {Driver}, {Dwelly}, {Eckert}, {Edge}, {Edvardsson}, {Youssoufi}, {Elhaddad},
  {Enke}, {Erfanianfar}, {Farrell}, {Fechner}, {Feiz}, {Feltzing}, {Ferreras},
  {Feuerstein}, {Feuillet}, {Finoguenov}, {Ford}, {Fotopoulou}, {Fouesneau},
  {Frenk}, {Frey}, {Gaessler}, {Geier}, {Gentile Fusillo}, {Gerhard},
  {Giannantonio}, {Giannone}, {Gibson}, {Gillingham},
  {Gonz{\'a}lez-Fern{\'a}ndez}, {Gonzalez-Solares}, {Gottloeber}, {Gould},
  {Grebel}, {Gueguen}, {Guiglion}, {Haehnelt}, {Hahn}, {Hansen}, {Hartman},
  {Hauptner}, {Hawkins}, {Haynes}, {Haynes}, {Heiter}, {Helmi}, {Aguayo},
  {Hewett}, {Hinton}, {Hobbs}, {Hoenig}, {Hofman}, {Hook}, {Hopgood},
  {Hopkins}, {Hourihane}, {Howes}, {Howlett}, {Huet}, {Irwin}, {Iwert},
  {Jablonka}, {Jahn}, {Jahnke}, {Jarno}, {Jin}, {Jofre}, {Johl}, {Jones},
  {J{\"o}nsson}, {Jordan}, {Karovicova}, {Khalatyan}, {Kelz}, {Kennicutt},
  {King}, {Kitaura}, {Klar}, {Klauser}, {Kneib}, {Koch}, {Koposov},
  {Kordopatis}, {Korn}, {Kosmalski}, {Kotak}, {Kovalev}, {Kreckel}, {Kripak},
  {Krumpe}, {Kuijken}, {Kunder}, {Kushniruk}, {Lam}, {Lamer}, {Laurent},
  {Lawrence}, {Lehmitz}, {Lemasle}, {Lewis}, {Li}, {Lidman}, {Lind}, {Liske},
  {Lizon}, {Loveday}, {Ludwig}, {McDermid}, {Maguire}, {Mainieri}, {Mali},
  {Mandel}, {Mandel}, {Mannering}, {Martell}, {Martinez Delgado}, {Matijevic},
  {McGregor}, {McMahon}, {McMillan}, {Mena}, {Merloni}, {Meyer}, {Michel},
  {Micheva}, {Migniau}, {Minchev}, {Monari}, {Muller}, {Murphy},
  {Muthukrishna}, {Nandra}, {Navarro}, {Ness}, {Nichani}, {Nichol}, {Nicklas},
  {Niederhofer}, {Norberg}, {Obreschkow}, {Oliver}, {Owers}, {Pai},
  {Pankratow}, {Parkinson}, {Paschke}, {Paterson}, {Pecontal}, {Parry},
  {Phillips}, {Pillepich}, {Pinard}, {Pirard}, {Piskunov}, {Plank},
  {Pl{\"u}schke}, {Pons}, {Popesso}, {Power}, {Pragt}, {Pramskiy}, {Pryer},
  {Quattri}, {Queiroz}, {Quirrenbach}, {Rahurkar}, {Raichoor}, {Ramstedt},
  {Rau}, {Recio-Blanco}, {Reiss}, {Renaud}, {Revaz}, {Rhode}, {Richard},
  {Richter}, {Rix}, {Robotham}, {Roelfsema}, {Romaniello}, {Rosario},
  {Rothmaier}, {Roukema}, {Ruchti}, {Rupprecht}, {Rybizki}, {Ryde}, {Saar},
  {Sadler}, {Sahl{\'e}n}, {Salvato}, {Sassolas}, {Saunders}, {Saviauk},
  {Sbordone}, {Schmidt}, {Schnurr}, {Scholz}, {Schwope}, {Seifert}, {Shanks},
  {Sheinis}, {Sivov}, {Sk{\'u}lad{\'o}ttir}, {Smartt}, {Smedley}, {Smith},
  {Smith}, {Sorce}, {Spitler}, {Starkenburg}, {Steinmetz}, {Stilz}, {Storm},
  {Sullivan}, {Sutherland}, {Swann}, {Tamone}, {Taylor}, {Teillon}, {Tempel},
  {ter Horst}, {Thi}, {Tolstoy}, {Trager}, {Traven}, {Tremblay}, {Tresse},
  {Valentini}, {van de Weygaert}, {van den Ancker}, {Veljanoski}, {Venkatesan},
  {Wagner}, {Wagner}, {Walcher}, {Waller}, {Walton}, {Wang}, {Winkler},
  {Wisotzki}, {Worley}, {Worseck}, {Xiang}, {Xu}, {Yong}, {Zhao}, {Zheng},
  {Zscheyge}, \& {Zucker}}]{jong19}
{de Jong}, R.~S., {Agertz}, O., {Berbel}, A.~A., {et~al.} 2019, The Messenger,
  175, 3

\bibitem[{{De Silva} {et~al.}(2015){De Silva}, {Freeman}, {Bland-Hawthorn},
  {Martell}, {de Boer}, {Asplund}, {Keller}, {Sharma}, {Zucker}, {Zwitter},
  {Anguiano}, {Bacigalupo}, {Bayliss}, {Beavis}, {Bergemann}, {Campbell},
  {Cannon}, {Carollo}, {Casagrande}, {Casey}, {Da Costa}, {D'Orazi}, {Dotter},
  {Duong}, {Heger}, {Ireland}, {Kafle}, {Kos}, {Lattanzio}, {Lewis}, {Lin},
  {Lind}, {Munari}, {Nataf}, {O'Toole}, {Parker}, {Reid}, {Schlesinger},
  {Sheinis}, {Simpson}, {Stello}, {Ting}, {Traven}, {Watson}, {Wittenmyer},
  {Yong}, \& {{\v{Z}}erjal}}]{desilva15}
{De Silva}, G.~M., {Freeman}, K.~C., {Bland-Hawthorn}, J., {et~al.} 2015,
  \mnras, 449, 2604

\bibitem[{{Dotter} {et~al.}(2017){Dotter}, {Conroy}, {Cargile}, \&
  {Asplund}}]{dotter17}
{Dotter}, A., {Conroy}, C., {Cargile}, P., \& {Asplund}, M. 2017, \apj, 840, 99

\bibitem[{{Dutra-Ferreira} {et~al.}(2016){Dutra-Ferreira}, {Pasquini},
  {Smiljanic}, {Porto de Mello}, \& {Steffen}}]{dutra16}
{Dutra-Ferreira}, L., {Pasquini}, L., {Smiljanic}, R., {Porto de Mello}, G.~F.,
  \& {Steffen}, M. 2016, \aap, 585, A75

\bibitem[{{Fabbro} {et~al.}(2018){Fabbro}, {Venn}, {O'Briain}, {Bialek},
  {Kielty}, {Jahandar}, \& {Monty}}]{fabbro18}
{Fabbro}, S., {Venn}, K.~A., {O'Briain}, T., {et~al.} 2018, \mnras, 475, 2978

\bibitem[{{Fu} {et~al.}(2022){Fu}, {Bragaglia}, {Liu}, {Zhang}, {Xu}, {Wang},
  {Zhang}, {Zhong}, {Chang}, {Li}, {Chen}, {Chen}, {Wang}, {Gjergo}, {Wang},
  {Yue}, \& {Zhang}}]{fu22}
{Fu}, X., {Bragaglia}, A., {Liu}, C., {et~al.} 2022, \aap, 668, A4

\bibitem[{{Gaia Collaboration} {et~al.}(2023){Gaia Collaboration},
  {Recio-Blanco}, {Kordopatis}, {de Laverny}, {Palicio}, {Spagna}, {Spina},
  {Katz}, {Re Fiorentin}, {Poggio}, {McMillan}, {Vallenari}, {Lattanzi},
  {Seabroke}, {Casamiquela}, {Bragaglia}, {Antoja}, {Bailer-Jones},
  {Schultheis}, {Andrae}, {Fouesneau}, {Cropper}, {Cantat-Gaudin}, {Bijaoui},
  {Heiter}, {Brown}, {Prusti}, {de Bruijne}, {Arenou}, {Babusiaux}, {Biermann},
  {Creevey}, {Ducourant}, {Evans}, {Eyer}, {Guerra}, {Hutton}, {Jordi},
  {Klioner}, {Lammers}, {Lindegren}, {Luri}, {Mignard}, {Panem}, {Pourbaix},
  {Randich}, {Sartoretti}, {Soubiran}, {Tanga}, {Walton}, {Bastian}, {Drimmel},
  {Jansen}, {van Leeuwen}, {Bakker}, {Cacciari}, {Casta{\~n}eda}, {De Angeli},
  {Fabricius}, {Fr{\'e}mat}, {Galluccio}, {Guerrier}, {Masana}, {Messineo},
  {Mowlavi}, {Nicolas}, {Nienartowicz}, {Pailler}, {Panuzzo}, {Riclet}, {Roux},
  {Sordo}, {Th{\'e}venin}, {Gracia-Abril}, {Portell}, {Teyssier}, {Altmann},
  {Audard}, {Bellas-Velidis}, {Benson}, {Berthier}, {Blomme}, {Burgess},
  {Busonero}, {Busso}, {C{\'a}novas}, {Carry}, {Cellino}, {Cheek},
  {Clementini}, {Damerdji}, {Davidson}, {de Teodoro}, {Nu{\~n}ez Campos},
  {Delchambre}, {Dell'Oro}, {Esquej}, {Fern{\'a}ndez-Hern{\'a}ndez}, {Fraile},
  {Garabato}, {Garc{\'\i}a-Lario}, {Gosset}, {Haigron}, {Halbwachs}, {Hambly},
  {Harrison}, {Hern{\'a}ndez}, {Hestroffer}, {Hodgkin}, {Holl}, {Jan{\ss}en},
  {Jevardat de Fombelle}, {Jordan}, {Krone-Martins}, {Lanzafame},
  {L{\"o}ffler}, {Marchal}, {Marrese}, {Moitinho}, {Muinonen}, {Osborne},
  {Pancino}, {Pauwels}, {Reyl{\'e}}, {Riello}, {Rimoldini}, {Roegiers},
  {Rybizki}, {Sarro}, {Siopis}, {Smith}, {Sozzetti}, {Utrilla}, {van Leeuwen},
  {Abbas}, {{\'A}brah{\'a}m}, {Abreu Aramburu}, {Aerts}, {Aguado}, {Ajaj},
  {Aldea-Montero}, {Altavilla}, {{\'A}lvarez}, {Alves}, {Anders}, {Anderson},
  {Anglada Varela}, {Baines}, {Baker}, {Balaguer-N{\'u}{\~n}ez}, {Balbinot},
  {Balog}, {Barache}, {Barbato}, {Barros}, {Barstow}, {Bartolom{\'e}},
  {Bassilana}, {Bauchet}, {Becciani}, {Bellazzini}, {Berihuete}, {Bernet},
  {Bertone}, {Bianchi}, {Binnenfeld}, {Blanco-Cuaresma}, {Boch}, {Bombrun},
  {Bossini}, {Bouquillon}, {Bramante}, {Breedt}, {Bressan}, {Brouillet},
  {Brugaletta}, {Bucciarelli}, {Burlacu}, {Butkevich}, {Buzzi}, {Caffau},
  {Cancelliere}, {Carballo}, {Carlucci}, {Carnerero}, {Carrasco}, {Castellani},
  {Castro-Ginard}, {Chaoul}, {Charlot}, {Chemin}, {Chiaramida}, {Chiavassa},
  {Chornay}, {Comoretto}, {Contursi}, {Cooper}, {Cornez}, {Cowell}, {Crifo},
  {Crosta}, {Crowley}, {Dafonte}, {Dapergolas}, {David}, {De Luise}, {De
  March}, {De Ridder}, {de Souza}, {de Torres}, {del Peloso}, {del Pozo},
  {Delbo}, {Delgado}, {Delisle}, {Demouchy}, {Dharmawardena}, {Di Matteo},
  {Diakite}, {Diener}, {Distefano}, {Dolding}, {Edvardsson}, {Enke}, {Fabre},
  {Fabrizio}, {Faigler}, {Fedorets}, {Fernique}, {Figueras}, {Fournier},
  {Fouron}, {Fragkoudi}, {Gai}, {Garcia-Gutierrez}, {Garcia-Reinaldos},
  {Garc{\'\i}a-Torres}, {Garofalo}, {Gavel}, {Gavras}, {Gerlach}, {Geyer},
  {Giacobbe}, {Gilmore}, {Girona}, {Giuffrida}, {Gomel}, {Gomez},
  {Gonz{\'a}lez-N{\'u}{\~n}ez}, {Gonz{\'a}lez-Santamar{\'\i}a},
  {Gonz{\'a}lez-Vidal}, {Granvik}, {Guillout}, {Guiraud},
  {Guti{\'e}rrez-S{\'a}nchez}, {Guy}, {Hatzidimitriou}, {Hauser}, {Haywood},
  {Helmer}, {Helmi}, {Sarmiento}, {Hidalgo}, {H{\l}adczuk}, {Hobbs}, {Holland},
  {Huckle}, {Jardine}, {Jasniewicz}, {Jean-Antoine Piccolo},
  {Jim{\'e}nez-Arranz}, {Juaristi Campillo}, {Julbe}, {Karbevska}, {Kervella},
  {Khanna}, {Korn}, {K{\'o}sp{\'a}l}, {Kostrzewa-Rutkowska}, {Kruszy{\'n}ska},
  {Kun}, {Laizeau}, {Lambert}, {Lanza}, {Lasne}, {Le Campion}, {Lebreton},
  {Lebzelter}, {Leccia}, {Leclerc}, {Lecoeur-Taibi}, {Liao}, {Licata},
  {Lindstr{\o}m}, {Lister}, {Livanou}, {Lobel}, {Lorca}, {Loup}, {Madrero
  Pardo}, {Magdaleno Romeo}, {Managau}, {Mann}, {Manteiga}, {Marchant},
  {Marconi}, {Marcos}, {Marcos Santos}, {Mar{\'\i}n Pina}, {Marinoni},
  {Marocco}, {Marshall}, {Martin Polo}, {Mart{\'\i}n-Fleitas}, {Marton},
  {Mary}, {Masip}, {Massari}, {Mastrobuono-Battisti}, {Mazeh}, {Messina},
  {Michalik}, {Millar}, {Mints}, {Molina}, {Molinaro}, {Moln{\'a}r}, {Monari},
  {Mongui{\'o}}, {Montegriffo}, {Montero}, {Mor}, {Mora}, {Morbidelli},
  {Morel}, {Morris}, {Muraveva}, {Murphy}, {Musella}, {Nagy}, {Noval},
  {Oca{\~n}a}, {Ogden}, {Ordenovic}, {Osinde}, {Pagani}, {Pagano}, {Palaversa},
  {Pallas-Quintela}, {Panahi}, {Payne-Wardenaar}, {Pe{\~n}alosa Esteller},
  {Penttil{\"a}}, {Pichon}, {Piersimoni}, {Pineau}, {Plachy}, {Plum},
  {Pr{\v{s}}a}, {Pulone}, {Racero}, {Ragaini}, {Rainer}, {Raiteri}, {Ramos},
  {Ramos-Lerate}, {Regibo}, {Richards}, {Rios Diaz}, {Ripepi}, {Riva}, {Rix},
  {Rixon}, {Robichon}, {Robin}, {Robin}, {Roelens}, {Rogues}, {Rohrbasser},
  {Romero-G{\'o}mez}, {Rowell}, {Royer}, {Ruz Mieres}, {Rybicki}, {Sadowski},
  {S{\'a}ez N{\'u}{\~n}ez}, {Sagrist{\`a} Sell{\'e}s}, {Sahlmann}, {Salguero},
  {Samaras}, {Sanchez Gimenez}, {Sanna}, {Santove{\~n}a}, {Sarasso}, {Sciacca},
  {Segol}, {Segovia}, {S{\'e}gransan}, {Semeux}, {Shahaf}, {Siddiqui},
  {Siebert}, {Siltala}, {Silvelo}, {Slezak}, {Slezak}, {Smart}, {Snaith},
  {Solano}, {Solitro}, {Souami}, {Souchay}, {Spoto}, {Steele},
  {Steidelm{\"u}ller}, {Stephenson}, {S{\"u}veges}, {Surdej}, {Szabados},
  {Szegedi-Elek}, {Taris}, {Taylor}, {Teixeira}, {Tolomei}, {Tonello}, {Torra},
  {Torra}, {Torralba Elipe}, {Trabucchi}, {Tsounis}, {Turon}, {Ulla}, {Unger},
  {Vaillant}, {van Dillen}, {van Reeven}, {Vanel}, {Vecchiato}, {Viala},
  {Vicente}, {Voutsinas}, {Weiler}, {Wevers}, {Wyrzykowski}, {Yoldas}, {Yvard},
  {Zhao}, {Zorec}, {Zucker}, \& {Zwitter}}]{recioblanco23}
{Gaia Collaboration}, {Recio-Blanco}, A., {Kordopatis}, G., {et~al.} 2023,
  \aap, 674, A38

\bibitem[{{Gerber} {et~al.}(2023){Gerber}, {Magg}, {Plez}, {Bergemann},
  {Heiter}, {Olander}, \& {Hoppe}}]{gerber23}
{Gerber}, J.~M., {Magg}, E., {Plez}, B., {et~al.} 2023, \aap, 669, A43

\bibitem[{{Grassitelli} {et~al.}(2015){Grassitelli}, {Fossati}, {Langer},
  {Miglio}, {Istrate}, \& {Sanyal}}]{grassitelli15}
{Grassitelli}, L., {Fossati}, L., {Langer}, N., {et~al.} 2015, \aap, 584, L2

\bibitem[{{Gratton} {et~al.}(2019){Gratton}, {Bragaglia}, {Carretta},
  {D'Orazi}, {Lucatello}, \& {Sollima}}]{gratton19}
{Gratton}, R., {Bragaglia}, A., {Carretta}, E., {et~al.} 2019, \aapr, 27, 8

\bibitem[{{Grevesse} {et~al.}(2007){Grevesse}, {Asplund}, \&
  {Sauval}}]{gravesse07}
{Grevesse}, N., {Asplund}, M., \& {Sauval}, A.~J. 2007, \ssr, 130, 105

\bibitem[{{Griggio} {et~al.}(2022){Griggio}, {Salaris}, {Cassisi},
  {Pietrinferni}, \& {Bedin}}]{griggio22}
{Griggio}, M., {Salaris}, M., {Cassisi}, S., {Pietrinferni}, A., \& {Bedin},
  L.~R. 2022, \mnras, 516, 3631

\bibitem[{{Grilo} {et~al.}(2024){Grilo}, {Souto}, {Cunha}, {Guer{\c{c}}o},
  {Vieira}, {Smith}, {Vilar}, {Silva-Andrade}, {Wanderley}, {Daflon}, \&
  {Silva}}]{grilo24}
{Grilo}, V., {Souto}, D., {Cunha}, K., {et~al.} 2024, \mnras, 534, 3005

\bibitem[{{Gustafsson} {et~al.}(2008){Gustafsson}, {Edvardsson}, {Eriksson},
  {J{\o}rgensen}, {Nordlund}, \& {Plez}}]{gustafsson08}
{Gustafsson}, B., {Edvardsson}, B., {Eriksson}, K., {et~al.} 2008, \aap, 486,
  951

\bibitem[{{Hayden} {et~al.}(2022){Hayden}, {Sharma}, {Bland-Hawthorn}, {Spina},
  {Buder}, {Ciuc{\u{a}}}, {Asplund}, {Casey}, {De Silva}, {D'Orazi}, {Freeman},
  {Kos}, {Lewis}, {Lin}, {Lind}, {Martell}, {Schlesinger}, {Simpson}, {Zucker},
  {Zwitter}, {Chen}, {{\v{C}}otar}, {Feuillet}, {Horner}, {Joyce},
  {Nordlander}, {Stello}, {Tepper-Garcia}, {Ting}, {Wang}, {Wittenmyer}, \&
  {Wyse}}]{hayden22}
{Hayden}, M.~R., {Sharma}, S., {Bland-Hawthorn}, J., {et~al.} 2022, \mnras,
  517, 5325

\bibitem[{{Heiter} {et~al.}(2015){Heiter}, {Jofr{\'e}}, {Gustafsson}, {Korn},
  {Soubiran}, \& {Th{\'e}venin}}]{heiter15}
{Heiter}, U., {Jofr{\'e}}, P., {Gustafsson}, B., {et~al.} 2015, \aap, 582, A49

\bibitem[{{Heiter} {et~al.}(2021){Heiter}, {Lind}, {Bergemann}, {Asplund},
  {Mikolaitis}, {Barklem}, {Masseron}, {de Laverny}, {Magrini}, {Edvardsson},
  {J{\"o}nsson}, {Pickering}, {Ryde}, {Bayo Ar{\'a}n}, {Bensby}, {Casey},
  {Feltzing}, {Jofr{\'e}}, {Korn}, {Pancino}, {Damiani}, {Lanzafame}, {Lardo},
  {Monaco}, {Morbidelli}, {Smiljanic}, {Worley}, {Zaggia}, {Randich}, \&
  {Gilmore}}]{heiter21}
{Heiter}, U., {Lind}, K., {Bergemann}, M., {et~al.} 2021, \aap, 645, A106

\bibitem[{{Hunt} \& {Reffert}(2024)}]{hunt24}
{Hunt}, E.~L. \& {Reffert}, S. 2024, \aap, 686, A42

\bibitem[{{Jin} {et~al.}(2024){Jin}, {Trager}, {Dalton}, {Aguerri}, {Drew},
  {Falc{\'o}n-Barroso}, {G{\"a}nsicke}, {Hill}, {Iovino}, {Pieri}, {Poggianti},
  {Smith}, {Vallenari}, {Abrams}, {Aguado}, {Antoja}, {Arag{\'o}n-Salamanca},
  {Ascasibar}, {Babusiaux}, {Balcells}, {Barrena}, {Battaglia}, {Belokurov},
  {Bensby}, {Bonifacio}, {Bragaglia}, {Carrasco}, {Carrera}, {Cornwell},
  {Dom{\'\i}nguez-Palmero}, {Duncan}, {Famaey}, {Fari{\~n}a}, {Gonzalez},
  {Guest}, {Hatch}, {Hess}, {Hoskin}, {Irwin}, {Knapen}, {Koposov}, {Kuchner},
  {Laigle}, {Lewis}, {Longhetti}, {Lucatello}, {M{\'e}ndez-Abreu}, {Mercurio},
  {Molaeinezhad}, {Mongui{\'o}}, {Morrison}, {Murphy}, {Peralta de Arriba},
  {P{\'e}rez}, {P{\'e}rez-R{\`a}fols}, {Pic{\'o}}, {Raddi}, {Romero-G{\'o}mez},
  {Royer}, {Siebert}, {Seabroke}, {Som}, {Terrett}, {Thomas}, {Wesson},
  {Worley}, {Alfaro}, {Allende Prieto}, {Alonso-Santiago}, {Amos}, {Ashley},
  {Balaguer-N{\'u}{\~n}ez}, {Balbinot}, {Bellazzini}, {Benn}, {Berlanas},
  {Bernard}, {Best}, {Bettoni}, {Bianco}, {Bishop}, {Blomqvist}, {Boeche},
  {Bolzonella}, {Bonoli}, {Bosma}, {Britavskiy}, {Busarello}, {Caffau},
  {Cantat-Gaudin}, {Castro-Ginard}, {Couto}, {Carbajo-Hijarrubia}, {Carter},
  {Casamiquela}, {Conrado}, {Corcho-Caballero}, {Costantin}, {Deason}, {de
  Burgos}, {De Grandi}, {Di Matteo}, {Dom{\'\i}nguez-G{\'o}mez}, {Dorda},
  {Drake}, {Dutta}, {Erkal}, {Feltzing}, {Ferr{\'e}-Mateu}, {Feuillet},
  {Figueras}, {Fossati}, {Franciosini}, {Frasca}, {Fumagalli}, {Gallazzi},
  {Garc{\'\i}a-Benito}, {Gentile Fusillo}, {Gebran}, {Gilbert}, {Gledhill},
  {Gonz{\'a}lez Delgado}, {Greimel}, {Guarcello}, {Guerra}, {Gullieuszik},
  {Haines}, {Hardcastle}, {Harris}, {Haywood}, {Helmi}, {Hernandez}, {Herrero},
  {Hughes}, {Ir{\v{s}}i{\v{c}}}, {Jablonka}, {Jarvis}, {Jordi}, {Kondapally},
  {Kordopatis}, {Krogager}, {La Barbera}, {Lam}, {Larsen}, {Lemasle}, {Lewis},
  {Lhom{\'e}}, {Lind}, {Lodi}, {Longobardi}, {Lonoce}, {Magrini}, {Ma{\'\i}z
  Apell{\'a}niz}, {Marchal}, {Marco}, {Martin}, {Matsuno}, {Maurogordato},
  {Merluzzi}, {Miralda-Escud{\'e}}, {Molinari}, {Monari}, {Morelli}, {Mottram},
  {Naylor}, {Negueruela}, {O{\~n}orbe}, {Pancino}, {Peirani}, {Peletier},
  {Pozzetti}, {Rainer}, {Ramos}, {Read}, {Rossi}, {R{\"o}ttgering},
  {Rubi{\~n}o-Mart{\'\i}n}, {Sabater}, {San Juan}, {Sanna}, {Schallig},
  {Schiavon}, {Schultheis}, {Serra}, {Shimwell}, {Sim{\'o}n-D{\'\i}az},
  {Smith}, {Sordo}, {Sorini}, {Soubiran}, {Starkenburg}, {Steele}, {Stott},
  {Stuik}, {Tolstoy}, {Tortora}, {Tsantaki}, {Van der Swaelmen}, {van Weeren},
  {Vergani}, {Verheijen}, {Verro}, {Vink}, {Vioque}, {Walcher}, {Walton},
  {Wegg}, {Weijmans}, {Williams}, {Wilson}, {Wright}, {Xylakis-Dornbusch},
  {Youakim}, {Zibetti}, \& {Zurita}}]{jin24}
{Jin}, S., {Trager}, S.~C., {Dalton}, G.~B., {et~al.} 2024, \mnras, 530, 2688

\bibitem[{Jofré {et~al.}(2018)Jofré, Heiter, Maia, Soubiran, Worley, Hawkins,
  Blanco-Cuaresma, \& Rodrigo}]{jofre18}
Jofré, P., Heiter, U., Maia, M.~T., {et~al.} 2018, Research Notes of the AAS,
  2, 152

\bibitem[{{Kos} {et~al.}(2019){Kos}, {Bland-Hawthorn}, {Asplund}, {Buder},
  {Lewis}, {Lin}, {Martell}, {Ness}, {Sharma}, {De Silva}, {Simpson}, {Zucker},
  {Zwitter}, {{\v{C}}otar}, \& {Spina}}]{kos19}
{Kos}, J., {Bland-Hawthorn}, J., {Asplund}, M., {et~al.} 2019, \aap, 631, A166

\bibitem[{{Kos} {et~al.}(2021){Kos}, {Bland-Hawthorn}, {Buder}, {Nordlander},
  {Spina}, {Beeson}, {Lind}, {Asplund}, {Freeman}, {Hayden}, {Lewis},
  {Martell}, {Sharma}, {De Silva}, {Simpson}, {Zucker}, {Zwitter},
  {{\v{C}}otar}, {Horner}, {Ting}, \& {Traven}}]{kos21}
{Kos}, J., {Bland-Hawthorn}, J., {Buder}, S., {et~al.} 2021, \mnras, 506, 4232

\bibitem[{{Kos} {et~al.}(2017){Kos}, {Lin}, {Zwitter}, {{\v{Z}}erjal},
  {Sharma}, {Bland-Hawthorn}, {Asplund}, {Casey}, {De Silva}, {Freeman},
  {Martell}, {Simpson}, {Schlesinger}, {Zucker}, {Anguiano}, {Bacigalupo},
  {Bedding}, {Betters}, {Da Costa}, {Duong}, {Hyde}, {Ireland}, {Kafle},
  {Leon-Saval}, {Lewis}, {Munari}, {Nataf}, {Stello}, {Tinney}, {Traven},
  {Watson}, \& {Wittenmyer}}]{kos17}
{Kos}, J., {Lin}, J., {Zwitter}, T., {et~al.} 2017, \mnras, 464, 1259

\bibitem[{{Krumholz} {et~al.}(2019){Krumholz}, {McKee}, \&
  {Bland-Hawthorn}}]{krumholz19}
{Krumholz}, M.~R., {McKee}, C.~F., \& {Bland-Hawthorn}, J. 2019, \araa, 57, 227

\bibitem[{{Larson}(1981)}]{larson81}
{Larson}, R.~B. 1981, \mnras, 194, 809

\bibitem[{{Leung} \& {Bovy}(2019)}]{leung19}
{Leung}, H.~W. \& {Bovy}, J. 2019, \mnras, 483, 3255

\bibitem[{{Lewis} {et~al.}(2002){Lewis}, {Cannon}, {Taylor}, {Glazebrook},
  {Bailey}, {Baldry}, {Barton}, {Bridges}, {Dalton}, {Farrell}, {Gray},
  {Lankshear}, {McCowage}, {Parry}, {Sharples}, {Shortridge}, {Smith},
  {Stevenson}, {Straede}, {Waller}, {Whittard}, {Wilcox}, \&
  {Willis}}]{lewis02}
{Lewis}, I.~J., {Cannon}, R.~D., {Taylor}, K., {et~al.} 2002, \mnras, 333, 279

\bibitem[{{Lin} {et~al.}(2018){Lin}, {Dotter}, {Ting}, \& {Asplund}}]{lin18}
{Lin}, J., {Dotter}, A., {Ting}, Y.-S., \& {Asplund}, M. 2018, \mnras, 477,
  2966

\bibitem[{{Lind} {et~al.}(2009){Lind}, {Asplund}, \& {Barklem}}]{lind09}
{Lind}, K., {Asplund}, M., \& {Barklem}, P.~S. 2009, \aap, 503, 541

\bibitem[{{Lind} {et~al.}(2011){Lind}, {Asplund}, {Barklem}, \&
  {Belyaev}}]{lind11}
{Lind}, K., {Asplund}, M., {Barklem}, P.~S., \& {Belyaev}, A.~K. 2011, \aap,
  528, A103

\bibitem[{{Lindegren} {et~al.}(2021){Lindegren}, {Klioner}, {Hern{\'a}ndez},
  {Bombrun}, {Ramos-Lerate}, {Steidelm{\"u}ller}, {Bastian}, {Biermann}, {de
  Torres}, {Gerlach}, {Geyer}, {Hilger}, {Hobbs}, {Lammers}, {McMillan},
  {Stephenson}, {Casta{\~n}eda}, {Davidson}, {Fabricius}, {Gracia-Abril},
  {Portell}, {Rowell}, {Teyssier}, {Torra}, {Bartolom{\'e}}, {Clotet},
  {Garralda}, {Gonz{\'a}lez-Vidal}, {Torra}, {Abbas}, {Altmann}, {Anglada
  Varela}, {Balaguer-N{\'u}{\~n}ez}, {Balog}, {Barache}, {Becciani}, {Bernet},
  {Bertone}, {Bianchi}, {Bouquillon}, {Brown}, {Bucciarelli}, {Busonero},
  {Butkevich}, {Buzzi}, {Cancelliere}, {Carlucci}, {Charlot}, {Cioni},
  {Crosta}, {Crowley}, {del Peloso}, {del Pozo}, {Drimmel}, {Esquej}, {Fienga},
  {Fraile}, {Gai}, {Garcia-Reinaldos}, {Guerra}, {Hambly}, {Hauser},
  {Jan{\ss}en}, {Jordan}, {Kostrzewa-Rutkowska}, {Lattanzi}, {Liao}, {Licata},
  {Lister}, {L{\"o}ffler}, {Marchant}, {Masip}, {Mignard}, {Mints}, {Molina},
  {Mora}, {Morbidelli}, {Murphy}, {Pagani}, {Panuzzo}, {Pe{\~n}alosa Esteller},
  {Poggio}, {Re Fiorentin}, {Riva}, {Sagrist{\`a} Sell{\'e}s}, {Sanchez
  Gimenez}, {Sarasso}, {Sciacca}, {Siddiqui}, {Smart}, {Souami}, {Spagna},
  {Steele}, {Taris}, {Utrilla}, {van Reeven}, \& {Vecchiato}}]{lindegren21}
{Lindegren}, L., {Klioner}, S.~A., {Hern{\'a}ndez}, J., {et~al.} 2021, \aap,
  649, A2

\bibitem[{Loaiza-Tacuri {et~al.}(2023)Loaiza-Tacuri, Cunha, Souto, Smith,
  Guerço, Chiappini, Sales-Silva, Horta, Prieto, Beaton, Bizyaev, Daflon,
  Frinchaboy, Hasselquist, Hayes, Holtzman, Jönsson, Majewski, Mészáros,
  Nidever, Pinsonneault, \& Zasowski}]{loaiza23}
Loaiza-Tacuri, V., Cunha, K., Souto, D., {et~al.} 2023, Monthly Notices of the
  Royal Astronomical Society, 526, 2378

\bibitem[{{Magic} {et~al.}(2013){Magic}, {Collet}, {Asplund}, {Trampedach},
  {Hayek}, {Chiavassa}, {Stein}, \& {Nordlund}}]{magic13}
{Magic}, Z., {Collet}, R., {Asplund}, M., {et~al.} 2013, \aap, 557, A26

\bibitem[{{Magrini} {et~al.}(2023){Magrini}, {Viscasillas V{\'a}zquez},
  {Spina}, {Randich}, {Romano}, {Franciosini}, {Recio-Blanco}, {Nordlander},
  {D'Orazi}, {Baratella}, {Smiljanic}, {Dantas}, {Pasquini}, {Spitoni},
  {Casali}, {Van der Swaelmen}, {Bensby}, {Stonkute}, {Feltzing}, {Sacco},
  {Bragaglia}, {Pancino}, {Heiter}, {Biazzo}, {Gilmore}, {Bergemann},
  {Tautvai{\v{s}}ien{\.{e}}}, {Worley}, {Hourihane}, {Gonneau}, \&
  {Morbidelli}}]{magrini23}
{Magrini}, L., {Viscasillas V{\'a}zquez}, C., {Spina}, L., {et~al.} 2023, \aap,
  669, A119

\bibitem[{{Marigo} {et~al.}(2017){Marigo}, {Girardi}, {Bressan}, {Rosenfield},
  {Aringer}, {Chen}, {Dussin}, {Nanni}, {Pastorelli}, {Rodrigues}, {Trabucchi},
  {Bladh}, {Dalcanton}, {Groenewegen}, {Montalb{\'a}n}, \& {Wood}}]{marigo17}
{Marigo}, P., {Girardi}, L., {Bressan}, A., {et~al.} 2017, \apj, 835, 77

\bibitem[{{Mar{\'\i}n-Franch} {et~al.}(2009){Mar{\'\i}n-Franch}, {Aparicio},
  {Piotto}, {Rosenberg}, {Chaboyer}, {Sarajedini}, {Siegel}, {Anderson},
  {Bedin}, {Dotter}, {Hempel}, {King}, {Majewski}, {Milone}, {Paust}, \&
  {Reid}}]{franch09}
{Mar{\'\i}n-Franch}, A., {Aparicio}, A., {Piotto}, G., {et~al.} 2009, \apj,
  694, 1498

\bibitem[{{Martell} {et~al.}(2017){Martell}, {Sharma}, {Buder}, {Duong},
  {Schlesinger}, {Simpson}, {Lind}, {Ness}, {Marshall}, {Asplund},
  {Bland-Hawthorn}, {Casey}, {De Silva}, {Freeman}, {Kos}, {Lin}, {Zucker},
  {Zwitter}, {Anguiano}, {Bacigalupo}, {Carollo}, {Casagrande}, {Da Costa},
  {Horner}, {Huber}, {Hyde}, {Kafle}, {Lewis}, {Nataf}, {Navin}, {Stello},
  {Tinney}, {Watson}, \& {Wittenmyer}}]{martell17}
{Martell}, S.~L., {Sharma}, S., {Buder}, S., {et~al.} 2017, \mnras, 465, 3203

\bibitem[{{Mashonkina} {et~al.}(1999){Mashonkina}, {Gehren}, \&
  {Bikmaev}}]{mashonkina99}
{Mashonkina}, L., {Gehren}, T., \& {Bikmaev}, I. 1999, \aap, 343, 519

\bibitem[{{Mashonkina} {et~al.}(2007){Mashonkina}, {Korn}, \&
  {Przybilla}}]{mashonkina07}
{Mashonkina}, L., {Korn}, A.~J., \& {Przybilla}, N. 2007, \aap, 461, 261

\bibitem[{Massari {et~al.}(2017)Massari, Mucciarelli, Dalessandro, Bellazzini,
  Cassisi, Fiorentino, Ibata, Lardo, \& Salaris}]{massari17}
Massari, D., Mucciarelli, A., Dalessandro, E., {et~al.} 2017, Monthly Notices
  of the Royal Astronomical Society, 468, 1249

\bibitem[{{Milone} \& {Marino}(2022)}]{milone22}
{Milone}, A.~P. \& {Marino}, A.~F. 2022, Universe, 8, 359

\bibitem[{{Moedas} {et~al.}(2022){Moedas}, {Deal}, {Bossini}, \&
  {Campilho}}]{moedas22}
{Moedas}, N., {Deal}, M., {Bossini}, D., \& {Campilho}, B. 2022, \aap, 666, A43

\bibitem[{{Monaco} {et~al.}(2018){Monaco}, {Villanova}, {Carraro},
  {Mucciarelli}, \& {Moni Bidin}}]{monaco18}
{Monaco}, L., {Villanova}, S., {Carraro}, G., {Mucciarelli}, A., \& {Moni
  Bidin}, C. 2018, \aap, 616, A181

\bibitem[{{Mu{\~n}oz} {et~al.}(2021){Mu{\~n}oz}, {Geisler}, {Villanova},
  {Sarajedini}, {Frelijj}, {Vargas}, {Monaco}, \& {O'Connell}}]{munoz21}
{Mu{\~n}oz}, C., {Geisler}, D., {Villanova}, S., {et~al.} 2021, \mnras, 506,
  4676

\bibitem[{Mészáros {et~al.}(2012)Mészáros, Prieto, Edvardsson, Castelli,
  Pérez, Gustafsson, Majewski, Plez, Schiavon, Shetrone, \&
  de~Vicente}]{meszaros12}
Mészáros, S., Prieto, C.~A., Edvardsson, B., {et~al.} 2012, The Astronomical
  Journal, 144, 120

\bibitem[{{Nepal} {et~al.}(2023){Nepal}, {Guiglion}, {de Jong}, {Valentini},
  {Chiappini}, {Steinmetz}, {Ambrosch}, {Pancino}, {Jeffries}, {Bensby},
  {Romano}, {Smiljanic}, {Dantas}, {Gilmore}, {Randich}, {Bayo}, {Bergemann},
  {Franciosini}, {Jim{\'e}nez-Esteban}, {Jofr{\'e}}, {Morbidelli}, {Sacco},
  {Tautvai{\v{s}}ien{\.{e}}}, \& {Zaggia}}]{nepal23}
{Nepal}, S., {Guiglion}, G., {de Jong}, R.~S., {et~al.} 2023, \aap, 671, A61

\bibitem[{{Ness} {et~al.}(2015){Ness}, {Hogg}, {Rix}, {Ho}, \&
  {Zasowski}}]{ness15}
{Ness}, M., {Hogg}, D.~W., {Rix}, H.~W., {Ho}, A. Y.~Q., \& {Zasowski}, G.
  2015, \apj, 808, 16

\bibitem[{{Osorio} \& {Barklem}(2016)}]{osorio16}
{Osorio}, Y. \& {Barklem}, P.~S. 2016, \aap, 586, A120

\bibitem[{O’Malley \& Chaboyer(2018)}]{omalley18}
O’Malley, E.~M. \& Chaboyer, B. 2018, The Astrophysical Journal, 856, 130

\bibitem[{{Pastorelli} {et~al.}(2019){Pastorelli}, {Marigo}, {Girardi}, {Chen},
  {Rubele}, {Trabucchi}, {Aringer}, {Bladh}, {Bressan}, {Montalb{\'a}n},
  {Boyer}, {Dalcanton}, {Eriksson}, {Groenewegen}, {H{\"o}fner}, {Lebzelter},
  {Nanni}, {Rosenfield}, {Wood}, \& {Cioni}}]{pastorelli19}
{Pastorelli}, G., {Marigo}, P., {Girardi}, L., {et~al.} 2019, \mnras, 485, 5666

\bibitem[{{Piskunov} \& {Valenti}(2017)}]{piskunov17}
{Piskunov}, N. \& {Valenti}, J.~A. 2017, \aap, 597, A16

\bibitem[{{Piskunov} {et~al.}(1995){Piskunov}, {Kupka}, {Ryabchikova}, {Weiss},
  \& {Jeffery}}]{piskunov95}
{Piskunov}, N.~E., {Kupka}, F., {Ryabchikova}, T.~A., {Weiss}, W.~W., \&
  {Jeffery}, C.~S. 1995, \aaps, 112, 525

\bibitem[{{Plez}(2012)}]{plez12}
{Plez}, B. 2012, {Turbospectrum: Code for spectral synthesis}, Astrophysics
  Source Code Library, record ascl:1205.004

\bibitem[{{Poovelil} {et~al.}(2020){Poovelil}, {Zasowski}, {Hasselquist},
  {Seth}, {Donor}, {Beaton}, {Cunha}, {Frinchaboy},
  {Garc{\'\i}a-Hern{\'a}ndez}, {Hawkins}, {Kratter}, {Lane}, \&
  {Nitschelm}}]{poovelil20}
{Poovelil}, V.~J., {Zasowski}, G., {Hasselquist}, S., {et~al.} 2020, \apj, 903,
  55

\bibitem[{{Ratzenb{\"o}ck} {et~al.}(2023){Ratzenb{\"o}ck}, {Gro{\ss}schedl},
  {M{\"o}ller}, {Alves}, {Bomze}, \& {Meingast}}]{ratzenbock23}
{Ratzenb{\"o}ck}, S., {Gro{\ss}schedl}, J.~E., {M{\"o}ller}, T., {et~al.} 2023,
  \aap, 677, A59

\bibitem[{Reyes {et~al.}(2024)Reyes, Stello, Hon, Trampedach, Sandquist, \&
  Pinsonneault}]{reyes24}
Reyes, C., Stello, D., Hon, M., {et~al.} 2024, Monthly Notices of the Royal
  Astronomical Society, 532, 2860

\bibitem[{{Ryabchikova} {et~al.}(2015){Ryabchikova}, {Piskunov}, {Kurucz},
  {Stempels}, {Heiter}, {Pakhomov}, \& {Barklem}}]{ryabchikova15}
{Ryabchikova}, T., {Piskunov}, N., {Kurucz}, R.~L., {et~al.} 2015, \physscr,
  90, 054005

\bibitem[{{Schiavon} {et~al.}(2024){Schiavon}, {Phillips}, {Myers}, {Horta},
  {Minniti}, {Allende Prieto}, {Anguiano}, {Beaton}, {Beers}, {Brownstein},
  {Cohen}, {Fern{\'a}ndez-Trincado}, {Frinchaboy}, {J{\"o}nsson}, {Kisku},
  {Lane}, {Majewski}, {Mason}, {M{\'e}sz{\'a}ros}, \&
  {Stringfellow}}]{schiavon24}
{Schiavon}, R.~P., {Phillips}, S.~G., {Myers}, N., {et~al.} 2024, \mnras, 528,
  1393

\bibitem[{{Sheinis} {et~al.}(2015){Sheinis}, {Anguiano}, {Asplund},
  {Bacigalupo}, {Barden}, {Birchall}, {Bland-Hawthorn}, {Brzeski}, {Cannon},
  {Carollo}, {Case}, {Casey}, {Churilov}, {Warrick}, {Dean}, {De Silva},
  {D'Orazi}, {Duong}, {Farrell}, {Fiegert}, {Freeman}, {Gabriella}, {Gers},
  {Goodwin}, {Gray}, {Green}, {Heald}, {Heijmans}, {Ireland}, {Jones}, {Kafle},
  {Keller}, {Klauser}, {Kondrat}, {Kos}, {Lawrence}, {Lee}, {Mali}, {Martell},
  {Mathews}, {Mayfield}, {Miziarski}, {Muller}, {Pai}, {Patterson}, {Penny},
  {Orr}, {Schlesinger}, {Sharma}, {Shortridge}, {Simpson}, {Smedley}, {Smith},
  {Stafford}, {Staszak}, {Vuong}, {Waller}, {de Boer}, {Xavier}, {Zheng},
  {Zhelem}, {Zucker}, \& {Zwitter}}]{sheinis15}
{Sheinis}, A., {Anguiano}, B., {Asplund}, M., {et~al.} 2015, Journal of
  Astronomical Telescopes, Instruments, and Systems, 1, 035002

\bibitem[{{Sinha} {et~al.}(2024){Sinha}, {Zasowski}, {Frinchaboy}, {Cunha},
  {Souto}, {Tayar}, \& {Stassun}}]{sinha24}
{Sinha}, A., {Zasowski}, G., {Frinchaboy}, P., {et~al.} 2024, \apj, 975, 89

\bibitem[{{Sitnova} {et~al.}(2016){Sitnova}, {Mashonkina}, \&
  {Ryabchikova}}]{sitnova16}
{Sitnova}, T.~M., {Mashonkina}, L.~I., \& {Ryabchikova}, T.~A. 2016, \mnras,
  461, 1000

\bibitem[{{Skrutskie} {et~al.}(2006){Skrutskie}, {Cutri}, {Stiening},
  {Weinberg}, {Schneider}, {Carpenter}, {Beichman}, {Capps}, {Chester},
  {Elias}, {Huchra}, {Liebert}, {Lonsdale}, {Monet}, {Price}, {Seitzer},
  {Jarrett}, {Kirkpatrick}, {Gizis}, {Howard}, {Evans}, {Fowler}, {Fullmer},
  {Hurt}, {Light}, {Kopan}, {Marsh}, {McCallon}, {Tam}, {Van Dyk}, \&
  {Wheelock}}]{skrutskie06}
{Skrutskie}, M.~F., {Cutri}, R.~M., {Stiening}, R., {et~al.} 2006, \aj, 131,
  1163

\bibitem[{{Smith} {et~al.}(2021){Smith}, {Bizyaev}, {Cunha}, {Shetrone},
  {Souto}, {Allende Prieto}, {Masseron}, {M{\'e}sz{\'a}ros}, {J{\"o}nsson},
  {Hasselquist}, {Osorio}, {Garc{\'\i}a-Hern{\'a}ndez}, {Plez}, {Beaton},
  {Holtzman}, {Majewski}, {Stringfellow}, \& {Sobeck}}]{smith21}
{Smith}, V.~V., {Bizyaev}, D., {Cunha}, K., {et~al.} 2021, \aj, 161, 254

\bibitem[{{Smith} \& {Lambert}(1990)}]{smith90}
{Smith}, V.~V. \& {Lambert}, D.~L. 1990, \apjs, 72, 387

\bibitem[{{Spina} {et~al.}(2022){Spina}, {Magrini}, {Sacco}, {Casali},
  {Vallenari}, {Tautvai{\v{s}}ien{\.{e}}}, {Jim{\'e}nez-Esteban}, {Gilmore},
  {Randich}, {Feltzing}, {Jeffries}, {Bensby}, {Bragaglia}, {Smiljanic},
  {Carraro}, {Morbidelli}, \& {Zaggia}}]{spina22}
{Spina}, L., {Magrini}, L., {Sacco}, G.~G., {et~al.} 2022, \aap, 668, A16

\bibitem[{{Spina} {et~al.}(2020){Spina}, {Nordlander}, {Casey}, {Bedell},
  {D'Orazi}, {Mel{\'e}ndez}, {Karakas}, {Desidera}, {Baratella}, {Yana
  Galarza}, \& {Casali}}]{spina20}
{Spina}, L., {Nordlander}, T., {Casey}, A.~R., {et~al.} 2020, \apj, 895, 52

\bibitem[{{Spina} {et~al.}(2021){Spina}, {Ting}, {De Silva}, {Frankel},
  {Sharma}, {Cantat-Gaudin}, {Joyce}, {Stello}, {Karakas}, {Asplund},
  {Nordlander}, {Casagrande}, {D'Orazi}, {Casey}, {Cottrell},
  {Tepper-Garc{\'\i}a}, {Baratella}, {Kos}, {{\v{C}}otar}, {Bland-Hawthorn},
  {Buder}, {Freeman}, {Hayden}, {Lewis}, {Lin}, {Lind}, {Martell},
  {Schlesinger}, {Simpson}, {Zucker}, \& {Zwitter}}]{spina21}
{Spina}, L., {Ting}, Y.~S., {De Silva}, G.~M., {et~al.} 2021, \mnras, 503, 3279

\bibitem[{{Storm} \& {Bergemann}(2023)}]{storm23}
{Storm}, N. \& {Bergemann}, M. 2023, \mnras, 525, 3718

\bibitem[{{Sun} {et~al.}(2018){Sun}, {de Grijs}, {Cioni}, {Rubele},
  {Subramanian}, {van Loon}, {Bekki}, {Bell}, {Ivanov}, {Marconi}, {Muraveva},
  {Oliveira}, \& {Ripepi}}]{sun18}
{Sun}, N.-C., {de Grijs}, R., {Cioni}, M.-R.~L., {et~al.} 2018, \apj, 858, 31

\bibitem[{{Tang} {et~al.}(2014){Tang}, {Bressan}, {Rosenfield}, {Slemer},
  {Marigo}, {Girardi}, \& {Bianchi}}]{tang14}
{Tang}, J., {Bressan}, A., {Rosenfield}, P., {et~al.} 2014, \mnras, 445, 4287

\bibitem[{{Ting} {et~al.}(2019){Ting}, {Conroy}, {Rix}, \& {Cargile}}]{ting19}
{Ting}, Y.-S., {Conroy}, C., {Rix}, H.-W., \& {Cargile}, P. 2019, \apj, 879, 69

\bibitem[{{Valenti} \& {Piskunov}(1996)}]{valenti96}
{Valenti}, J.~A. \& {Piskunov}, N. 1996, \aaps, 118, 595

\bibitem[{{Vasiliev} \& {Baumgardt}(2021)}]{vasiliev21}
{Vasiliev}, E. \& {Baumgardt}, H. 2021, \mnras, 505, 5978

\bibitem[{{V{\'a}zquez-Semadeni} {et~al.}(2017){V{\'a}zquez-Semadeni},
  {Gonz{\'a}lez-Samaniego}, \& {Col{\'\i}n}}]{vazques17}
{V{\'a}zquez-Semadeni}, E., {Gonz{\'a}lez-Samaniego}, A., \& {Col{\'\i}n}, P.
  2017, \mnras, 467, 1313

\bibitem[{Virtanen {et~al.}(2020)Virtanen, Gommers, Oliphant, Haberland, Reddy,
  Cournapeau, Burovski, Peterson, Weckesser, Bright, {van der Walt}, Brett,
  Wilson, Millman, Mayorov, Nelson, Jones, Kern, Larson, Carey, Polat, Feng,
  Moore, {VanderPlas}, Laxalde, Perktold, Cimrman, Henriksen, Quintero, Harris,
  Archibald, Ribeiro, Pedregosa, {van Mulbregt}, \& {SciPy 1.0
  Contributors}}]{virtanen20}
Virtanen, P., Gommers, R., Oliphant, T.~E., {et~al.} 2020, Nature Methods, 17,
  261

\bibitem[{{{\v{Z}}erjal} {et~al.}(2023){{\v{Z}}erjal}, {Ireland}, {Crundall},
  {Krumholz}, \& {Rains}}]{zerjal23}
{{\v{Z}}erjal}, M., {Ireland}, M.~J., {Crundall}, T.~D., {Krumholz}, M.~R., \&
  {Rains}, A.~D. 2023, \mnras, 519, 3992

\bibitem[{{Wehrhahn} {et~al.}(2023){Wehrhahn}, {Piskunov}, \&
  {Ryabchikova}}]{wehrhahn23}
{Wehrhahn}, A., {Piskunov}, N., \& {Ryabchikova}, T. 2023, \aap, 671, A171

\bibitem[{Wheeler {et~al.}(2023)Wheeler, Abruzzo, Casey, \& Ness}]{wheeler23}
Wheeler, A.~J., Abruzzo, M.~W., Casey, A.~R., \& Ness, M.~K. 2023, The
  Astronomical Journal, 165, 68

\bibitem[{{Wheeler} {et~al.}(2024){Wheeler}, {Casey}, \& {Abruzzo}}]{wheeler24}
{Wheeler}, A.~J., {Casey}, A.~R., \& {Abruzzo}, M.~W. 2024, \aj, 167, 83

\bibitem[{{Zari} {et~al.}(2019){Zari}, {Brown}, \& {de Zeeuw}}]{zari19}
{Zari}, E., {Brown}, A.~G.~A., \& {de Zeeuw}, P.~T. 2019, \aap, 628, A123

\bibitem[{{Zari} {et~al.}(2018){Zari}, {Hashemi}, {Brown}, {Jardine}, \& {de
  Zeeuw}}]{zari18}
{Zari}, E., {Hashemi}, H., {Brown}, A.~G.~A., {Jardine}, K., \& {de Zeeuw},
  P.~T. 2018, \aap, 620, A172

\bibitem[{{Zhang} {et~al.}(2022){Zhang}, {Lucatello}, {Bragaglia},
  {Alonso-Santiago}, {Andreuzzi}, {Casali}, {Carrera}, {Carretta}, {D'Orazi},
  {Frasca}, {Fu}, {Magrini}, {Minchev}, {Origlia}, {Spina}, \&
  {Vallenari}}]{zhang22}
{Zhang}, R., {Lucatello}, S., {Bragaglia}, A., {et~al.} 2022, \aap, 667, A103

\end{thebibliography}

\begin{appendix}

\section{Trends}
\label{sec:aptrends}

Tables \ref{tab:coef_dwarfs} and \ref{tab:coef_giants} provide the knots and coefficients to reproduce the trends derived in Section \ref{sec:deriving_trends}. For elements not included in the tables, there were not enough data to produce meaningful trends. Trends are valid between temperatures given by the first and last knot (in kelvin). Trends are well constrained between $3500<T_\mathrm{eff}<6500\ \mathrm{K}$, and estimated to our best ability outside of these regions. We leave it to the readers discretion to decide in which temperature range they will use our trends. Instructions on how to evaluate the splines given by a list of knots and coefficients are given in the following section.

\begin{landscape}
\begin{table}
\caption{Table of knots and coefficients for trends fitted to dwarfs.}
    \centering
    \tiny
    \renewcommand{\arraystretch}{1.3}
    \begin{tabular}{lll}
         \hline
         Element & Knots / K & Coefficients\\\hline
         {[Fe/H]} & 3000.0, 3000.0, 3000.0, 3000.0, 3800.0, 4200.0, 4750.0, 5700.0, 7983.6, 7983.6, 7983.6, 7983.6 & 0.457653, 0.075255, -0.713959, -0.157331, -0.046431, 0.177822, -0.418889, 0.206850, 0.0, 0.0, 0.0, 0.0 \\
{[O/Fe]} & 3000.0, 3000.0, 3000.0, 3000.0, 3800.0, 4200.0, 4750.0, 5700.0, 7987.6, 7987.6, 7987.6, 7987.6 & -0.117032, -0.308347, 0.004433, 0.007909, -0.309385, 0.211845, 0.858807, -0.247306, 0.0, 0.0, 0.0, 0.0 \\
{[Na/Fe]} & 3000.0, 3000.0, 3000.0, 3000.0, 3800.0, 4200.0, 4750.0, 5700.0, 7985.0, 7985.0, 7985.0, 7985.0 & -0.185673, -0.085814, -0.027853, 0.001826, -0.022231, 0.046801, 0.159053, -0.147338, 0.0, 0.0, 0.0, 0.0 \\
{[Mg/Fe]} & 3000.0, 3000.0, 3000.0, 3000.0, 3800.0, 4200.0, 4750.0, 5700.0, 7987.6, 7987.6, 7987.6, 7987.6 & -0.232248, 0.277913, -0.153326, 0.067829, 0.169221, -0.269772, 0.121204, -0.197248, 0.0, 0.0, 0.0, 0.0 \\
{[Al/Fe]} & 3000.0, 3000.0, 3000.0, 3000.0, 3800.0, 4200.0, 4750.0, 5700.0, 7945.5, 7945.5, 7945.5, 7945.5 & -0.091229, -0.313054, 0.223800, 0.015607, -0.012922, 0.014934, -0.237792, -0.053959, 0.0, 0.0, 0.0, 0.0 \\
{[Si/Fe]} & 3000.0, 3000.0, 3000.0, 3000.0, 3800.0, 4200.0, 4750.0, 5700.0, 7985.0, 7985.0, 7985.0, 7985.0 & 0.038251, -0.199766, 0.524489, 0.311998, 0.106612, -0.185635, 0.204907, -0.102458, 0.0, 0.0, 0.0, 0.0 \\
{[K/Fe]} & 3000.0, 3000.0, 3000.0, 3000.0, 3800.0, 4200.0, 4750.0, 5700.0, 7983.6, 7983.6, 7983.6, 7983.6 & -0.412584, -0.292760, -0.281450, 0.078132, 0.136972, -0.188183, 0.044897, -0.420229, 0.0, 0.0, 0.0, 0.0 \\
{[Ca/Fe]} & 3000.0, 3000.0, 3000.0, 3000.0, 3800.0, 4200.0, 4750.0, 5700.0, 7985.0, 7985.0, 7985.0, 7985.0 & -0.065189, 0.032561, 0.085793, 0.014333, -0.017823, -0.008922, -0.189601, -0.193591, 0.0, 0.0, 0.0, 0.0 \\
{[Sc/Fe]} & 3000.0, 3000.0, 3000.0, 3000.0, 3800.0, 4200.0, 4750.0, 5700.0, 7985.0, 7985.0, 7985.0, 7985.0 & -0.001195, -0.128857, 0.125831, -0.119705, -0.024846, -0.023888, 0.316865, -0.044381, 0.0, 0.0, 0.0, 0.0 \\
{[Ti/Fe]} & 3000.0, 3000.0, 3000.0, 3000.0, 3800.0, 4200.0, 4750.0, 5700.0, 7987.6, 7987.6, 7987.6, 7987.6 & 0.098994, -0.078315, 0.428725, 0.024144, 0.043078, -0.099250, 0.191933, 0.213194, 0.0, 0.0, 0.0, 0.0 \\
{[V/Fe]} & 3000.0, 3000.0, 3000.0, 3000.0, 3800.0, 4200.0, 4750.0, 5700.0, 7985.0, 7985.0, 7985.0, 7985.0 & 0.270231, -0.145948, 0.335026, 0.152033, 0.035496, -0.095696, -0.134248, 0.469337, 0.0, 0.0, 0.0, 0.0 \\
{[Cr/Fe]} & 3000.0, 3000.0, 3000.0, 3000.0, 3800.0, 4200.0, 4750.0, 5700.0, 7985.0, 7985.0, 7985.0, 7985.0 & 0.052245, -0.274822, 0.156974, -0.053392, -0.055980, 0.096809, -0.111390, 0.333791, 0.0, 0.0, 0.0, 0.0 \\
{[Mn/Fe]} & 3000.0, 3000.0, 3000.0, 3000.0, 3800.0, 4200.0, 4750.0, 5700.0, 7987.6, 7987.6, 7987.6, 7987.6 & -0.165288, -0.071191, -0.117457, 0.030573, 0.117208, -0.170064, 0.194840, -0.161654, 0.0, 0.0, 0.0, 0.0 \\
{[Co/Fe]} & 3000.0, 3000.0, 3000.0, 3000.0, 3800.0, 4200.0, 4750.0, 5700.0, 7985.0, 7985.0, 7985.0, 7985.0 & 0.225778, 0.091096, 0.237710, 0.173320, -0.003998, -0.020661, 0.054034, -0.292333, 0.0, 0.0, 0.0, 0.0 \\
{[Ni/Fe]} & 3000.0, 3000.0, 3000.0, 3000.0, 3800.0, 4200.0, 4750.0, 5700.0, 7987.6, 7987.6, 7987.6, 7987.6 & -0.057951, 0.100797, -0.069027, 0.084350, 0.017405, 0.013622, -0.092709, 0.188395, 0.0, 0.0, 0.0, 0.0 \\
{[Cu/Fe]} & 3000.0, 3000.0, 3000.0, 3000.0, 3800.0, 4200.0, 4750.0, 5700.0, 7985.0, 7985.0, 7985.0, 7985.0 & 0.235171, 0.416619, 0.151908, 0.283906, -0.001337, 0.010870, 0.567931, -0.086920, 0.0, 0.0, 0.0, 0.0 \\
{[Zn/Fe]} & 3007.2, 3007.2, 3007.2, 3007.2, 3800.0, 4200.0, 4750.0, 5700.0, 7985.0, 7985.0, 7985.0, 7985.0 & -0.313631, 0.331193, -0.134803, 0.125649, 0.233005, -0.198277, 0.073113, 0.028699, 0.0, 0.0, 0.0, 0.0 \\
{[Rb/Fe]} & 3000.0, 3000.0, 3000.0, 3000.0, 3800.0, 4200.0, 4750.0, 5587.0, 5587.0, 5587.0, 5587.0 & 0.661036, 1.341827, 0.884854, -0.780483, -0.531555, -0.181466, 0.242901, 0.0, 0.0, 0.0, 0.0 \\
{[Sr/Fe]} & 3370.4, 3370.4, 3370.4, 3370.4, 3800.0, 4200.0, 4750.0, 5103.8, 5103.8, 5103.8, 5103.8 & -0.131555, 2.903368, -1.323664, 0.101044, -0.516985, 0.367483, -0.724400, 0.0, 0.0, 0.0, 0.0 \\
{[Y/Fe]} & 3000.0, 3000.0, 3000.0, 3000.0, 3800.0, 4200.0, 4750.0, 5700.0, 7987.6, 7987.6, 7987.6, 7987.6 & -0.219949, -0.216443, -0.243339, -0.067555, -0.102378, 0.115648, -0.318702, -0.045202, 0.0, 0.0, 0.0, 0.0 \\
{[Zr/Fe]} & 3000.0, 3000.0, 3000.0, 3000.0, 3800.0, 4200.0, 4750.0, 5700.0, 6295.9, 6295.9, 6295.9, 6295.9 & 0.101085, -0.108220, 0.097312, 0.377988, 0.033459, -0.023260, -0.257840, 0.020917, 0.0, 0.0, 0.0, 0.0 \\
{[Ba/Fe]} & 3000.0, 3000.0, 3000.0, 3000.0, 3800.0, 4200.0, 4750.0, 5700.0, 7985.0, 7985.0, 7985.0, 7985.0 & -0.646093, -0.383449, -0.538306, -0.001086, -0.040931, 0.182715, -0.245585, 0.503925, 0.0, 0.0, 0.0, 0.0 \\
{[Nd/Fe]} & 3000.0, 3000.0, 3000.0, 3000.0, 3800.0, 4200.0, 4750.0, 5700.0, 7945.5, 7945.5, 7945.5, 7945.5 & -0.201715, -0.582559, -0.175004, -0.503311, -0.062051, -0.234755, 0.764007, 0.801122, 0.0, 0.0, 0.0, 0.0 \\
{[Sm/Fe]} & 3000.0, 3000.0, 3000.0, 3000.0, 3800.0, 4200.0, 4750.0, 5700.0, 6266.1, 6266.1, 6266.1, 6266.1 & -0.543578, -0.776561, -0.420207, -0.371978, -0.687730, 0.447647, -0.495802, -0.618170, 0.0, 0.0, 0.0, 0.0 \\
        \hline
    \end{tabular}
    \label{tab:coef_dwarfs}
\end{table}
\end{landscape}

\begin{landscape}
\begin{table}
\caption{Table of knots and coefficients for trends fitted to giants. }
\tiny
    \centering
    \renewcommand{\arraystretch}{1.3}
    \begin{tabular}{lll}
         \hline
         Element & Knots / K & Coefficients\\\hline
         {[Fe/H]} & 3507.0, 3507.0, 3507.0, 3507.0, 3850.0, 4200.0, 4750.0, 5700.0, 7988.5, 7988.5, 7988.5, 7988.5 & 0.795987, 0.323125, -0.074245, 0.037688, -0.018581, 0.028052, -0.085221, -0.112136, 0.0, 0.0, 0.0, 0.0 \\
{[C/Fe]} & 3487.3, 3487.3, 3487.3, 3487.3, 3850.0, 4200.0, 4750.0, 5700.0, 7988.5, 7988.5, 7988.5, 7988.5 & 0.019854, -0.200670, -0.136232, -0.069324, -0.004219, 0.263794, -1.246279, -0.246377, 0.0, 0.0, 0.0, 0.0 \\
{[N/Fe]} & 3487.3, 3487.3, 3487.3, 3487.3, 3850.0, 4200.0, 4750.0, 5700.0, 6284.9, 6284.9, 6284.9, 6284.9 & -0.725755, -0.929724, -0.237811, -0.526673, 0.239244, 0.017337, 0.485075, 1.207699, 0.0, 0.0, 0.0, 0.0 \\
{[O/Fe]} & 3487.3, 3487.3, 3487.3, 3487.3, 4750.0, 5700.0, 7988.5, 7988.5, 7988.5, 7988.5 & -0.042376, 0.079526, -0.137399, 0.256680, 0.172891, -0.071140, 0.0, 0.0, 0.0, 0.0 \\
{[Na/Fe]} & 3487.3, 3487.3, 3487.3, 3487.3, 3850.0, 4200.0, 4750.0, 5700.0, 6451.8, 6451.8, 6451.8, 6451.8 & -0.222259, -0.428405, -0.211048, -0.063481, 0.085994, -0.213751, 0.043912, -0.167365, 0.0, 0.0, 0.0, 0.0 \\
{[Mg/Fe]} & 3487.3, 3487.3, 3487.3, 3487.3, 3850.0, 4200.0, 4750.0, 5700.0, 7988.5, 7988.5, 7988.5, 7988.5 & 0.040850, -0.190672, 0.109246, 0.032076, -0.010416, 0.035698, -0.061631, -0.232001, 0.0, 0.0, 0.0, 0.0 \\
{[Al/Fe]} & 3487.3, 3487.3, 3487.3, 3487.3, 3850.0, 4200.0, 4750.0, 5700.0, 6288.6, 6288.6, 6288.6, 6288.6 & -0.347546, -0.649171, -0.198878, 0.004306, 0.014533, 0.077579, -0.144337, -0.134961, 0.0, 0.0, 0.0, 0.0 \\
{[Si/Fe]} & 3487.3, 3487.3, 3487.3, 3487.3, 3850.0, 4200.0, 4750.0, 5700.0, 6451.8, 6451.8, 6451.8, 6451.8 & 0.167560, -0.008642, 0.189092, 0.063737, -0.066852, 0.133388, -0.137641, 0.126858, 0.0, 0.0, 0.0, 0.0 \\
{[K/Fe]} & 3487.3, 3487.3, 3487.3, 3487.3, 3850.0, 4200.0, 4750.0, 5700.0, 6481.0, 6481.0, 6481.0, 6481.0 & -0.309664, -0.289162, -0.072563, -0.100415, 0.030573, 0.063428, 0.246290, -0.076469, 0.0, 0.0, 0.0, 0.0 \\
{[Ca/Fe]} & 3487.3, 3487.3, 3487.3, 3487.3, 3850.0, 4200.0, 4750.0, 5700.0, 7988.5, 7988.5, 7988.5, 7988.5 & -0.106630, -0.157397, -0.057981, 0.043664, -0.045594, 0.142581, -0.398818, -0.405393, 0.0, 0.0, 0.0, 0.0 \\
{[Sc/Fe]} & 3487.3, 3487.3, 3487.3, 3487.3, 3850.0, 4200.0, 4750.0, 5700.0, 7988.5, 7988.5, 7988.5, 7988.5 & -0.201797, -0.117606, -0.102580, -0.069934, 0.003113, 0.175084, -0.211031, -0.084936, 0.0, 0.0, 0.0, 0.0 \\
{[Ti/Fe]} & 3487.3, 3487.3, 3487.3, 3487.3, 3850.0, 4200.0, 4750.0, 5700.0, 7988.5, 7988.5, 7988.5, 7988.5 & -0.281792, -0.189488, 0.011903, -0.023557, 0.016979, -0.037364, 0.269079, -0.254281, 0.0, 0.0, 0.0, 0.0 \\
{[V/Fe]} & 3487.3, 3487.3, 3487.3, 3487.3, 3850.0, 4200.0, 4750.0, 5700.0, 6288.6, 6288.6, 6288.6, 6288.6 & -0.205457, -0.155564, 0.009446, 0.126830, -0.082037, 0.002102, -0.258481, -0.046194, 0.0, 0.0, 0.0, 0.0 \\
{[Cr/Fe]} & 3487.3, 3487.3, 3487.3, 3487.3, 3850.0, 4200.0, 4750.0, 5700.0, 7988.5, 7988.5, 7988.5, 7988.5 & -0.033090, -0.013155, -0.079645, -0.050574, 0.012455, 0.053022, 0.049132, -0.049605, 0.0, 0.0, 0.0, 0.0 \\
{[Mn/Fe]} & 3487.3, 3487.3, 3487.3, 3487.3, 3850.0, 4200.0, 4750.0, 5700.0, 7871.7, 7871.7, 7871.7, 7871.7 & -0.019169, 0.284409, -0.065138, -0.000808, 0.007048, -0.071185, 0.044374, -0.175110, 0.0, 0.0, 0.0, 0.0 \\
{[Co/Fe]} & 3487.3, 3487.3, 3487.3, 3487.3, 3850.0, 4200.0, 4750.0, 5700.0, 6288.6, 6288.6, 6288.6, 6288.6 & 0.020023, 0.017998, -0.012003, 0.114879, -0.132135, 0.269998, -0.242511, -0.055271, 0.0, 0.0, 0.0, 0.0 \\
{[Ni/Fe]} & 3487.3, 3487.3, 3487.3, 3487.3, 3850.0, 4200.0, 4750.0, 5700.0, 7988.5, 7988.5, 7988.5, 7988.5 & -0.028563, -0.007236, 0.018166, 0.050022, -0.044344, 0.114811, -0.565099, 0.246756, 0.0, 0.0, 0.0, 0.0 \\
{[Cu/Fe]} & 3487.3, 3487.3, 3487.3, 3487.3, 3850.0, 4200.0, 4750.0, 5700.0, 6288.6, 6288.6, 6288.6, 6288.6 & 0.155441, 0.364094, 0.072602, -0.075157, -0.018064, 0.112507, -0.095726, 0.081581, 0.0, 0.0, 0.0, 0.0 \\
{[Zn/Fe]} & 3507.0, 3507.0, 3507.0, 3507.0, 3850.0, 4200.0, 4750.0, 5700.0, 6456.4, 6456.4, 6456.4, 6456.4 & -0.178851, -0.212781, -0.014498, -0.188704, 0.126466, -0.234194, 0.238270, -0.279677, 0.0, 0.0, 0.0, 0.0 \\
{[Rb/Fe]} & 3487.3, 3487.3, 3487.3, 3487.3, 3850.0, 4200.0, 4750.0, 5362.7, 5362.7, 5362.7, 5362.7 & -0.263860, -0.984955, -0.030933, -0.768906, 0.427089, -0.399146, 0.618583, 0.0, 0.0, 0.0, 0.0 \\
{[Sr/Fe]} & 3487.3, 3487.3, 3487.3, 3487.3, 3850.0, 4200.0, 4750.0, 5296.9, 5296.9, 5296.9, 5296.9 & 0.027300, 0.534891, -0.179765, 0.289023, -0.355190, 0.418015, -0.558731, 0.0, 0.0, 0.0, 0.0 \\
{[Y/Fe]} & 3487.3, 3487.3, 3487.3, 3487.3, 3850.0, 4200.0, 4750.0, 5700.0, 7988.5, 7988.5, 7988.5, 7988.5 & 0.065946, -0.036114, -0.098239, 0.023564, -0.040620, 0.171624, -0.070719, 0.116063, 0.0, 0.0, 0.0, 0.0 \\
{[Zr/Fe]} & 3487.3, 3487.3, 3487.3, 3487.3, 3850.0, 4200.0, 4750.0, 5700.0, 6266.1, 6266.1, 6266.1, 6266.1 & 0.026935, -0.149707, 0.074734, 0.097865, -0.123637, 0.189784, -0.368704, -0.046153, 0.0, 0.0, 0.0, 0.0 \\
{[Mo/Fe]} & 3487.3, 3487.3, 3487.3, 3487.3, 4750.0, 4929.1, 4929.1, 4929.1, 4929.1 & 0.210454, -1.332938, 0.814734, -0.217671, 1.589361, 0.0, 0.0, 0.0, 0.0 \\
{[Ru/Fe]} & 3487.3, 3487.3, 3487.3, 3487.3, 4750.0, 5180.9, 5180.9, 5180.9, 5180.9 & -0.645576, 0.023434, 0.277086, -0.157062, 0.035613, 0.0, 0.0, 0.0, 0.0 \\
{[Ba/Fe]} & 3487.3, 3487.3, 3487.3, 3487.3, 3850.0, 4200.0, 4750.0, 5700.0, 7988.5, 7988.5, 7988.5, 7988.5 & -0.112103, -0.262118, 0.031794, 0.091302, -0.073759, 0.392188, -0.797714, -0.239590, 0.0, 0.0, 0.0, 0.0 \\
{[La/Fe]} & 3487.3, 3487.3, 3487.3, 3487.3, 3850.0, 4200.0, 4750.0, 5700.0, 6284.9, 6284.9, 6284.9, 6284.9 & -0.024910, -0.524414, -0.033054, -0.005053, 0.004407, -0.017983, 0.103852, -0.127880, 0.0, 0.0, 0.0, 0.0 \\
{[Ce/Fe]} & 3487.3, 3487.3, 3487.3, 3487.3, 3850.0, 4200.0, 4750.0, 5700.0, 6284.9, 6284.9, 6284.9, 6284.9 & 0.274445, 0.307072, -0.280165, -0.158664, 0.012134, 0.310152, 0.315819, 0.021108, 0.0, 0.0, 0.0, 0.0 \\
{[Nd/Fe]} & 3487.3, 3487.3, 3487.3, 3487.3, 3850.0, 4200.0, 4750.0, 5700.0, 6288.6, 6288.6, 6288.6, 6288.6 & -0.068811, -0.501599, -0.022263, -0.109350, 0.056175, -0.004964, -0.048993, -0.073611, 0.0, 0.0, 0.0, 0.0 \\
{[Sm/Fe]} & 3487.3, 3487.3, 3487.3, 3487.3, 3850.0, 4200.0, 4750.0, 5700.0, 6266.1, 6266.1, 6266.1, 6266.1 & -0.346007, -0.653015, -0.175442, -0.003262, -0.093413, 0.363895, 0.420202, 0.178723, 0.0, 0.0, 0.0, 0.0 \\
{[Eu/Fe]} & 3487.3, 3487.3, 3487.3, 3487.3, 3850.0, 4200.0, 4750.0, 5856.9, 5856.9, 5856.9, 5856.9 & -0.358864, -0.279614, -0.602531, -0.034860, -0.009517, 0.024117, -0.074516, 0.0, 0.0, 0.0, 0.0 \\
         \hline
    \end{tabular}
    \label{tab:coef_giants}
\end{table}
\end{landscape}

\newpage

\section{Reproducing fitted trends}
\label{sec:splines}

We represented the trends with cubic plines, as described in Section \ref{sec:deriving_trends}. Tables \ref{tab:coef_dwarfs} and \ref{tab:coef_giants} provide a set of coefficients $c_i$ and knots $t_i$ for each element. The function describing a trend can be constructed as a sum of weighted base functions:
\begin{equation}
    S(x)=\sum_i c_i B_{i,n}(x),
\end{equation}
where base functions $B_{i,n}(x)$ are defined recursively:
\begin{align}
B_{i,0}(x) = &\begin{cases}
 1, & \text{if } t_i \leq x < t_{i+1}, \\
 0, & \text{otherwise}.
\end{cases}\\
B_{i,k}(x) = &\frac{x - t_i}{t_{i+k} - t_i} B_{i,k-1}(x) + \frac{t_{i+k+1} - x}{t_{i+k+1} - t_{i+1}} B_{i+1,k-1}(x),
\end{align}
and $k$ is the order of the spline ($k=3$ for a cubic spline in our case).

Below is a \texttt{Python} code to produce trends from a list of coefficients and a list of knots, as given in tables \ref{tab:coef_dwarfs} and \ref{tab:coef_giants}. The code is adapted from \url{https://docs.scipy.org/doc/scipy/reference/generated/scipy.interpolate.BSpline.html}. Variable \texttt{x} is a proxy for temperature in Kelvins. Function \texttt{bspline} returns the value for a trend as a function of temperature. \texttt{t} is a list of knots, and \texttt{c} is a list of coefficients.\\[0.5cm]
\begin{minipage}{\textwidth}
\begin{python}
def B(x, k, i, t):
    if t[i+k] == t[i]:
        c1 = 0.0
    else:
        c1 = (x - t[i])/(t[i+k] - t[i]) * B(x, k-1, i, t)
    if t[i+k+1] == t[i+1]:
        c2 = 0.0
    else:
        c2 = (t[i+k+1] - x)/(t[i+k+1] - t[i+1]) * B(x, k-1, i+1, t)
    return c1 + c2

def bspline(x, t, c):
    k = 3 # spline order (cubic spline)
    n = len(t) - k - 1
    assert (n >= k+1) and (len(c) >= n)
    return sum(c[i] * B(x, k, i, t) for i in range(n))
\end{python}
\vspace{1ex}
\end{minipage}

\newpage

\section{Trends for all elements}
\label{sec:all_trends}

\begin{figure*}
    \centering
    \includegraphics[width=\textwidth]{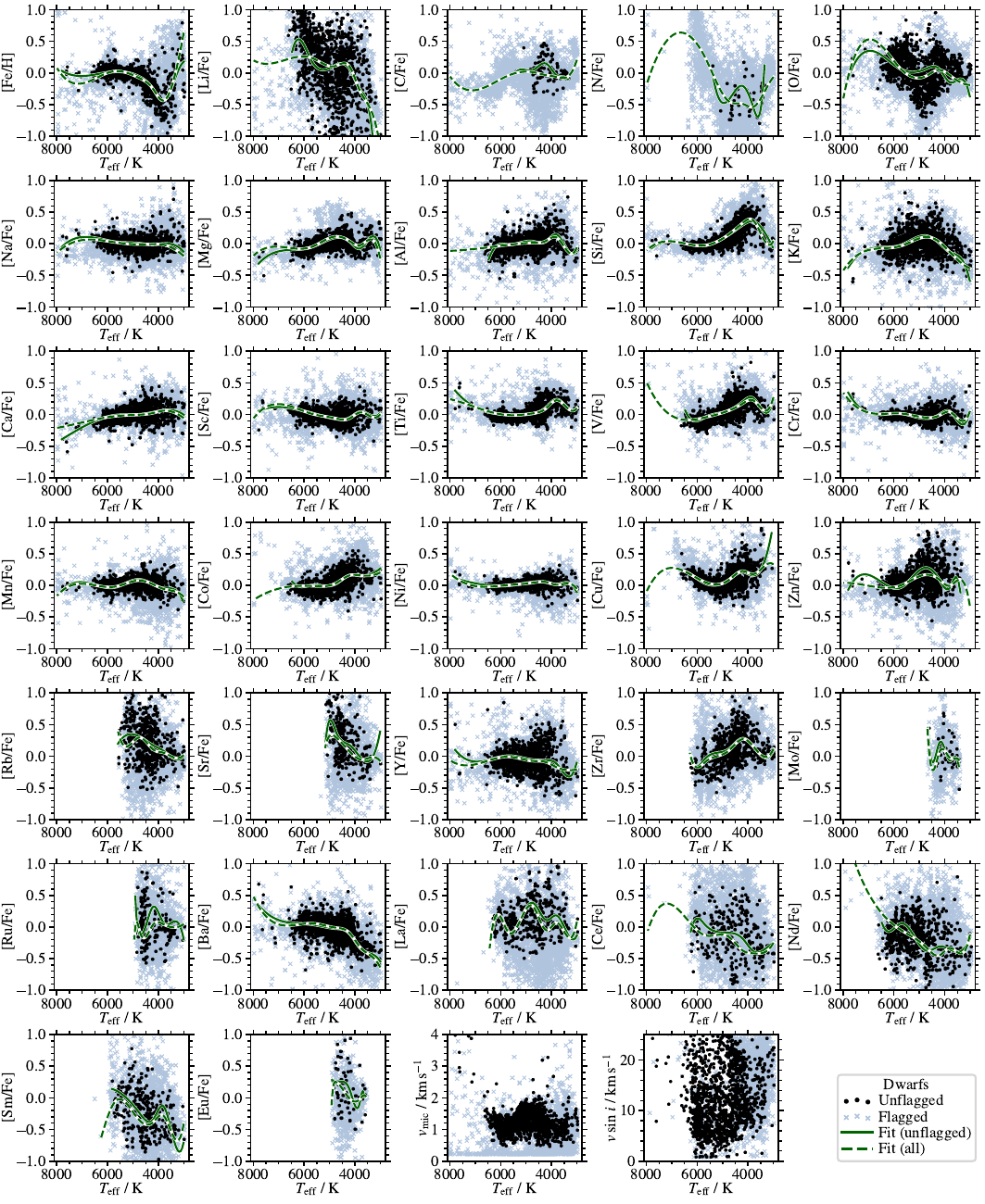}
    \caption{Trends of abundances of 32 chemical elements in dwarfs. Each panel shows the trends of abundances vs. effective temperature for one element, starting with Fe, and then continuing from small to large atomic numbers. Last couple of panels show the relations between microturbulence velocity ($v_\mathrm{mic}$), and rotational velocity ($v\,\sin i$) vs. effective temperature. Black points represent unflagged stars, grey crosses represent flagged stars. Solid lines are trends fitted on unflagged stars only, and dashed lines are trends fitted to all stars.}
    \label{fig:trends_all_dwarfs}
\end{figure*}

\begin{figure*}
    \centering
    \includegraphics[width=\textwidth]{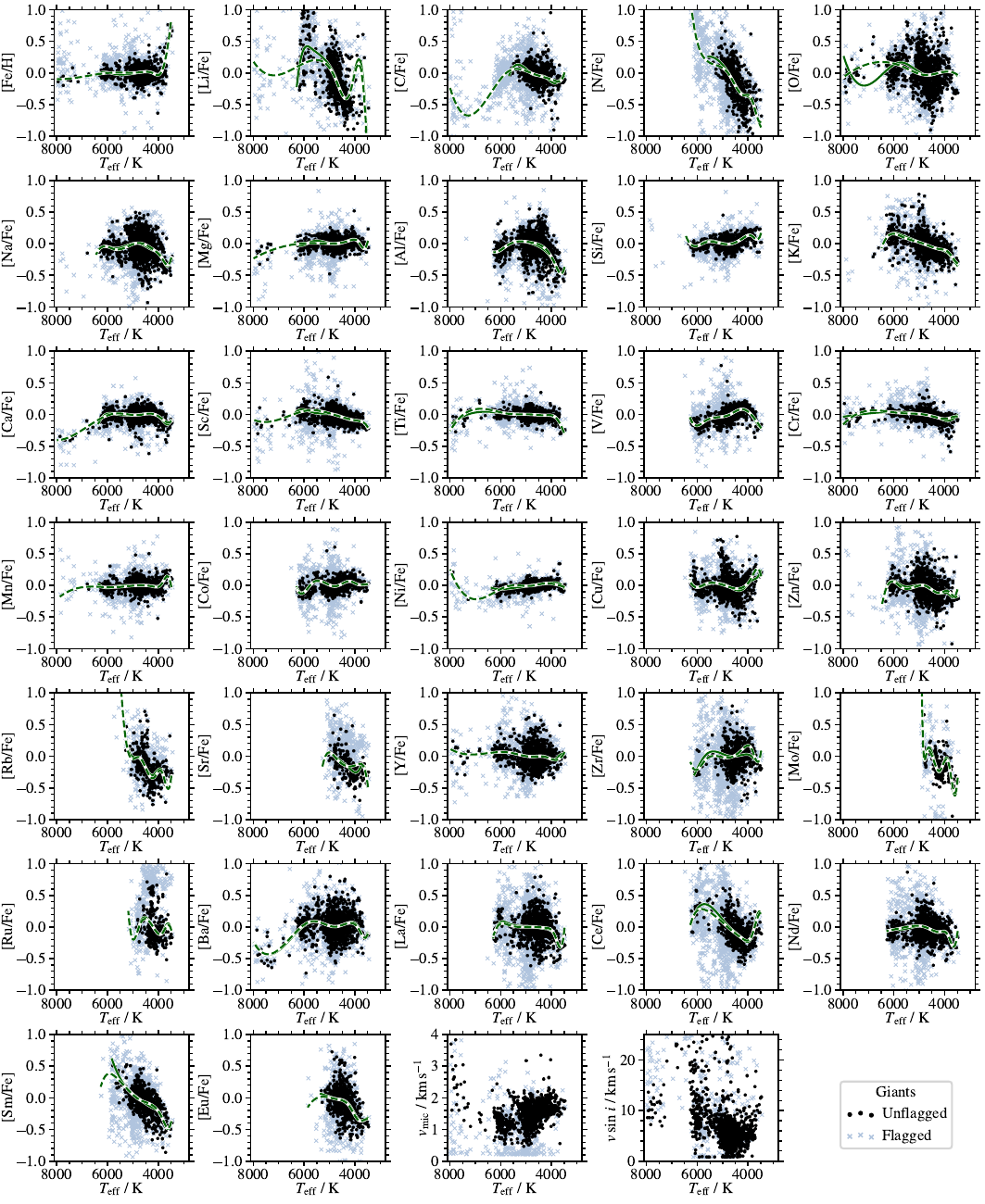}
    \caption{Trends of abundances of 32 chemical elements in giants. Each panel shows the trends of abundances vs. effective temperature for one element, starting with Fe, and then continuing from small to large atomic numbers. Last couple of panels show the relations between microturbulence velocity ($v_\mathrm{mic}$), and rotational velocity ($v\,\sin i$) vs. effective temperature. Black points represent unflagged stars, grey crosses represent flagged stars. Solid lines are trends fitted on unflagged stars only, and dashed lines are trends fitted to all stars.}
    \label{fig:trends_all_giants}
\end{figure*}

Here we plot the trends for all elements. Trends are shown separately for dwarfs and giants (Figures \ref{fig:trends_all_dwarfs} and \ref{fig:trends_all_giants}, respectively), for unflagged stars only and for all stars together (solid and dashed lines, respectively). Note that the lines showing the trends for flagged and unflagged stars often overlap. Microturbulence and rotational velocity as a function of temperature is also shown in the last two panels.

\newpage

\section{Age verification}
\label{sec:age_ap}

\begin{figure}[!h]
    \centering
    \includegraphics[width=\columnwidth]{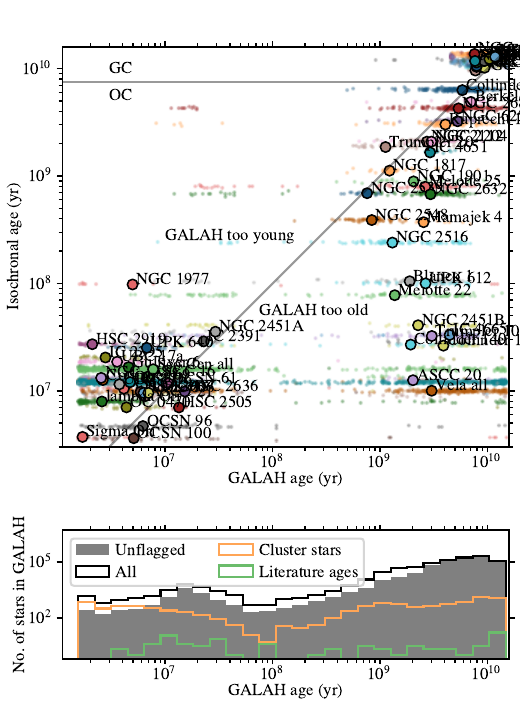}
    \caption{Same as Figure \ref{fig:ages_comp} but with marked names of some prominent clusters.}
    \label{fig:ages_comp_names}
\end{figure}

\begin{figure}[!h]
    \centering
    \includegraphics[width=\columnwidth]{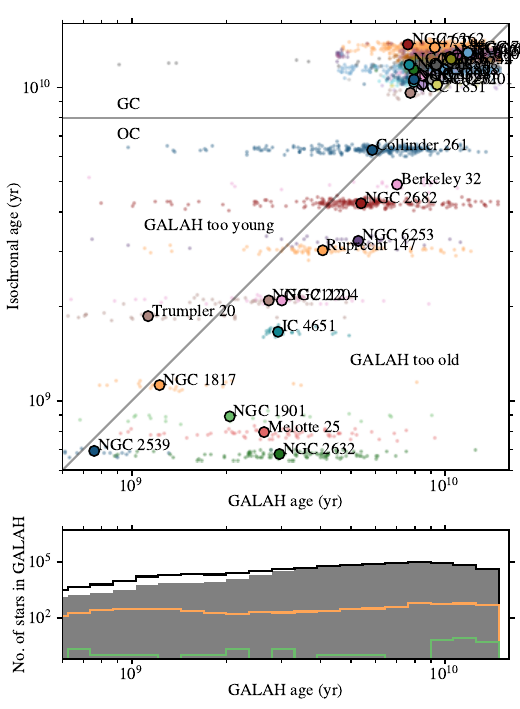}
    \caption{Same as Figure \ref{fig:ages_old_comp} but with marked names of some prominent clusters.}
    \label{fig:ages_comp_old_names}
\end{figure}

Figures \ref{fig:ages_comp_names} and \ref{fig:ages_comp_old_names} are the same as Figures \ref{fig:ages_comp} and \ref{fig:ages_old_comp}, except that they show the names of the clusters as well. 

\newpage

\section{Stars and clusters used in tests}
\label{sec:rejection}

To search for the origin of the trends, we re-analysed a small selection of stars while varying parameters and models used in spectral fitting. Because the spectral fitting as done in this work is a slow process, we did not perform it for all clusters. Instead we selected three clusters (Melotte 22, Melotte 25, and NGC 2632), that show all the trends that we would like to explore. Additionally, these three clusters have good quality spectra over a large range of temperatures and complement each other at temperature ranges where one cluster might not have many stars. Stars in these three clusters have well determined age, and have a very similar metallicity. 

\begin{figure*}
    \includegraphics[width=\textwidth]{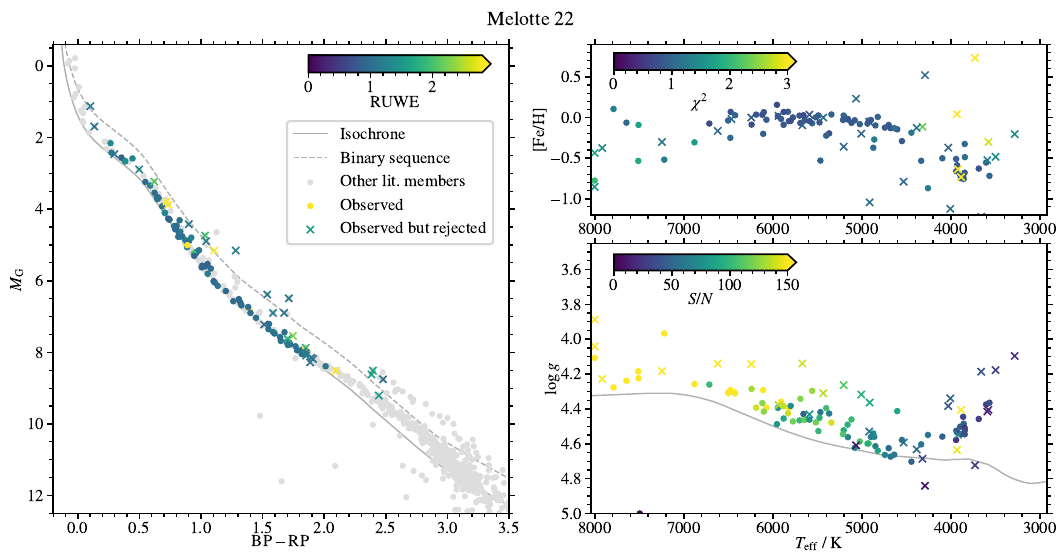}
    \caption{Stars observed in Melotte 22 and suitable for the analysis of the origin of the trends. Left panel shows the CMD of literature members with stars observed in GALAH marked with colored symbols. The right two panels show $T_\mathrm{eff}$, $\log g$, and iron abundance from the DR4. Additionally, the $\chi^2$ of the spectral fitting in DR4 and $S/N$ of the spectra are shown with colours. The isochrone in the left and bottom panels is a Padova isochrones with parameters from \citet{cantat20}. Stars that might not be reliable for the re-analysis in the search of the origin of the trends are marked with crosses. The reasons why each star was rejected are given Table \ref{tab:stars3}.}
    \label{fig:mel22_detail}
\end{figure*}

\begin{figure*}[!h]
    \includegraphics[width=\textwidth]{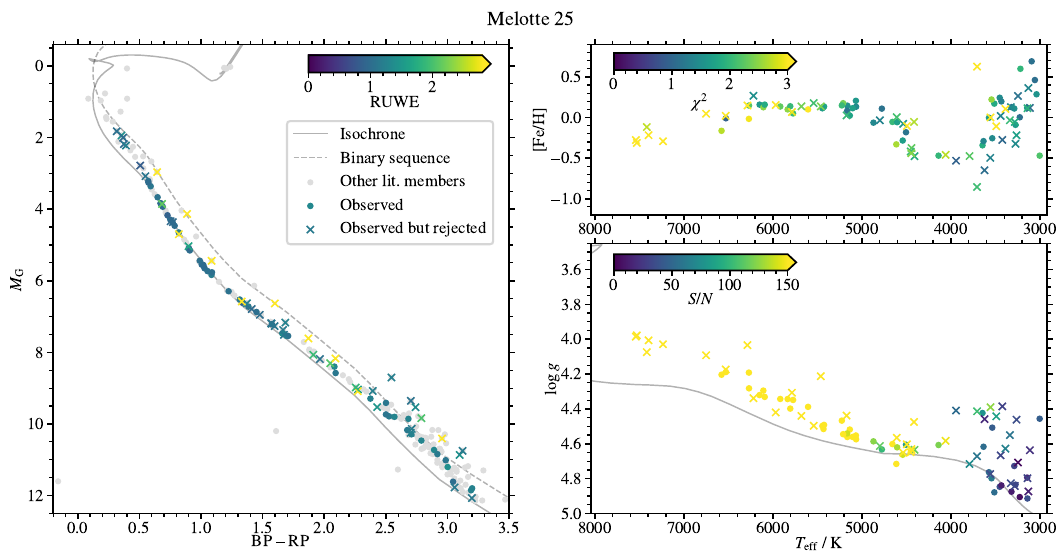}
    \caption{Same as Figure \ref{fig:mel22_detail} but for Melotte 25.}
    \label{fig:mel25_detail}
\end{figure*}

\begin{figure*}[!h]
    \includegraphics[width=\textwidth]{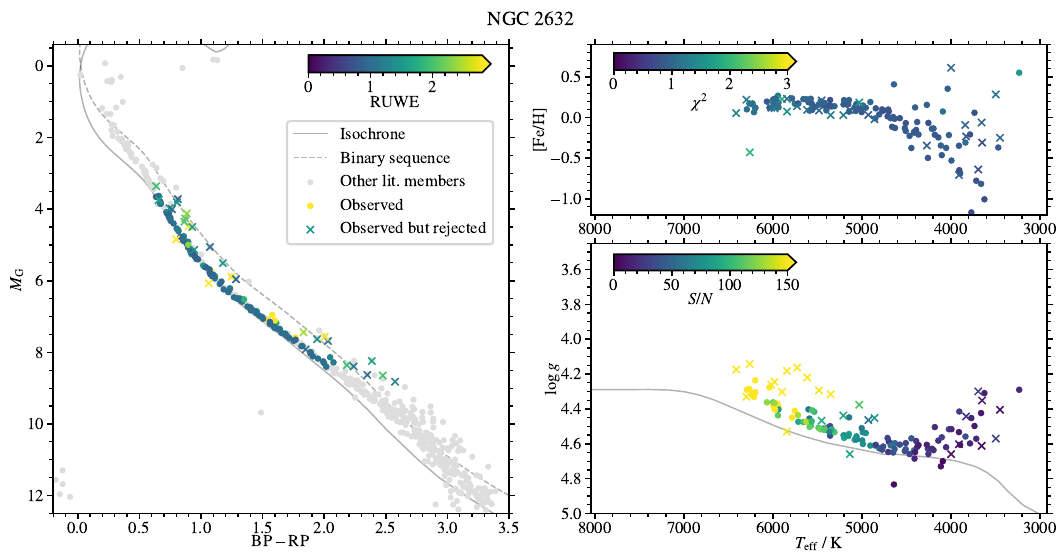}
    \caption{Same as Figure \ref{fig:mel22_detail} but for NGC 2632.}
    \label{fig:ngc2632_detail}
\end{figure*}

Because we re-analysed a small subsample of all cluster stars, we took more care about which stars are suitable for the re-analysis. The selection is described in Section \ref{sec:refitting}. In this appendix we illustrate the final selection of stars in Melotte 22, Melotte 25, and NGC 2632 (Figures \ref{fig:mel22_detail} to \ref{fig:ngc2632_detail}), and provide a table of basic parameters and reasons for the rejection of stars (Table \ref{tab:stars3}).

\newpage

\begin{table*}
\tiny
    \setlength{\tabcolsep}{5pt}
    \renewcommand{\arraystretch}{0.9}
    \caption{List of observed stars in clusters Melotte 22, Melotte 25, and NGC 2632. Last column gives remarks on stars that were observed and analysed in DR4, but rejected for the analysis regarding the origin of the trends. Column C-G marks members from \citet{cantat20}, and column H\&R those from \citet{hunt24}.}
\begin{tabular}{llccccccccccc}
\hline
\texttt{sobject\_id} & \textit{Gaia} DR3 \texttt{source\_id} & C-G & H\&R & $M_\mathrm{G}$ & BP-RP & $RUWE$ & $T_\mathrm{eff}$ & $\log g$ & $S/N_\mathrm{CCD 3}$ & $\chi^2$ & [Fe/H] & Rejected \\
\hline
\multicolumn{13}{c}{Melotte 22}\\
150109001001324 & 66733552578791296 &   & \checkmark & 6.794 & 1.393 & 3.34 & 4602.5 & 4.41 & 86.7 & 0.89 & -0.15 &  \\
150109001001329 & 65224442806459008 & \checkmark & \checkmark & 7.943 & 1.811 & 1.03 & 3939.7 & 4.53 & 46.0 & 0.88 & -0.66 &  \\
160106001601238 & 65292234570088064 &   & \checkmark & 2.654 & 0.383 & 0.96 & 7511.0 & 4.18 & 349.3 & 1.8 & -0.53 &  \\
150109001001004 & 66500013736315264 &   & \checkmark & 9.206 & 2.443 & 1.51 & 4290.7 & 4.84 & 7.0 & 1.01 & 0.52 & 3,5 \\
160106001601346 & 66729880383767168 & \checkmark & \checkmark & 1.124 & 0.101 & 1.14 & 8000.0 & 3.89 & 489.8 & 1.71 & -0.44 & 1 \\
150109001001091 & 64856312568645248 & \checkmark & \checkmark & 7.831 & 1.768 & 1.08 & 3604.3 & 4.37 & 29.6 & 0.86 & -0.55 &  \\
160106001601260 & 68310359628169088 & \checkmark & \checkmark & 5.157 & 1.108 & 5.59 & 5006.3 & 4.32 & 96.5 & 0.94 & -0.2 & 1 \\
150109001001244 & 65248151025861120 & \checkmark & \checkmark & 7.54 & 1.746 & 2.25 & 4006.6 & 4.34 & 50.1 & 1.03 & -1.12 & 3 \\
160106001601242 & 68296890611072384 & \checkmark & \checkmark & 4.883 & 0.904 & 1.05 & 5690.2 & 4.34 & 122.5 & 0.81 & -0.07 &  \\
160108002001089 & 65787289680918528 & \checkmark & \checkmark & 6.146 & 1.136 & 1.07 & 5079.3 & 4.62 & 71.3 & 0.85 & -0.03 &  \\
150109001001142 & 65072435323295872 & \checkmark & \checkmark & 5.603 & 1.033 & 1.0 & 5159.8 & 4.49 & 95.4 & 0.88 & -0.02 &  \\
150109001001039 & 64990586130069504 & \checkmark & \checkmark & 7.435 & 1.586 & 1.11 & 4333.4 & 4.6 & 57.9 & 1.0 & -0.53 &  \\
150109001001196 & 65151531440914560 & \checkmark & \checkmark & 8.279 & 1.886 & 1.04 & 3728.9 & 4.72 & 24.4 & 5.82 & 0.73 & 4,5 \\
160106001601384 & 66713044112564864 &   & \checkmark & 3.79 & 0.719 & 9.69 & 6241.9 & 4.14 & 151.7 & 0.82 & 0.0 & 1 \\
150109001001263 & 68295481861500928 & \checkmark & \checkmark & 8.184 & 1.909 & 1.09 & 3931.6 & 4.64 & 146.6 & 4.3 & 0.04 & 4,5 \\
160108002001157 & 66612473157921024 & \checkmark & \checkmark & 6.897 & 1.675 & 1.05 & 4029.3 & 4.38 & 37.7 & 0.99 & -0.37 & 1 \\
160106001601205 & 65249250535404928 & \checkmark & \checkmark & 5.157 & 1.281 & 1.18 & 5667.9 & 4.14 & 138.5 & 0.93 & -0.1 & 1 \\
160108002001115 & 65864427293364864 & \checkmark & \checkmark & 6.689 & 1.328 & 1.0 & 4847.4 & 4.61 & 64.0 & 0.91 & -0.1 &  \\
150109001001311 & 66800592725851776 & \checkmark & \checkmark & 5.805 & 1.054 & 1.2 & 5156.0 & 4.56 & 66.4 & 0.76 & -0.02 &  \\
150109001001310 & 66811450402457216 & \checkmark & \checkmark & 7.712 & 1.695 & 1.08 & 7493.4 & 5.0 & 3.5 & 0.92 & 1.07 &  \\
160106001601027 & 66503449709270400 & \checkmark & \checkmark & 4.738 & 1.032 & 1.93 & 5204.3 & 4.26 & 111.1 & 0.96 & -0.36 & 1 \\
150109001001163 & 65089473457126784 & \checkmark & \checkmark & 7.731 & 1.717 & 1.04 & 3855.8 & 4.48 & 45.1 & 0.95 & -0.76 &  \\
160923003701001 & 64671903852618624 & \checkmark & \checkmark & 4.294 & 0.744 & 0.87 & 5906.7 & 4.36 & 163.8 & 0.87 & -0.02 &  \\
160106001601024 & 66517468482370304 & \checkmark & \checkmark & 4.569 & 0.798 & 0.9 & 5829.8 & 4.42 & 137.8 & 0.89 & 0.07 &  \\
160106001601048 & 64930495244783616 & \checkmark & \checkmark & 3.341 & 0.587 & 1.2 & 6879.4 & 4.26 & 277.6 & 1.63 & -0.31 &  \\
160108002001143 & 66581961710329984 & \checkmark & \checkmark & 5.19 & 0.965 & 1.15 & 5465.3 & 4.42 & 58.1 & 0.94 & -0.53 &  \\
150109001001126 & 65161976802151680 & \checkmark & \checkmark & 7.776 & 1.722 & 0.95 & 3895.5 & 4.5 & 44.0 & 0.98 & -0.71 &  \\
160106001601208 & 65222759179728640 & \checkmark & \checkmark & 4.312 & 0.826 & 1.02 & 6107.7 & 4.29 & 161.8 & 0.92 & -0.2 &  \\
160106001601097 & 64952829074688896 & \checkmark & \checkmark & 3.649 & 0.642 & 1.13 & 6422.6 & 4.31 & 168.9 & 0.89 & 0.03 &  \\
160106001601230 & 68310561489710336 & \checkmark & \checkmark & 3.688 & 0.656 & 1.02 & 6501.4 & 4.31 & 216.3 & 0.93 & -0.04 &  \\
150109001001314 & 66808869124393600 & \checkmark & \checkmark & 6.012 & 1.102 & 0.81 & 4898.3 & 4.56 & 63.1 & 0.81 & -0.03 &  \\
150109001001299 & 65222518661019520 & \checkmark & \checkmark & 8.029 & 1.83 & 1.12 & 3845.6 & 4.51 & 23.2 & 0.85 & -0.68 &  \\
160106001601129 & 64879402312818944 & \checkmark & \checkmark & 3.241 & 0.548 & 0.98 & 6712.9 & 4.26 & 101.7 & 0.7 & -0.07 &  \\
150109001001342 & 65226401312036864 & \checkmark & \checkmark & 7.893 & 1.847 & 7.65 & 3862.5 & 4.45 & 28.7 & 0.78 & -0.73 &  \\
160106001601330 & 65225611037551360 & \checkmark & \checkmark & 3.544 & 0.623 & 1.0 & 6472.8 & 4.3 & 143.4 & 0.8 & 0.07 &  \\
150109001001309 & 66799768088857856 & \checkmark & \checkmark & 7.037 & 1.447 & 1.05 & 4676.9 & 4.67 & 56.2 & 0.8 & -0.13 &  \\
160106001601052 & 64999588381514496 & \checkmark & \checkmark & 4.287 & 0.74 & 0.94 & 6071.5 & 4.39 & 157.3 & 0.84 & -0.03 &  \\
160106001601268 & 65296357738731008 & \checkmark & \checkmark & 2.892 & 0.498 & 1.23 & 7244.3 & 4.18 & 219.7 & 1.13 & -0.3 & 2 \\
150109001001012 & 65011412428446592 & \checkmark & \checkmark & 8.51 & 2.395 & 1.61 & 3283.5 & 4.1 & 20.8 & 1.47 & -0.2 & 1,3 \\
150109001001175 & 65131121756446592 & \checkmark & \checkmark & 7.872 & 1.851 & 1.85 & 3885.5 & 4.41 & 202.9 & 3.9 & -0.73 & 2,3,4 \\
160106001601095 & 65199978672758272 & \checkmark & \checkmark & 4.422 & 0.901 & 1.11 & 5433.6 & 4.31 & 133.4 & 1.06 & 0.0 & 1 \\
160106001601178 & 65196370900147968 & \checkmark & \checkmark & 4.895 & 1.042 & 1.15 & 5935.1 & 4.38 & 133.0 & 0.88 & 0.02 & 1 \\
150109001001049 & 64985127228608128 & \checkmark & \checkmark & 6.9 & 1.584 & 1.18 & 4916.6 & 4.53 & 58.6 & 0.82 & -0.05 & 1 \\
160106001601296 & 66781832308018176 & \checkmark & \checkmark & 5.222 & 0.961 & 0.99 & 5367.2 & 4.43 & 62.7 & 0.75 & -0.2 &  \\
150109001001242 & 65244749411766528 & \checkmark & \checkmark & 8.018 & 1.822 & 1.15 & 3844.3 & 4.55 & 40.0 & 0.98 & -0.33 &  \\
160106001601022 & 66501181966524800 & \checkmark & \checkmark & 5.063 & 0.887 & 1.14 & 5516.5 & 4.48 & 116.9 & 0.92 & 0.0 &  \\
150109001001355 & 65223618172733952 & \checkmark & \checkmark & 6.287 & 1.201 & 0.93 & 4921.9 & 4.6 & 113.2 & 0.93 & -0.04 &  \\
160106001601051 & 64928605459180416 & \checkmark & \checkmark & 5.235 & 0.942 & 1.96 & 5475.3 & 4.45 & 95.8 & 0.8 & -0.06 &  \\
150109001001256 & 68294554148575232 & \checkmark & \checkmark & 7.995 & 1.826 & 1.05 & 4319.2 & 4.69 & 31.5 & 2.33 & -0.11 & 5 \\
160106001601007 & 66462939577861248 & \checkmark & \checkmark & 4.474 & 0.78 & 1.2 & 5878.3 & 4.38 & 135.7 & 0.78 & -0.09 &  \\
150109001001246 & 65266494828710400 & \checkmark & \checkmark & 8.514 & 2.096 & 4.69 & 5064.8 & 4.61 & 6.3 & 1.08 & 0.23 & 3,5 \\
150109001001347 & 66730120901930240 & \checkmark & \checkmark & 7.624 & 1.704 & 1.75 & 4530.6 & 4.59 & 55.2 & 1.03 & -0.79 & 3 \\
150109001001312 & 66808220587759104 & \checkmark & \checkmark & 8.028 & 1.848 & 1.1 & 3918.6 & 4.55 & 69.0 & 4.46 & -0.64 & 4 \\
150109001001213 & 65241313437941504 & \checkmark & \checkmark & 8.753 & 2.478 & 0.96 & 3498.4 & 4.18 & 29.1 & 1.47 & -0.48 & 1,2 \\
160106001601280 & 69819404977607168 & \checkmark & \checkmark & 3.231 & 0.621 & 2.07 & 6618.5 & 4.14 & 182.2 & 0.98 & -0.16 & 1 \\
150109001001289 & 66802276352348416 & \checkmark & \checkmark & 7.794 & 1.705 & 1.08 & 3842.5 & 4.52 & 33.1 & 1.07 & -0.49 &  \\
160106001601271 & 65230764996027776 & \checkmark & \checkmark & 2.585 & 0.444 & 1.25 & 7218.9 & 3.97 & 253.3 & 1.24 & -0.52 &  \\
150109001001043 & 64934893291213952 & \checkmark & \checkmark & 6.611 & 1.319 & 1.0 & 4860.4 & 4.56 & 65.7 & 1.26 & -0.27 &  \\
160106001601246 & 65276703968959488 & \checkmark & \checkmark & 4.797 & 0.921 & 1.23 & 5557.8 & 4.3 & 126.2 & 0.93 & -0.07 &  \\
160106001601197 & 65247704349267584 & \checkmark & \checkmark & 5.58 & 1.01 & 0.95 & 5203.7 & 4.52 & 94.1 & 0.97 & 0.02 &  \\
150109001001308 & 66801623514684416 & \checkmark & \checkmark & 8.618 & 2.384 & 1.62 & 3578.2 & 4.4 & 20.0 & 2.82 & -0.3 & 1,3,5 \\
150109001001114 & 65164171528305792 & \checkmark & \checkmark & 6.514 & 1.307 & 0.94 & 4935.9 & 4.6 & 94.9 & 0.93 & -0.07 &  \\
150109001001006 & 65013474012743936 & \checkmark & \checkmark & 7.681 & 1.648 & 1.03 & 4443.0 & 4.7 & 46.7 & 1.16 & -0.19 &  \\
160106001601301 & 66838452861270272 & \checkmark & \checkmark & 4.466 & 0.802 & 0.96 & 5942.7 & 4.39 & 78.4 & 0.68 & 0.02 &  \\
160106001601267 & 69811948914407168 & \checkmark & \checkmark & 4.097 & 0.706 & 1.2 & 6117.2 & 4.37 & 123.5 & 0.72 & 0.03 &  \\
160106001601306 & 66809491897360896 & \checkmark & \checkmark & 4.766 & 0.834 & 1.08 & 5655.5 & 4.43 & 48.5 & 0.7 & 0.03 &  \\
150109001001045 & 64979732749686016 & \checkmark & \checkmark & 5.627 & 1.005 & 0.91 & 5211.4 & 4.54 & 108.0 & 0.89 & -0.03 &  \\
160106001601016 & 66506331628024832 & \checkmark & \checkmark & 2.463 & 0.296 & 1.09 & 7914.6 & 4.23 & 316.4 & 1.43 & -0.38 & 2 \\
150109001001104 & 64981137202021376 & \checkmark & \checkmark & 6.733 & 1.382 & 1.04 & 4871.2 & 4.54 & 74.4 & 1.04 & -0.37 &  \\
160106001601371 & 66715273197982848 & \checkmark & \checkmark & 2.562 & 0.354 & 0.87 & 7643.7 & 4.24 & 260.1 & 1.02 & -0.06 &  \\
150109001001031 & 65201039527567232 & \checkmark & \checkmark & 6.491 & 1.713 & 1.1 & 5588.6 & 4.43 & 77.4 & 0.73 & 0.04 & 1 \\
160106001601021 & 65020414679879296 & \checkmark & \checkmark & 4.11 & 0.708 & 1.31 & 6186.5 & 4.39 & 138.3 & 0.8 & 0.04 &  \\
160106001601261 & 69811639676769536 & \checkmark & \checkmark & 4.861 & 0.871 & 1.24 & 5400.1 & 4.4 & 104.5 & 0.93 & -0.22 &  \\
150109001001356 & 65211145587780992 & \checkmark & \checkmark & 8.046 & 1.853 & 1.03 & 3930.2 & 4.53 & 38.6 & 0.85 & -0.59 &  \\
150109001001071 & 64964717544035200 & \checkmark & \checkmark & 8.388 & 2.011 & 1.06 & 3566.7 & 4.36 & 38.1 & 1.0 & -0.72 &  \\
150109001001029 & 64989490915460608 & \checkmark & \checkmark & 8.047 & 1.817 & 1.06 & 3686.9 & 4.46 & 27.1 & 0.78 & -0.63 &  \\
160106001601343 & 66737095928811648 & \checkmark & \checkmark & 4.757 & 0.847 & 1.02 & 5953.3 & 4.49 & 80.8 & 0.69 & 0.16 &  \\
150109001001038 & 64922283267331968 & \checkmark & \checkmark & 8.007 & 1.818 & 1.13 & 3941.6 & 4.58 & 21.8 & 0.92 & -0.51 &  \\
160106001601089 & 65004712279475712 & \checkmark & \checkmark & 4.772 & 0.815 & 1.0 & 5773.8 & 4.47 & 117.8 & 0.84 & 0.0 &  \\
150109001001245 & 65214409762926720 & \checkmark & \checkmark & 7.943 & 1.78 & 0.99 & 3886.8 & 4.55 & 36.5 & 0.8 & -0.75 &  \\
\hline
\end{tabular}
\label{tab:stars3}
\end{table*}

\begin{table*}
\addtocounter{table}{-1}
\tiny
    \setlength{\tabcolsep}{5pt}
    \renewcommand{\arraystretch}{0.9}
    \caption{cont.}
\begin{tabular}{llccccccccccc}
\hline
\texttt{sobject\_id} & \textit{Gaia} DR3 \texttt{source\_id} & C-G & H\&R & $M_\mathrm{G}$ & BP-RP & $RUWE$ & $T_\mathrm{eff}$ & $\log g$ & $S/N_\mathrm{CCD 3}$ & $\chi^2$ & [Fe/H] & Rejected \\
\hline
150109001001376 & 65208396808627840 & \checkmark & \checkmark & 7.353 & 1.547 & 1.05 & 4636.0 & 4.66 & 60.2 & 1.1 & -0.13 &  \\
150109001001278 & 68303487680516224 & \checkmark & \checkmark & 5.659 & 1.08 & 0.99 & 5116.1 & 4.46 & 102.0 & 0.81 & -0.07 &  \\
150109001001134 & 65190220507070208 & \checkmark & \checkmark & 7.208 & 1.552 & 1.01 & 4740.7 & 4.66 & 68.5 & 0.86 & -0.12 &  \\
150109001001035 & 65214856439472640 & \checkmark & \checkmark & 7.438 & 1.677 & 1.02 & 4549.4 & 4.58 & 54.0 & 0.83 & -0.29 &  \\
150109001001254 & 65289623229979136 & \checkmark & \checkmark & 5.532 & 1.053 & 0.91 & 5341.6 & 4.48 & 140.1 & 0.96 & 0.04 &  \\
160106001601259 & 68322145018429952 & \checkmark & \checkmark & 4.936 & 0.887 & 0.97 & 5700.1 & 4.47 & 100.6 & 0.76 & 0.02 &  \\
160106001601335 & 66729261908482048 & \checkmark & \checkmark & 3.72 & 0.645 & 1.01 & 6476.7 & 4.29 & 155.0 & 1.0 & -0.25 &  \\
160106001601058 & 64941211186036352 & \checkmark & \checkmark & 4.56 & 0.79 & 0.96 & 5819.8 & 4.39 & 147.7 & 0.86 & -0.05 &  \\
160106001601049 & 64929391436043264 & \checkmark & \checkmark & 2.426 & 0.273 & 1.21 & 7788.4 & 4.28 & 367.5 & 1.28 & 0.11 &  \\
150109001001344 & 66739982146803456 & \checkmark & \checkmark & 8.09 & 1.853 & 1.21 & 3590.7 & 4.42 & 16.8 & 1.31 & -0.53 & 2 \\
160106001601078 & 64913452814489216 & \checkmark & \checkmark & 5.306 & 0.945 & 0.91 & 5488.0 & 4.53 & 99.1 & 0.85 & 0.03 &  \\
160106001601292 & 65231486552803328 & \checkmark & \checkmark & 1.694 & 0.134 & 1.09 & 7999.8 & 4.04 & 340.9 & 1.61 & -0.85 & 2 \\
160106001601294 & 66788291938818304 & \checkmark & \checkmark & 4.651 & 0.822 & 0.85 & 5877.6 & 4.45 & 89.9 & 0.69 & 0.07 &  \\
150109001001066 & 64959662365552384 & \checkmark & \checkmark & 7.476 & 1.597 & 1.07 & 4257.4 & 4.55 & 58.4 & 1.27 & -0.87 &  \\
160106001601167 & 65232105028172160 & \checkmark & \checkmark & 5.007 & 0.889 & 2.92 & 5644.2 & 4.46 & 114.7 & 0.89 & -0.04 &  \\
150109001001346 & 66727234683931520 & \checkmark & \checkmark & 8.149 & 1.895 & 1.03 & 3659.8 & 4.19 & 38.6 & 0.86 & -1.22 & 5 \\
160106001601063 & 64933759417769984 & \checkmark & \checkmark & 2.67 & 0.385 & 1.39 & 7507.6 & 4.22 & 335.5 & 1.69 & -0.09 &  \\
160106001601153 & 65086591535491584 & \checkmark & \checkmark & 5.149 & 0.984 & 1.11 & 5588.3 & 4.46 & 74.4 & 0.76 & -0.04 &  \\
160106001601181 & 65242069352190976 & \checkmark & \checkmark & 3.876 & 0.735 & 3.06 & 6464.3 & 4.31 & 205.8 & 1.09 & -0.02 & 1,3 \\
160106001601192 & 65221487868870784 & \checkmark & \checkmark & 2.16 & 0.266 & 1.34 & 8000.0 & 4.11 & 412.1 & 2.06 & -0.78 &  \\
160106001601249 & 65295430026352512 & \checkmark & \checkmark & 4.631 & 0.832 & 1.14 & 5798.8 & 4.38 & 72.8 & 0.68 & -0.04 &  \\
160106001601399 & 66523893753437824 & \checkmark & \checkmark & 4.265 & 0.734 & 0.87 & 6100.8 & 4.41 & 123.2 & 0.78 & 0.07 &  \\
160106001601282 & 69819027020489088 & \checkmark & \checkmark & 3.922 & 0.685 & 1.02 & 6238.1 & 4.29 & 129.1 & 0.73 & -0.09 &  \\
150109001001008 & 65010549138050944 & \checkmark & \checkmark & 6.572 & 1.262 & 0.98 & 4820.8 & 4.64 & 86.9 & 0.96 & 0.01 &  \\
150109001001080 & 64971933089093760 & \checkmark & \checkmark & 7.573 & 1.641 & 1.02 & 4095.9 & 4.56 & 46.0 & 0.97 & -0.5 &  \\
150109001001133 & 65190048708380160 & \checkmark & \checkmark & 6.386 & 1.538 & 1.1 & 4912.4 & 4.36 & 90.1 & 1.09 & -1.04 & 1,5 \\
150109001001186 & 65194584193758336 & \checkmark & \checkmark & 6.052 & 1.111 & 1.04 & 5022.6 & 4.59 & 108.0 & 0.92 & -0.08 &  \\
150109001001317 & 66787119410915072 & \checkmark & \checkmark & 7.225 & 1.513 & 0.93 & 4384.4 & 4.63 & 52.7 & 1.19 & -0.13 & 2 \\
150109001001328 & 66816295125702144 & \checkmark & \checkmark & 6.898 & 1.401 & 0.99 & 4698.6 & 4.63 & 52.1 & 0.82 & -0.09 &  \\
\multicolumn{13}{c}{Melotte 25}\\
161106003601390 & 3287041726938931968 & \checkmark & \checkmark & 10.725 & 2.891 & 1.31 & 3226.2 & 4.91 & 5.8 & 1.09 & 0.0 &  \\
160111001601292 & 51902691204229888 & \checkmark & \checkmark & 12.068 & 3.198 & 1.14 & nan & nan & -5.0 & nan & nan & 4,5 \\
160108002001373 & 66286704180780672 & \checkmark & \checkmark & 9.529 & 2.429 & 1.47 & 3557.5 & 4.77 & 34.5 & 1.34 & -0.5 & 3 \\
161104003801145 & 3287167277423046272 & \checkmark & \checkmark & 10.246 & 2.701 & 1.22 & 3282.0 & 4.84 & 31.6 & 1.81 & -0.06 & 2 \\
170131002301192 & 3419656057748048512 & \checkmark & \checkmark & 10.17 & 2.674 & 1.13 & 3319.5 & 4.87 & 11.2 & 1.13 & 0.13 &  \\
170106002601393 & 3282226793722500224 & \checkmark & \checkmark & 10.136 & 2.71 & 1.18 & 3326.5 & 4.83 & 35.1 & 1.44 & 0.13 & 2 \\
181226002601346 & 3281410509417809152 & \checkmark & \checkmark & 9.787 & 2.526 & 1.16 & 3512.0 & 4.88 & 53.0 & 1.81 & 0.17 &  \\
151109002601113 & 64552022725337856 & \checkmark & \checkmark & 11.851 & 3.192 & 1.23 & 3219.0 & 5.13 & 40.1 & 1.97 & 0.6 &  \\
151109003101345 & 47917511308704128 & \checkmark & \checkmark & 9.734 & 2.501 & 1.08 & 3451.5 & 4.85 & 37.5 & 1.44 & 0.24 &  \\
170117002101360 & 3313070293905997824 & \checkmark & \checkmark & 6.634 & 1.599 & 23.68 & 4413.9 & 4.47 & 216.3 & 2.65 & -0.05 & 1,2,3 \\
170828004401029 & 3404850790083594368 & \checkmark & \checkmark & 3.841 & 0.674 & 0.93 & 6148.2 & 4.32 & 313.9 & 1.61 & 0.16 &  \\
161006005901272 & 57406365376348928 & \checkmark & \checkmark & 10.842 & 2.918 & 1.15 & 3002.8 & 4.46 & 44.5 & 2.16 & -0.47 &  \\
170828004401279 & 3405127244241184256 & \checkmark & \checkmark & 5.548 & 1.005 & 0.98 & 5135.1 & 4.52 & 167.3 & 1.74 & 0.02 &  \\
151109003101313 & 47881811540636800 & \checkmark & \checkmark & 11.613 & 3.046 & 1.0 & 3038.0 & 5.02 & 19.0 & 1.55 & 0.29 &  \\
160111001601116 & 47247938791363200 & \checkmark & \checkmark & 10.865 & 3.099 & 1.43 & 3129.6 & 4.87 & 13.5 & 1.37 & 0.36 & 1,3 \\
151111003601164 & 53257461329357312 & \checkmark & \checkmark & 9.038 & 2.274 & 1.57 & 3440.9 & 4.67 & 32.0 & 1.38 & 0.02 & 3 \\
160109002001367 & 38329666836450304 & \checkmark & \checkmark & 11.797 & 3.2 & 1.23 & 3056.5 & 5.05 & 5.2 & 1.08 & 0.46 &  \\
191107003601277 & 48000730596272128 & \checkmark & \checkmark & 6.951 & 1.477 & 1.01 & 4535.4 & 4.65 & 155.2 & 1.59 & -0.08 & 2 \\
170828003901187 & 3313630082762446208 & \checkmark & \checkmark & 4.295 & 0.751 & 0.98 & 5833.2 & 4.34 & 341.7 & 2.48 & 0.1 &  \\
151009004601341 & 3305524242524308352 & \checkmark & \checkmark & 9.085 & 2.271 & 11.87 & 3262.6 & 4.46 & 35.5 & 1.7 & -0.22 & 3 \\
170128002601082 & 3308501926171656832 & \checkmark & \checkmark & 11.012 & 2.986 & 1.1 & 3140.2 & 4.91 & 36.9 & 2.31 & 0.12 &  \\
170828003901253 & 3313759000500605056 & \checkmark & \checkmark & 6.293 & 1.224 & 1.06 & 4869.4 & 4.59 & 123.4 & 1.41 & -0.06 &  \\
151008004001343 & 47917816253918720 & \checkmark & \checkmark & 8.303 & 2.049 & 2.04 & 3422.2 & 4.39 & 29.8 & 1.02 & -0.28 & 3 \\
170117002101072 & 3309541170817293824 &   & \checkmark & 8.171 & 2.091 & 3.34 & 3705.7 & 4.41 & 103.4 & 2.07 & -0.86 & 2,3 \\
181225002101151 & 50298125783225088 & \checkmark & \checkmark & 7.472 & 1.66 & 1.36 & 4612.9 & 4.72 & 180.7 & 2.6 & -0.33 &  \\
170119002601077 & 3413146914553161728 & \checkmark & \checkmark & 6.612 & 1.334 & 0.93 & 4656.4 & 4.6 & 182.8 & 2.01 & -0.05 &  \\
170119002601308 & 146989143968434688 & \checkmark & \checkmark & 8.186 & 1.965 & 1.06 & 3705.9 & 4.67 & 59.1 & 3.18 & 0.63 & 2,4 \\
160106002601123 & 49968306654050944 & \checkmark & \checkmark & 9.8 & 2.57 & 1.28 & 3433.1 & 4.84 & 12.6 & 1.31 & 0.2 &  \\
170130001601354 & 3404798597640071296 & \checkmark & \checkmark & 7.611 & 1.872 & 20.37 & 3554.9 & 4.39 & 124.6 & 4.29 & 0.0 & 2,3,4 \\
170220001601140 & 3411811317161811328 & \checkmark & \checkmark & 9.857 & 2.683 & 1.26 & 3288.2 & 4.73 & 31.6 & 1.49 & 0.11 &  \\
161228001501254 & 46458115782368384 & \checkmark & \checkmark & 11.202 & 3.002 & 1.38 & 3092.7 & 4.98 & nan & 1.12 & 0.69 &  7\\
170828004401068 & 3404812685132622592 & \checkmark & \checkmark & 4.135 & 0.887 & 7.12 & 5460.8 & 4.21 & 350.2 & 2.41 & 0.14 & 1,3 \\
170829003401390 & 145391484855481344 & \checkmark & \checkmark & 2.184 & 0.374 & 0.97 & 7414.8 & 4.07 & 502.6 & 2.71 & -0.12 & 2 \\
151109003101241 & 47760521664503296 & \checkmark & \checkmark & 10.464 & 2.788 & 1.28 & 3142.7 & 4.79 & 17.2 & 1.61 & 0.12 &  \\
181225002601329 & 145218857232329472 & \checkmark & \checkmark & 9.839 & 2.788 & 2.0 & 3246.5 & 4.71 & 5.4 & 1.14 & 0.27 & 1,3 \\
140820003701047 & 45142206521351552 & \checkmark & \checkmark & 2.783 & 0.507 & 0.82 & 6748.9 & 4.09 & 1839.9 & 6.72 & 0.05 & 4 \\
170829003401320 & 145484634106814592 & \checkmark & \checkmark & 1.943 & 0.355 & 1.39 & 7526.6 & 3.98 & 894.3 & 4.86 & -0.31 & 4 \\
170830005101052 & 3410640887035452928 & \checkmark & \checkmark & 6.526 & 1.315 & 1.14 & 4775.2 & 4.63 & 100.2 & 1.14 & 0.11 &  \\
170103001801022 & 3410154456217987584 & \checkmark & \checkmark & 10.289 & 2.707 & 1.23 & 3274.0 & 4.83 & 33.0 & 1.64 & -0.09 &  \\
181226002101346 & 3411031385460853248 & \checkmark & \checkmark & 8.703 & 2.545 & 1.2 & 3621.9 & 4.46 & 17.6 & 1.09 & -0.65 & 1 \\
181226002101119 & 3410505784542875008 & \checkmark & \checkmark & 9.412 & 2.501 & 1.11 & 3572.3 & 4.76 & 51.9 & 1.68 & -0.0 &  \\
170103001801202 & 3410787568757738752 & \checkmark & \checkmark & 10.405 & 2.956 & 11.55 & 3112.6 & 4.64 & 22.2 & 1.09 & 0.12 & 1,3 \\
191107003601231 & 3314597408180024960 & \checkmark & \checkmark & 9.354 & 2.703 & 1.07 & 3337.1 & 4.55 & 58.2 & 2.02 & -0.03 & 1,2 \\
170829002901128 & 3312602348628348032 & \checkmark & \checkmark & 5.836 & 1.085 & 1.18 & 5064.3 & 4.57 & 168.8 & 1.44 & 0.13 &  \\
181226002101085 & 3410529355323453312 & \checkmark & \checkmark & 9.292 & 2.375 & 1.22 & 3543.5 & 4.8 & 61.7 & 2.51 & 0.22 &  \\
170829003401169 & 145325548516513280 & \checkmark & \checkmark & 5.154 & 0.909 & 1.05 & 5430.5 & 4.52 & 221.7 & 1.73 & 0.13 &  \\
170829002901201 & 3312652513846244096 &   & \checkmark & 5.443 & 1.086 & 16.33 & 5171.2 & 4.44 & 218.2 & 2.02 & 0.03 & 1,3 \\
170828004401341 & 3405220088550711808 & \checkmark & \checkmark & 3.668 & 0.648 & 1.07 & 6267.0 & 4.28 & 459.6 & 2.21 & 0.15 &  \\
140820003701270 & 45367056650753280 & \checkmark & \checkmark & 4.338 & 0.754 & 0.97 & 5967.0 & 4.42 & 906.3 & 4.08 & 0.15 & 4 \\
140820003701223 & 45159901786885632 &   & \checkmark & 6.79 & 1.411 & 0.98 & 4493.8 & 4.6 & 323.6 & 4.64 & -0.1 & 2,4 \\
170830005101301 & 144171233106399104 & \checkmark & \checkmark & 5.091 & 0.901 & 0.9 & 5424.0 & 4.49 & 261.1 & 1.78 & 0.13 &  \\
170829003401291 & 145373377272257664 & \checkmark & \checkmark & 3.838 & 0.671 & 0.98 & 6106.1 & 4.29 & 453.1 & 2.46 & 0.13 &  \\
170828004401178 & 3309006602007842048 & \checkmark & \checkmark & 4.701 & 0.821 & 3.3 & 5678.5 & 4.44 & 305.3 & 2.36 & 0.14 & 3 \\
\hline
\end{tabular}
\end{table*}

\begin{table*}
\addtocounter{table}{-1}
\tiny
    \setlength{\tabcolsep}{5pt}
    \renewcommand{\arraystretch}{0.9}
    \caption{cont.}
\begin{tabular}{llccccccccccc}
\hline
\texttt{sobject\_id} & \textit{Gaia} DR3 \texttt{source\_id} & C-G & H\&R & $M_\mathrm{G}$ & BP-RP & $RUWE$ & $T_\mathrm{eff}$ & $\log g$ & $S/N_\mathrm{CCD 3}$ & $\chi^2$ & [Fe/H] & Rejected \\
\hline
170828004401316 & 3405988677241799040 & \checkmark & \checkmark & 6.724 & 1.385 & 1.09 & 4603.6 & 4.62 & 130.3 & 1.7 & -0.06 &  \\
170828004401325 & 3405113740864365440 & \checkmark & \checkmark & 4.171 & 0.728 & 0.87 & 5916.2 & 4.34 & 382.5 & 2.27 & 0.14 &  \\
170130001601276 & 3308896238528440448 & \checkmark & \checkmark & 7.535 & 1.69 & 1.25 & 4138.6 & 4.61 & 120.2 & 1.99 & -0.47 &  \\
170829003401026 & 144534724778235392 & \checkmark & \checkmark & 7.511 & 1.673 & 1.11 & 4058.6 & 4.58 & 146.0 & 2.68 & -0.46 & 2 \\
170830005101223 & 48061409893621248 & \checkmark & \checkmark & 5.037 & 0.899 & 1.64 & 5540.9 & 4.5 & 261.3 & 2.1 & 0.18 & 3 \\
151008004001152 & 47485609397541632 & \checkmark & \checkmark & 8.981 & 2.256 & 1.87 & 3387.2 & 4.63 & 66.0 & 3.03 & 0.11 & 2,3,4 \\
170829003401003 & 145293181643038336 & \checkmark & \checkmark & 3.856 & 0.684 & 2.12 & 6217.4 & 4.34 & 294.5 & 1.49 & 0.27 & 3 \\
151109003101031 & 3314393590508675456 & \checkmark & \checkmark & 10.75 & 3.123 & 1.15 & 3139.9 & 4.8 & 30.8 & 1.93 & 0.11 & 1,2 \\
170828003901251 & 3313947704182762240 & \checkmark & \checkmark & 3.361 & 0.591 & 1.11 & 6268.4 & 4.19 & 521.8 & 2.95 & -0.02 &  \\
170830004001040 & 3307528033746225664 & \checkmark & \checkmark & 3.215 & 0.572 & 0.99 & 6526.6 & 4.19 & 176.9 & 0.93 & -0.01 &  \\
151009005101086 & 3310640648085373824 & \checkmark & \checkmark & 8.398 & 2.083 & 1.3 & 3535.5 & 4.51 & 43.7 & 1.42 & -0.26 &  \\
170828003901334 & 3314109916508904064 & \checkmark & \checkmark & 4.373 & 0.77 & 1.0 & 5772.2 & 4.35 & 348.1 & 2.06 & 0.05 &  \\
170829003401292 & 145664777917536512 & \checkmark & \checkmark & 7.213 & 1.575 & 1.05 & 4540.8 & 4.66 & 139.4 & 1.68 & -0.28 &  \\
170829002901052 & 3309956850635519488 & \checkmark & \checkmark & 5.651 & 1.028 & 0.99 & 5163.1 & 4.56 & 207.4 & 1.5 & 0.15 &  \\
170830005101043 & 3410453489023420416 & \checkmark & \checkmark & 6.873 & 1.45 & 1.05 & 4503.9 & 4.61 & 83.1 & 1.16 & -0.18 &  \\
170829003401161 & 49231668222673920 & \checkmark & \checkmark & 3.258 & 0.573 & 0.97 & 6575.5 & 4.2 & 381.0 & 2.68 & -0.16 &  \\
151109003101320 & 47982794811614080 & \checkmark & \checkmark & 11.772 & 3.058 & 1.14 & nan & nan & -2.4 & nan & nan & 1,4,5 \\
151009005101168 & 3311492803955469696 & \checkmark & \checkmark & 7.542 & 1.707 & 1.05 & 3644.6 & 4.42 & 76.2 & 1.87 & -0.34 &  \\
191107003601005 & 3314581984953794816 & \checkmark & \checkmark & 9.077 & 2.394 & 1.21 & 3790.3 & 4.72 & 72.0 & 2.22 & -0.47 & 2 \\
170829003901072 & 3310820624394671488 & \checkmark & \checkmark & 7.187 & 1.569 & 1.06 & 4442.4 & 4.64 & 161.2 & 2.27 & -0.38 &  \\
170828003901005 & 3312921378798754304 & \checkmark & \checkmark & 7.375 & 1.666 & 1.15 & 4408.4 & 4.63 & 114.0 & 1.83 & -0.48 &  \\
170828003901129 & 3312783561888068352 &   & \checkmark & 2.968 & 0.644 & 23.99 & 6284.7 & 4.03 & 907.2 & 6.09 & 0.15 & 1,3,4 \\
170117002101213 & 3312899491645515776 &   & \checkmark & 6.63 & 1.373 & 1.23 & 4607.9 & 4.57 & 169.6 & 2.12 & 0.0 & 2 \\
170828003901291 & 3313689422030650496 & \checkmark & \checkmark & 4.652 & 0.82 & 1.19 & 5603.8 & 4.39 & 388.4 & 2.99 & 0.1 &  \\
170829003901184 & 3311024789960504576 & \checkmark & \checkmark & 3.081 & 0.549 & 1.09 & 6528.5 & 4.17 & 893.2 & 5.21 & 0.02 & 4 \\
151008004001047 & 3314063908819076352 & \checkmark & \checkmark & 7.268 & 1.597 & 1.1 & 4451.3 & 4.65 & 148.9 & 2.04 & -0.42 & 2 \\
170828003901237 & 3313662896313355008 & \checkmark & \checkmark & 4.364 & 0.772 & 0.97 & 5785.1 & 4.31 & 401.3 & 2.94 & 0.07 & 2 \\
151009005101033 & 3310702770492165120 & \checkmark & \checkmark & 8.578 & 2.092 & 1.11 & 3633.4 & 4.62 & 53.2 & 1.38 & -0.29 &  \\
170830004001231 & 3310876802567157248 &   & \checkmark & 6.58 & 1.331 & 12.67 & 4797.3 & 4.61 & 113.2 & 1.34 & -0.03 & 3 \\
170828003901086 & 3312709379213017728 & \checkmark & \checkmark & 3.933 & 0.689 & 1.0 & 6090.5 & 4.33 & 331.1 & 1.84 & 0.16 &  \\
170828003901384 & 3312951748510907648 & \checkmark & \checkmark & 5.567 & 1.02 & 0.96 & 5164.6 & 4.54 & 138.6 & 1.18 & 0.08 &  \\
170117002101130 & 3312564037520033792 & \checkmark & \checkmark & 7.201 & 1.573 & 1.1 & 4443.5 & 4.63 & 157.3 & 2.45 & -0.44 &  \\
170829002901108 & 3312575685471393664 & \checkmark & \checkmark & 4.465 & 0.786 & 1.0 & 5803.9 & 4.4 & 362.4 & 2.01 & 0.14 &  \\
170829002901175 & 3312644885984344704 & \checkmark & \checkmark & 5.127 & 0.904 & 0.96 & 5443.3 & 4.5 & 225.3 & 1.66 & 0.11 &  \\
170829002901275 & 3313259169388356608 & \checkmark & \checkmark & 5.732 & 1.053 & 1.04 & 5103.8 & 4.56 & 182.3 & 1.71 & 0.11 &  \\
170829003901164 & 3310634944368705664 & \checkmark & \checkmark & 2.225 & 0.391 & 1.21 & 7232.8 & 4.03 & 1037.8 & 7.59 & -0.29 & 4,5 \\
170829003901249 & 3311883817778820480 & \checkmark & \checkmark & 1.826 & 0.316 & 1.03 & 7537.7 & 3.99 & 712.2 & 3.9 & -0.28 & 4 \\
170829003901358 & 3311179340063437952 & \checkmark & \checkmark & 8.067 & 1.912 & 1.89 & 3490.0 & 4.44 & 86.1 & 3.17 & -0.11 & 2,3,4 \\
170829003901336 & 3311148828615843328 & \checkmark & \checkmark & 5.55 & 1.013 & 0.98 & 5209.8 & 4.55 & 329.8 & 2.36 & 0.18 &  \\
170830004001065 & 3307504222447587584 & \checkmark & \checkmark & 7.168 & 1.684 & 1.23 & 3942.0 & 4.41 & 52.4 & 0.94 & -0.53 & 1 \\
170830004001244 & 3310903740601986432 & \checkmark & \checkmark & 5.765 & 1.089 & 1.03 & 5066.5 & 4.55 & 153.8 & 1.6 & 0.21 &  \\
170830004001205 & 3307844864893938304 & \checkmark & \checkmark & 5.441 & 0.99 & 1.02 & 5224.7 & 4.47 & 232.5 & 2.03 & 0.15 &  \\
170830004001131 & 3307782570688378496 & \checkmark & \checkmark & 2.006 & 0.368 & 0.97 & 7394.9 & 4.01 & 1016.6 & 7.02 & -0.22 & 2,4 \\
160923004201201 & 42716408993806080 & \checkmark &   & 9.534 & 2.741 & 1.35 & 3274.2 & 4.45 & nan & 1.43 & -0.32 & 1 \\
\multicolumn{13}{c}{NGC 2632}\\
160401002101227 & 664334035130360192 & \checkmark & \checkmark & 5.766 & 1.089 & 6.13 & 5143.1 & 4.45 & 60.0 & 0.82 & 0.07 &  \\
160113002301143 & 659704266543218816 & \checkmark & \checkmark & 6.954 & 1.578 & 4.01 & 4180.2 & 4.5 & 18.9 & 0.85 & -0.21 &  \\
160113002301045 & 661018904492612224 & \checkmark & \checkmark & 6.584 & 1.334 & 1.09 & 4659.9 & 4.6 & 27.6 & 0.77 & -0.11 &  \\
160401002101212 & 661297115295806592 & \checkmark & \checkmark & 3.976 & 0.744 & 1.52 & 6017.1 & 4.26 & 184.9 & 1.1 & 0.12 & 3 \\
160113002301256 & 664343415338898944 & \checkmark & \checkmark & 7.182 & 1.612 & 1.76 & 4022.0 & 4.52 & 28.0 & 0.9 & -0.22 &  \\
160113002301129 & 661216404270424832 & \checkmark & \checkmark & 7.587 & 1.766 & 2.76 & 4194.8 & 4.64 & 19.6 & 0.93 & -0.13 &  \\
160401002101020 & 661293537585522816 & \checkmark & \checkmark & 4.483 & 0.898 & 8.29 & 5481.8 & 4.29 & 147.2 & 1.03 & 0.09 & 1,3 \\
160401002101265 & 661294332158553856 & \checkmark & \checkmark & 4.352 & 0.865 & 2.0 & 5613.9 & 4.22 & 178.6 & 1.17 & 0.09 & 1,3 \\
170102001901023 & 662925629454594944 & \checkmark & \checkmark & 4.4 & 0.777 & 1.0 & 5941.9 & 4.44 & 112.9 & 1.13 & 0.27 &  \\
160113002301074 & 661220080762501120 & \checkmark & \checkmark & 7.439 & 1.832 & 2.42 & 3692.6 & 4.3 & 27.0 & 0.73 & -0.64 & 1 \\
160401002101397 & 661148268907314432 & \checkmark & \checkmark & 4.152 & 0.737 & 6.38 & 6037.3 & 4.28 & 155.8 & 0.86 & 0.12 &  \\
160403002101243 & 661528184534528896 &   & \checkmark & 7.992 & 1.89 & 1.06 & 4090.2 & 4.7 & 4.3 & 1.15 & 0.07 &  \\
160113002301216 & 661313504890920448 & \checkmark & \checkmark & 8.647 & 2.475 & 1.99 & 3448.2 & 4.41 & 7.8 & 0.99 & -0.25 & 1 \\
160110003101265 & 663781564897655296 &   & \checkmark & 7.209 & 1.574 & 1.03 & 4414.1 & 4.64 & 54.4 & 0.94 & -0.18 &  \\
160109003801204 & 661470460174102144 & \checkmark & \checkmark & 5.394 & 0.962 & 1.1 & 5356.2 & 4.55 & 33.3 & 0.73 & 0.22 &  \\
160113002301336 & 661410360695218048 & \checkmark & \checkmark & 6.494 & 1.316 & 1.08 & 4697.0 & 4.59 & 22.8 & 0.77 & 0.07 &  \\
160113002301073 & 661221828810801408 & \checkmark & \checkmark & 6.464 & 1.267 & 1.04 & 4804.8 & 4.63 & 49.8 & 0.86 & 0.04 &  \\
160401002101330 & 661306319407420160 & \checkmark & \checkmark & 4.123 & 0.885 & 2.34 & 5839.7 & 4.18 & 173.6 & 1.55 & 0.07 & 1,2,3 \\
160113002301245 & 664323761568567808 & \checkmark & \checkmark & 7.54 & 1.697 & 0.99 & 4258.0 & 4.63 & 28.1 & 0.84 & -0.5 &  \\
160401002101357 & 661386244455690752 & \checkmark & \checkmark & 3.744 & 0.657 & 1.65 & 6295.9 & 4.33 & 209.0 & 1.14 & 0.22 & 3 \\
160401002101392 & 661338858079663488 & \checkmark & \checkmark & 5.841 & 1.087 & 1.47 & 5081.1 & 4.58 & 89.8 & 0.88 & 0.12 &  \\
160401002101077 & 661239390935360512 & \checkmark & \checkmark & 5.884 & 1.092 & 0.96 & 5054.8 & 4.57 & 81.6 & 0.92 & 0.11 &  \\
160110003101302 & 663011734958828416 &   & \checkmark & 8.82 & 2.575 & 1.19 & 3495.9 & 4.57 & 29.1 & 1.24 & 0.28 & 1 \\
161229002701319 & 658607950370340608 & \checkmark & \checkmark & 8.173 & 1.949 & 1.07 & 3895.9 & 4.61 & 11.2 & 0.8 & -0.69 &  \\
160401002101195 & 664283079638402688 & \checkmark & \checkmark & 5.134 & 0.945 & 1.43 & 5451.2 & 4.47 & 95.7 & 1.0 & 0.21 & 3 \\
160113002301058 & 661219840244243200 & \checkmark & \checkmark & 7.113 & 1.599 & 3.81 & 4443.6 & 4.57 & 40.6 & 0.89 & -0.38 &  \\
160401002101112 & 661211727047722752 & \checkmark & \checkmark & 6.251 & 1.231 & 1.05 & 4904.5 & 4.61 & 71.9 & 1.01 & 0.19 &  \\
160401002101260 & 664344824088164352 & \checkmark & \checkmark & 4.757 & 0.86 & 1.28 & 5599.3 & 4.4 & 62.3 & 0.75 & 0.19 &  \\
160113002301385 & 661333708415842304 & \checkmark & \checkmark & 6.53 & 1.333 & 1.02 & 4760.8 & 4.62 & 35.7 & 0.77 & 0.08 &  \\
160110003101037 & 662912813272290688 & \checkmark & \checkmark & 7.827 & 1.814 & 1.06 & 3836.2 & 4.56 & 37.7 & 1.08 & -0.31 &  \\
160401002101055 & 661028662658249344 & \checkmark & \checkmark & 5.181 & 0.92 & 1.47 & 5433.2 & 4.51 & 78.9 & 0.81 & 0.15 &  \\
170122003601105 & 665129291274749696 & \checkmark & \checkmark & 5.197 & 0.915 & 1.07 & 5387.5 & 4.51 & 59.7 & 0.76 & 0.13 &  \\
160113002301210 & 664293387559886464 & \checkmark & \checkmark & 7.63 & 1.942 & 1.34 & 3649.3 & 4.35 & 13.9 & 0.7 & -0.31 & 1 \\
160113002301258 & 664399078114921600 & \checkmark & \checkmark & 8.255 & 1.991 & 1.06 & 4112.8 & 4.73 & 7.7 & 0.92 & -0.34 &  \\
160401002101122 & 659687498990893056 & \checkmark & \checkmark & 4.277 & 0.749 & 0.94 & 5935.3 & 4.39 & 170.4 & 1.14 & 0.12 &  \\
160110003101374 & 663069150081902592 & \checkmark & \checkmark & 6.566 & 1.326 & 1.1 & 4627.3 & 4.6 & 51.4 & 0.85 & -0.13 &  \\
160401002101192 & 664283526315590656 & \checkmark & \checkmark & 5.819 & 1.076 & 0.87 & 5066.8 & 4.57 & 73.9 & 0.96 & 0.08 &  \\
160401002101172 & 661265951013145344 & \checkmark & \checkmark & 6.025 & 1.166 & 0.96 & 4955.3 & 4.54 & 76.1 & 0.98 & 0.12 &  \\
\hline
\end{tabular}
\end{table*}

\begin{table*}
\addtocounter{table}{-1}
\tiny
    \setlength{\tabcolsep}{5pt}
    \renewcommand{\arraystretch}{0.9}
    \caption{cont.}
\begin{tabular}{llccccccccccc}
\hline
\texttt{sobject\_id} & \textit{Gaia} DR3 \texttt{source\_id} & C-G & H\&R & $M_\mathrm{G}$ & BP-RP & $RUWE$ & $T_\mathrm{eff}$ & $\log g$ & $S/N_\mathrm{CCD 3}$ & $\chi^2$ & [Fe/H] & Rejected \\
\hline
160113002301239 & 664388357876585344 & \checkmark & \checkmark & 8.357 & 2.183 & 1.92 & 3909.1 & 4.6 & 7.4 & 0.93 & -0.71 & 1 \\
160401002101328 & 661453692623304704 & \checkmark & \checkmark & 5.873 & 1.113 & 1.61 & 4989.1 & 4.53 & 72.0 & 0.85 & 0.13 &  \\
160401002101067 & 661221764391246080 & \checkmark & \checkmark & 3.795 & 0.668 & 0.94 & 6227.7 & 4.31 & 210.3 & 1.12 & 0.17 &  \\
160113002301286 & 664442233945917440 & \checkmark & \checkmark & 8.393 & 2.239 & 1.18 & 3839.0 & 4.63 & nan & 0.96 & -0.09 & 1,5 \\
160113002301303 & 664437320503310336 & \checkmark & \checkmark & 8.031 & 1.888 & 1.14 & 4036.4 & 4.74 & nan & 0.92 & 0.35 &  \\
160401002101298 & 661319483485360000 & \checkmark & \checkmark & 4.543 & 0.803 & 1.06 & 5752.7 & 4.41 & 171.6 & 1.12 & 0.17 &  \\
160110003101347 & 663140171661012480 & \checkmark & \checkmark & 7.684 & 2.03 & 1.08 & 3829.3 & 4.44 & 22.7 & 0.74 & -0.23 & 1 \\
160113002301177 & 661282031370596608 & \checkmark & \checkmark & 6.975 & 1.472 & 0.93 & 4517.3 & 4.64 & 32.4 & 0.85 & -0.13 &  \\
160401002101229 & 661313023854578560 & \checkmark & \checkmark & 5.884 & 1.248 & 4.02 & 4857.4 & 4.45 & 82.9 & 0.98 & -0.02 & 1 \\
160113002301395 & 661335288963844736 &   & \checkmark & 6.52 & 1.349 & 1.83 & 4711.0 & 4.57 & 29.6 & 0.76 & 0.01 & 7 \\
160113002301268 & 664444914005442304 & \checkmark & \checkmark & 7.479 & 1.684 & 1.06 & 3909.2 & 4.53 & 11.1 & 0.86 & -0.35 &  \\
160401002101101 & 661212933936844416 & \checkmark & \checkmark & 4.612 & 0.817 & 1.07 & 5716.8 & 4.44 & 130.5 & 1.02 & 0.17 &  \\
160401002101329 & 661300993647979776 & \checkmark & \checkmark & 5.607 & 1.027 & 0.92 & 5185.1 & 4.55 & 73.9 & 0.83 & 0.17 &  \\
160113002301291 & 661297458893183232 & \checkmark & \checkmark & 8.138 & 2.048 & 1.27 & 3623.1 & 4.31 & 13.1 & 0.72 & -1.01 &  \\
160113002301329 & 661414213282680960 & \checkmark & \checkmark & 7.455 & 1.653 & 0.99 & 4408.3 & 4.69 & 16.6 & 0.86 & -0.19 &  \\
160401002101289 & 664422683254834304 & \checkmark & \checkmark & 5.075 & 0.895 & 1.05 & 5487.7 & 4.51 & 111.3 & 0.92 & 0.16 &  \\
160113002301110 & 661189260077235072 & \checkmark & \checkmark & 8.244 & 2.386 & 1.6 & 3997.5 & 4.66 & 7.6 & 1.03 & 0.61 & 1,3 \\
160401002101215 & 661295324291154816 & \checkmark & \checkmark & 5.4 & 0.975 & 0.87 & 5306.9 & 4.53 & 107.5 & 1.02 & 0.14 &  \\
160401002101323 & 661424074528836480 & \checkmark & \checkmark & 4.086 & 0.719 & 1.16 & 6065.8 & 4.36 & 125.1 & 0.85 & 0.2 &  \\
160113002301379 & 661342710667273856 & \checkmark & \checkmark & 7.354 & 1.635 & 1.09 & 4140.8 & 4.58 & 19.9 & 0.83 & -0.25 &  \\
160113002301292 & 664455118847781760 & \checkmark & \checkmark & 7.919 & 1.85 & 1.01 & 4269.7 & 4.7 & nan & 0.94 & -0.35 & 5 \\
160401002101052 & 661029384212747392 & \checkmark & \checkmark & 5.577 & 1.018 & 1.49 & 5205.8 & 4.55 & 74.9 & 0.87 & 0.16 &  \\
160401002101072 & 661244270018200192 & \checkmark & \checkmark & 3.64 & 0.656 & 0.94 & 6264.0 & 4.29 & 233.7 & 1.2 & 0.17 &  \\
160401002101275 & 661317250102374528 & \checkmark & \checkmark & 5.113 & 0.911 & 1.02 & 5462.6 & 4.52 & 117.5 & 0.99 & 0.2 &  \\
160401002101103 & 660998975844267264 & \checkmark & \checkmark & 4.917 & 0.894 & 1.58 & 5534.2 & 4.42 & 99.2 & 0.9 & 0.13 &  \\
160113002301267 & 664325170317925120 & \checkmark & \checkmark & 8.282 & 2.075 & 1.15 & 3660.7 & 4.42 & 10.1 & 0.87 & -0.82 &  \\
160401002101066 & 660986331460669312 & \checkmark & \checkmark & 4.261 & 0.872 & 2.13 & 5896.5 & 4.3 & 181.1 & 1.8 & 0.23 & 1,2,3 \\
160113002301384 & 661362055200029056 & \checkmark & \checkmark & 7.604 & 1.717 & 1.0 & 3232.7 & 4.29 & 17.3 & 1.59 & 0.55 &  \\
160401002101060 & 661222279785743616 & \checkmark & \checkmark & 5.617 & 1.034 & 1.3 & 5199.7 & 4.57 & 85.0 & 0.9 & 0.21 &  \\
160401002101155 & 661258872906952320 & \checkmark & \checkmark & 5.837 & 1.101 & 0.92 & 5062.0 & 4.58 & 69.9 & 0.9 & 0.17 &  \\
160401002101086 & 660995230632845184 & \checkmark & \checkmark & 6.212 & 1.235 & 1.11 & 5001.4 & 4.51 & 60.5 & 0.84 & 0.07 &  \\
160401002101341 & 661438913637806208 & \checkmark & \checkmark & 4.953 & 0.882 & 1.16 & 5573.5 & 4.5 & 102.4 & 0.92 & 0.23 &  \\
160113002301159 & 659750514750888192 & \checkmark & \checkmark & 7.091 & 1.543 & 4.56 & 4269.9 & 4.61 & 14.7 & 0.9 & 0.16 & 7 \\
160113002301387 & 661351850357695616 & \checkmark & \checkmark & 6.852 & 1.453 & 1.59 & 4506.3 & 4.6 & 31.5 & 0.76 & -0.07 & 7 \\
160401002101234 & 664387808120781440 & \checkmark & \checkmark & 4.884 & 0.855 & 0.83 & 5588.6 & 4.46 & 59.0 & 0.74 & 0.09 &  \\
160401002101225 & 664311392062657920 & \checkmark & \checkmark & 4.771 & 0.832 & 0.97 & 5713.2 & 4.47 & 88.2 & 0.83 & 0.17 &  \\
160401002101303 & 664437286143573760 & \checkmark & \checkmark & 5.756 & 1.07 & 0.89 & 5059.6 & 4.56 & 76.7 & 0.89 & 0.11 &  \\
160113002301161 & 659750892708009088 & \checkmark & \checkmark & 7.826 & 1.824 & 1.06 & 4639.4 & 4.83 & 6.3 & 0.96 & 0.41 &  \\
160401002101157 & 659756493345357056 & \checkmark & \checkmark & 3.36 & 0.636 & 1.94 & 6411.3 & 4.17 & 211.0 & 1.33 & 0.05 & 3 \\
160113002301331 & 661401431460014976 & \checkmark & \checkmark & 6.864 & 1.44 & 1.05 & 4390.5 & 4.58 & 25.6 & 0.78 & -0.15 &  \\
160401002101018 & 661325908755275648 & \checkmark & \checkmark & 5.257 & 0.922 & 1.28 & 5398.3 & 4.52 & 123.3 & 0.95 & 0.09 &  \\
160113002301079 & 661008527851529728 & \checkmark & \checkmark & 7.281 & 1.593 & 1.02 & 4463.5 & 4.66 & 30.9 & 0.83 & -0.23 &  \\
160113002301139 & 661210666194127360 & \checkmark & \checkmark & 7.505 & 1.687 & 1.02 & 4252.0 & 4.66 & 23.9 & 0.75 & -0.07 &  \\
160113002301060 & 661031480156796544 & \checkmark & \checkmark & 6.746 & 1.402 & 1.02 & 4573.2 & 4.6 & 35.5 & 0.79 & -0.01 &  \\
160401002101208 & 664302424170984576 & \checkmark & \checkmark & 4.82 & 0.852 & 0.79 & 5656.3 & 4.48 & 105.1 & 0.92 & 0.23 &  \\
160401002101110 & 661190153430426880 & \checkmark & \checkmark & 5.959 & 1.284 & 0.96 & 4926.7 & 4.46 & 55.9 & 0.9 & 0.03 & 1 \\
160401002101263 & 664327438061177344 & \checkmark & \checkmark & 3.658 & 0.635 & 0.78 & 6291.1 & 4.29 & 229.3 & 1.11 & 0.1 &  \\
160113002301168 & 659770340319909760 & \checkmark & \checkmark & 6.721 & 1.378 & 0.99 & 4703.9 & 4.65 & 24.6 & 0.78 & 0.1 &  \\
160113002301180 & 659777074828577024 & \checkmark & \checkmark & 8.631 & 2.35 & 1.02 & 3654.5 & 4.61 & 3.5 & 1.0 & -0.06 & 1 \\
160401002101221 & 661294705815866240 & \checkmark & \checkmark & 6.242 & 1.184 & 1.03 & 4848.7 & 4.61 & 77.4 & 0.95 & -0.01 &  \\
160113002301342 & 661409609077800960 & \checkmark & \checkmark & 7.35 & 1.64 & 1.02 & 3734.8 & 4.5 & 10.7 & 0.77 & 0.06 &  \\
160401002101241 & 664324684984105728 & \checkmark & \checkmark & 4.332 & 0.755 & 1.04 & 5935.7 & 4.4 & 82.7 & 0.76 & 0.19 &  \\
160401002101079 & 661293812462666240 & \checkmark & \checkmark & 5.629 & 1.041 & 0.85 & 5144.4 & 4.54 & 87.9 & 0.94 & 0.21 &  \\
160401002101335 & 661401225301656192 & \checkmark & \checkmark & 5.224 & 0.929 & 1.0 & 5421.2 & 4.52 & 109.5 & 0.95 & 0.18 &  \\
160113002301028 & 661251966599576576 & \checkmark & \checkmark & 6.828 & 1.404 & 1.01 & 4647.8 & 4.65 & 37.2 & 0.8 & 0.01 &  \\
160113002301138 & 661209566682500736 & \checkmark & \checkmark & 8.298 & 1.998 & 1.02 & 3770.4 & 4.45 & 14.6 & 0.73 & -1.17 & 7 \\
160401002101240 & 664337230586013312 & \checkmark & \checkmark & 4.114 & 0.739 & 0.95 & 5996.0 & 4.36 & 111.9 & 0.85 & 0.22 &  \\
160113002301102 & 660996020906770560 & \checkmark & \checkmark & 7.319 & 1.596 & 1.05 & 4213.2 & 4.63 & 18.6 & 0.89 & -0.14 & 7 \\
160113002301269 & 661310756111835392 & \checkmark & \checkmark & 6.506 & 1.3 & 0.93 & 4772.3 & 4.63 & 34.5 & 0.76 & 0.1 & 7 \\
160401002101255 & 664401960036152960 & \checkmark & \checkmark & 5.201 & 0.933 & 0.86 & 5356.7 & 4.5 & 35.4 & 0.71 & 0.15 &  \\
160113002301207 & 661294052980833792 & \checkmark & \checkmark & 6.49 & 1.292 & 1.05 & 4675.5 & 4.61 & 53.1 & 0.9 & -0.01 &  \\
160113002301204 & 661283199601699328 & \checkmark & \checkmark & 7.34 & 1.619 & 0.99 & 4102.0 & 4.56 & 28.6 & 0.84 & -0.46 &  \\
160401002101004 & 661331131435476608 & \checkmark & \checkmark & 5.507 & 1.177 & 1.49 & 5207.6 & 4.44 & 90.8 & 1.28 & 0.03 & 1,2,3 \\
160113002301059 & 661034155919816960 & \checkmark & \checkmark & 6.913 & 1.452 & 0.99 & 4507.0 & 4.63 & 38.9 & 0.86 & -0.17 &  \\
160113002301030 & 661246675199973504 & \checkmark & \checkmark & 6.82 & 1.419 & 1.02 & 4632.6 & 4.63 & 42.2 & 0.83 & -0.02 &  \\
160113002301170 & 661263305313199360 & \checkmark & \checkmark & 7.678 & 1.752 & 0.99 & 4053.6 & 4.62 & 19.3 & 0.88 & -0.37 &  \\
160113002301167 & 661270486498500352 & \checkmark & \checkmark & 8.113 & 1.908 & 1.04 & 3785.2 & 4.6 & 17.4 & 0.77 & -0.21 & 7 \\
160113002301105 & 661211147230556160 & \checkmark & \checkmark & 7.647 & 1.768 & 0.93 & 3905.5 & 4.41 & 27.0 & 0.72 & -0.65 &  \\
160113002301067 & 661222073627319296 & \checkmark & \checkmark & 8.401 & 2.016 & 1.05 & 3714.8 & 4.57 & 17.0 & 0.73 & -0.78 &  \\
160113002301109 & 661186683096853248 & \checkmark & \checkmark & 7.015 & 1.495 & 1.02 & 4378.5 & 4.62 & 19.7 & 0.86 & -0.06 & 7 \\
160113002301184 & 664281018054115968 & \checkmark & \checkmark & 7.063 & 1.504 & 0.97 & 4363.3 & 4.61 & 12.8 & 0.87 & -0.21 &  \\
160113002301224 & 664297064051880832 & \checkmark & \checkmark & 7.211 & 1.558 & 1.03 & 3993.2 & 4.52 & 17.6 & 0.88 & -0.53 &  \\
160401002101071 & 661242895630246016 & \checkmark & \checkmark & 3.846 & 0.673 & 0.96 & 6230.6 & 4.32 & 204.1 & 1.13 & 0.16 &  \\
160401002101064 & 661220969817342848 & \checkmark & \checkmark & 4.919 & 0.896 & 1.23 & 5557.3 & 4.41 & 106.8 & 0.9 & 0.09 &  \\
160113002301296 & 664429005446645376 & \checkmark & \checkmark & 8.04 & 1.891 & 0.96 & 3467.7 & 4.38 & nan & 0.87 & -0.37 &  \\
160401002101080 & 661292511090872960 & \checkmark & \checkmark & 4.844 & 0.86 & 1.08 & 5623.6 & 4.47 & 132.9 & 1.04 & 0.23 &  \\
160401002101092 & 661243273585807872 & \checkmark & \checkmark & 4.224 & 0.74 & 1.01 & 5973.9 & 4.4 & 156.8 & 1.09 & 0.21 &  \\
160401002101156 & 659755630055476992 & \checkmark & \checkmark & 5.685 & 1.07 & 0.93 & 5152.5 & 4.55 & 79.1 & 0.89 & 0.16 &  \\
160113002301320 & 661427063824818816 & \checkmark & \checkmark & 7.567 & 2.004 & 2.98 & 5339.0 & 4.65 & nan & 0.9 & -1.66 & 1,5 \\
160401002101088 & 661004400386393984 & \checkmark & \checkmark & 6.318 & 1.222 & 0.99 & 4843.5 & 4.61 & 29.0 & 0.82 & 0.11 &  \\
160113002301346 & 661408818803811200 & \checkmark & \checkmark & 7.15 & 1.559 & 0.92 & 4401.6 & 4.65 & 10.6 & 0.89 & 0.12 &  \\
160401002101164 & 661268940310277888 & \checkmark & \checkmark & 4.001 & 0.764 & 0.98 & 5979.7 & 4.24 & 150.3 & 1.16 & 0.11 & 1 \\
160401002101146 & 661264439184662016 & \checkmark & \checkmark & 4.218 & 0.742 & 0.82 & 5951.9 & 4.37 & 179.2 & 1.17 & 0.15 &  \\
\hline
\end{tabular}
\end{table*}

\begin{table*}
\addtocounter{table}{-1}
\tiny
    \setlength{\tabcolsep}{5pt}
    \renewcommand{\arraystretch}{0.9}
    \caption{cont.}
\begin{tabular}{llccccccccccc}
\hline
\texttt{sobject\_id} & \textit{Gaia} DR3 \texttt{source\_id} & C-G & H\&R & $M_\mathrm{G}$ & BP-RP & $RUWE$ & $T_\mathrm{eff}$ & $\log g$ & $S/N_\mathrm{CCD 3}$ & $\chi^2$ & [Fe/H] & Rejected \\
\hline
160401002101180 & 661277461525418752 & \checkmark & \checkmark & 4.177 & 0.738 & 0.95 & 5995.4 & 4.38 & 170.5 & 1.2 & 0.18 &  \\
160401002101173 & 661271238116475648 & \checkmark & \checkmark & 5.323 & 0.95 & 0.84 & 5338.0 & 4.52 & 96.3 & 0.94 & 0.13 &  \\
160401002101130 & 661216743570426240 & \checkmark & \checkmark & 3.801 & 0.671 & 0.89 & 6196.0 & 4.3 & 200.8 & 1.18 & 0.12 &  \\
160401002101126 & 661216369910684544 & \checkmark & \checkmark & 4.498 & 0.929 & 1.57 & 5350.0 & 4.32 & 172.0 & 1.29 & 0.03 & 1,3 \\
160401002101163 & 659758967246506880 & \checkmark & \checkmark & 5.307 & 0.948 & 0.85 & 5364.8 & 4.53 & 66.6 & 0.81 & 0.15 &  \\
160401002101118 & 661190015991474688 & \checkmark & \checkmark & 3.72 & 0.812 & 0.89 & 6257.6 & 4.14 & 185.5 & 1.96 & -0.43 & 1,2 \\
160401002101138 & 661207024061875456 & \checkmark & \checkmark & 4.13 & 0.726 & 0.93 & 5983.4 & 4.37 & 126.8 & 1.0 & 0.15 &  \\
160401002101201 & 664288748995228160 & \checkmark & \checkmark & 4.971 & 0.885 & 0.81 & 5587.0 & 4.5 & 114.0 & 1.0 & 0.21 &  \\
160401002101228 & 661315837054947840 & \checkmark & \checkmark & 5.932 & 1.11 & 0.99 & 4995.9 & 4.58 & 70.1 & 0.87 & 0.13 &  \\
160401002101249 & 661311752544249088 & \checkmark & \checkmark & 3.851 & 0.669 & 1.09 & 6210.9 & 4.33 & 203.0 & 1.07 & 0.07 &  \\
160401002101237 & 661314638762274176 & \checkmark & \checkmark & 6.087 & 1.153 & 0.92 & 4987.3 & 4.6 & 77.6 & 0.95 & 0.15 &  \\
160401002101188 & 661269043386545536 & \checkmark & \checkmark & 5.057 & 1.072 & 0.98 & 5028.8 & 4.38 & 108.1 & 1.59 & 0.19 & 1,2 \\
160401002101399 & 661253478428045568 & \checkmark & \checkmark & 4.582 & 0.814 & 1.18 & 5784.0 & 4.45 & 154.9 & 1.01 & 0.24 &  \\
160401002101268 & 664326922664482944 & \checkmark & \checkmark & 3.809 & 0.805 & 1.42 & 5727.4 & 4.16 & 196.8 & 1.17 & 0.14 & 1,3 \\
160401002101279 & 661419264165477504 & \checkmark & \checkmark & 3.635 & 0.657 & 1.01 & 6197.1 & 4.24 & 227.9 & 1.15 & 0.12 &  \\
160401002101332 & 661401122222444288 & \checkmark & \checkmark & 4.242 & 0.752 & 1.0 & 5981.1 & 4.4 & 152.3 & 0.99 & 0.23 &  \\
160401002101324 & 661302887731820416 & \checkmark & \checkmark & 4.995 & 0.896 & 2.32 & 5562.7 & 4.47 & 98.2 & 0.88 & 0.17 &  \\
160401002101032 & 661229564050262528 & \checkmark &   & 4.844 & 0.798 & 4.74 & 5840.6 & 4.53 & 146.5 & 1.06 & 0.23 & 1,3 \\
160401002101338 & 661402423595631232 & \checkmark &   & 6.072 & 1.063 & 5.92 & 5135.1 & 4.66 & 84.9 & 0.91 & 0.09 & 1 \\
\hline\\[-2pt]
    \end{tabular}
    
1: rejected for being a photometric binary.\\
2: rejected for being a spectroscopic binary.\\
3: rejected for $RUWE > 1.4$.\\
4: rejected for spectral fitting giving $\chi^2 > 3.0$.\\
5: rejected for large $v_\mathrm{r}$ error.\\
6: rejected for low $S/N$.\\
7: rejected after visual inspection.
\end{table*}

\end{appendix}

\end{document}